\documentclass[fleqn,usenatbib]{mnras}
\usepackage{newtxtext,newtxmath}

\usepackage[T1]{fontenc}

\DeclareRobustCommand{\VAN}[3]{#2}
\let\VANthebibliography\thebibliography
\def\thebibliography{\DeclareRobustCommand{\VAN}[3]{##3}\VANthebibliography}

\usepackage{graphicx}	
\usepackage{amsmath}	

\usepackage{amssymb}	
\usepackage{multirow}
\usepackage{graphicx}
\usepackage{gensymb}
\usepackage{enumerate}
\usepackage{algorithm}
\usepackage{algpseudocode} 
\usepackage{amsmath}
\usepackage{pgfplots}
\usepackage{threeparttable}
\usepackage{bbding}
\usepackage[figuresright]{rotating}
\usepackage{tabularx}
\usepackage[]{hyperref}

\title[Clustering methods on astronomical spectra]{Data mining techniques on astronomical spectra data. I : Clustering Analysis}

\author[H.Yang, C.Shi, J.Cai et al.]{
Haifeng Yang$^{1}$,
Chenhui Shi$^{1}$,
Jianghui Cai$^{1,2}$\thanks{E-mail: hfyang@tyust.edu.cn (HY); shichenhui@stu.tyust.edu.cn (CS); jianghui@tyust.edu.cn},
Lichan Zhou$^{1}$,
Yuqing Yang$^{1}$,
Xujun Zhao$^{1}$,
Yanting He$^{1}$ \and and
Jing Hao$^{1}$   
\\
$^{1}$School of Computer Science and Technology, Taiyuan University of Science and Technology, Taiyuan 030024, China\\
$^{2}$School of Computer Science and Technology, North University of China, Taiyuan 030051, China
}

\date{Accepted XXX. Received YYY; in original form ZZZ}

\pubyear{2022}

\begin{document}
\label{firstpage}
\pagerange{\pageref{firstpage}--\pageref{lastpage}}
\maketitle

\begin{abstract}
Clustering is an effective tool for astronomical spectral analysis, to mine clustering patterns among data. With the implementation of large sky surveys, many clustering methods have been applied to tackle spectroscopic and photometric data effectively and automatically. Meanwhile, the performance of clustering methods under different data characteristics varies greatly. With the aim of summarizing astronomical spectral clustering algorithms and laying the foundation for further research, this work gives a review of clustering methods applied to astronomical spectra data in three parts. First, many clustering methods for astronomical spectra are investigated and analysed theoretically, looking at algorithmic ideas, applications, and features. Secondly, experiments are carried out on unified datasets constructed using three criteria (spectra data type, spectra quality, and data volume) to compare the performance of typical algorithms; spectra data are selected from the Large Sky Area Multi-Object Fibre Spectroscopic Telescope (LAMOST) survey and Sloan Digital Sky Survey (SDSS). Finally, source codes of the comparison clustering algorithms and manuals for usage and improvement are provided on GitHub.
\end{abstract}

\begin{keywords}
methods: data analysis-techniques: spectroscopic-software: data analysis
\end{keywords}



\section{Introduction}
Astronomical spectral clustering has attracted increasing attention in astronomy. It helps us to analyse the birth, formation and evolution of the cosmic and astronomical objects \citep{2012A&A...545A..80F, 2012ApJ...750...91C}. Nowadays, rapid developing sky survey projects have obtained a large amount of  astronomical data, which poses great challenges for effective astronomical clustering \citep{2008SPIE.7019E..35L}. In view of the above challenges, a wide range of clustering methods (partition-based clustering methods, density-based clustering methods and so on) \citep{10.1145/3522592, 2012Learning, 2010Data, 1997Data, LAM20141115} have been applied to diverse astronomical tasks. However, clustering methods perform differently on various data sources, making it difficult to evaluate the performance and determine their applicable scenarios.

To explore the advantages of each type of clustering method, we investigate a large number of clustering methods applied to astronomical spectra data and analyse their applications, core ideas, merits and caveats. We then compare the performance of classical clustering methods on unified datasets and give objective appraisals. The datasets are constructed considering four aspects, as follows.

(i) Different data types of some spectral feature extraction methods. Four types of data including 1D spectra, PCA (Principal Component Analysis, a widely used dimensionality reduction method) features of spectra, line indices and stellar parameters are used to test different methods.\\
(ii) The quality of spectra. Three signal-to-noise ratios (S/Ns) are used to analyse the robustness of clustering algorithms.\\
(iii) Data volume. Four sizes of spectral datasets are constructed including 8000, 20000, 40000 and 80000, however some algorithms cannot run on the size of 80000.\\
(iv) Outliers detection. We know that some clustering methods can be used to detect outliers, so datasets containing normal spectra and rare objects are constructed to test the ability of some algorithms to detect outliers.

After each experiment, an objective analysis is given. Moreover, experimental source codes in this paper and a brief manual about usage of the source codes are available to readers.

This work is organized as follows: In Section 2, we briefly describe the clustering methods from applications and theory analysis. In Section 3, experiments on four astronomical tasks - A/F/G/K stars classification, star/galaxy/quasar classification, subclasses of A-type star classification and outliers detection - are carried out. Section 4 puts forward source codes of the aforementioned experiments and a manual about the usage of the source codes. Finally, a discussion and our future work are presented in Section 5.

\section{Investigation of Clustering Methods on Astronomical Spectra Data}

As a prevalent task, clustering has been applied in many fields, such as imaging processing  \citep{10.5555/839284.841388, forsyth:hal-01063327}, social networks \citep{2011Introduction, everton_2012}, finance security,  biological fields  \citep{Kaplan2013-KAPPOA, 3ac78af0492d473c8e1492890c612e8e}, and others  \citep{2012Multiple, fotheringham1998geographically, 1985sdae.book.....U, Connell98learningprototypes}. It attempts to divide data into different groups according to certain criteria. Data in the same group appears to be more similar than data in different groups. The classical clustering process can be divided into the following steps \citep{2015A}.

\begin{enumerate}[(1)]
\item Data preprocessing: extract and choose the most representative features;
\item Clustering: design suitable clustering algorithms and train models with unlabelled data to cluster the real data;
\item Result analysis: explain the clustering result and evaluate the clustering method.
\end{enumerate}

Clustering methods play a vital role in astronomical spectral data analysis  \citep{Rebbapragada_2008, 1996A&A...311..145B, 2012MNRAS.427.1153S, 2000ApJ...532.1215S, Li.18.P-Cygni}. Traditional clustering algorithms can be divided into partition-based, density-based, hierarchical, grid-based, and model-based. In recent years, lots of new clustering algorithms have appeared, like fuzzy theory-based clustering algorithm, kernel-based clustering algorithm \citep{8.2015arXiv151003547C}, etc. The next subsections will introduce the above typical clustering algorithms, considering applicable scenarios, core ideas, merits, and caveats.

\subsection{Partition-Based Clustering Algorithm}



\begin{table*}
\caption{Investigations of partition-based clustering algorithms on astronomical spectra.}
\label{table_partitioned_related}
\resizebox{\linewidth}{!}{
\begin{tabular}{lll}
\hline
Merits & Caveats & References \\ \hline
\multirow{6}{*}{\begin{tabular}[c]{@{}l@{}}Some improved partition-based methods classify \\ boundary spectra effectively.\\ K-means is simple and fast, it can deal with \\ large dataset.\end{tabular}} & \multirow{6}{*}{\begin{tabular}[c]{@{}l@{}}In flux space, clustering result is sensitive \\ to the number of clusters.\\ K-means can not cluster arbitrary shape of groups.\\ On classifying light curves, the classification \\ algorithm is not robust because the data amount is small.\end{tabular}} & \citealt{8732318}, 
\citealt{Kheirdastan_2016},\\
 &  & {\citealt{2014A&A...565A..53O}},\\
 &  & \citealt{2020JApA...41...15C}, 
\citealt{2010ApJ...714..487S},\\
 &  & \citealt{2013ApJ...763...50S},\\
 &  & {\citealt{2018A&A...612A..98G}},
\citealt{10.1093/mnras/stw1228},\\
 &  &  \citealt{10.1093/mnras/staa978}\\
\hline
\multirow{1}{*}{On detecting special objects, K-means is also efficient.} & \multirow{1}{*}{} & \citealt{2018RAA....18...73C}, \citealt{2011ApJ...743...77M}\\
\hline
\multirow{6}{*}{\begin{tabular}[c]{@{}l@{}}Space and morphological structure can be \\ identified by K-means.\\ After clustering some data, lots of meaningful \\ information can be analysed from the clustering result.\end{tabular}} & \multirow{6}{*}{\begin{tabular}[c]{@{}l@{}}Some distinct features in colour space which \\ are apparent in visual examinations can not be \\ recognized as clusters in colour space.\end{tabular}} & \citealt{Rahmani_2018},
\citealt{7984705}\\

 &  & \citealt{Hogg_2016}, \citealt{2012ApJ...756..163S},\\
 &  & \citealt{2018ApJ...861...62P}, \citealt{Rubin_2016},
\\
 &  & \citealt{2016MNRAS.457..362B}, \citealt{2017MNRAS.469.3374C},\\

 &  & \citealt{10.1093/mnras/staa978},{\citealt{2015A&A...577A..47B}},\\
  &  & \citealt{2015MNRAS.447.1638M},\citealt{M82017}\\
   \hline
\end{tabular}
}
\end{table*}

Partition-based clustering methods classify the data into different clusters by finding optimal cluster centers and K-means is a widely used partition-based algorithm because of its simplicity and efficiency. Astronomical investigations based on partition-based clustering algorithm are shown in Table \ref{table_partitioned_related}. 

\citet{2010ApJ...714..487S} used K-means to classify galaxies from SDSS DR7. However, some spectra appeared to be between classes, so they proposed a K-means-based method to identify marginal galaxies. \citet{2018A&A...612A..98G} analysed the application of K-means to massive Apache Point Observatory Galactic Evolution Experiment (APOGEE) high resolution spectra. The results showed that K-means was able to separate the bulge and halo populations and distinguish dwarfs, subgiants, red clump (RC) stars, and red giant branch (RGB) stars. Clustering of star, galaxy, quasar and subclasses have been carried out extensively to analyse physical properties, chemical abundances, boundary spectra, etc ( \citealt{2018RAA....18...73C, 7984705, Hogg_2016, Kheirdastan_2016, 2011ApJ...743...77M},  \citealt{2013ApJ...763...50S}).  Furthermore, clustering results contain valuable information, such as structure of Galaxy and evolutionary stages of the Universe, which provides convenience for astronomers  \citep{Rahmani_2018, 2010ApJ...714..487S, 2015MNRAS.447.1638M, 2012ApJ...756..163S, M82017, 2016MNRAS.457..362B}.

Initialization of clusters centers and the number of clusters are two factors that affect K-means results greatly, and many improved methods have been proposed \citep{2020JApA...41...15C, 2014A&A...565A..53O, 8732318}. In order to optimize clustering, \citet{2020JApA...41...15C}, \citet{2018A&A...612A..98G} and \citet{8732318} used new approaches to initialize cluster centers and results suggested that clustering with better initial centers are of high quality. \citet{YANG2022117018} optimized K-means by influence space and applied ISBFK-means on astronomical spectral data of LAMOST. Results presented a good performance on stability and speed.

Another common application is special objects identification (e.g. rare objects, light curves, star clusters) and retrieval tasks  \citep{7984705, 2018ApJ...861...62P, 2017MNRAS.469.3374C, 10.1093/mnras/stw1228, 10.1093/mnras/staa978, Rubin_2016, chemical_abundance, 2015A&A...577A..47B}, and the detected special objects are a  significant supplement to the astronomy category. \citet{Jin_2022} used K-means clustering to successfully divide the confirmed exoplanets from NASA dataset into different clusters.

In astronomical applications, effective feature selection not only affects clustering accuracy and time, but is also closely related to celestial properties. \citet{2018RAA....18...73C} and \citet{7984705} adopt line indices as spectral features to improve data quality and results indicated that they were consistent with internal properties of stars. Optimal feature extraction (color space, chemical pattern, etc.) can provide more information for clustering without additional prior knowledge  \citep{chemical_abundance, 2015A&A...577A..47B, Hogg_2016, 2010ApJ...714..487S, 2016MNRAS.459.1659T}.

\begin{figure}
\centering
\includegraphics[width=\columnwidth]{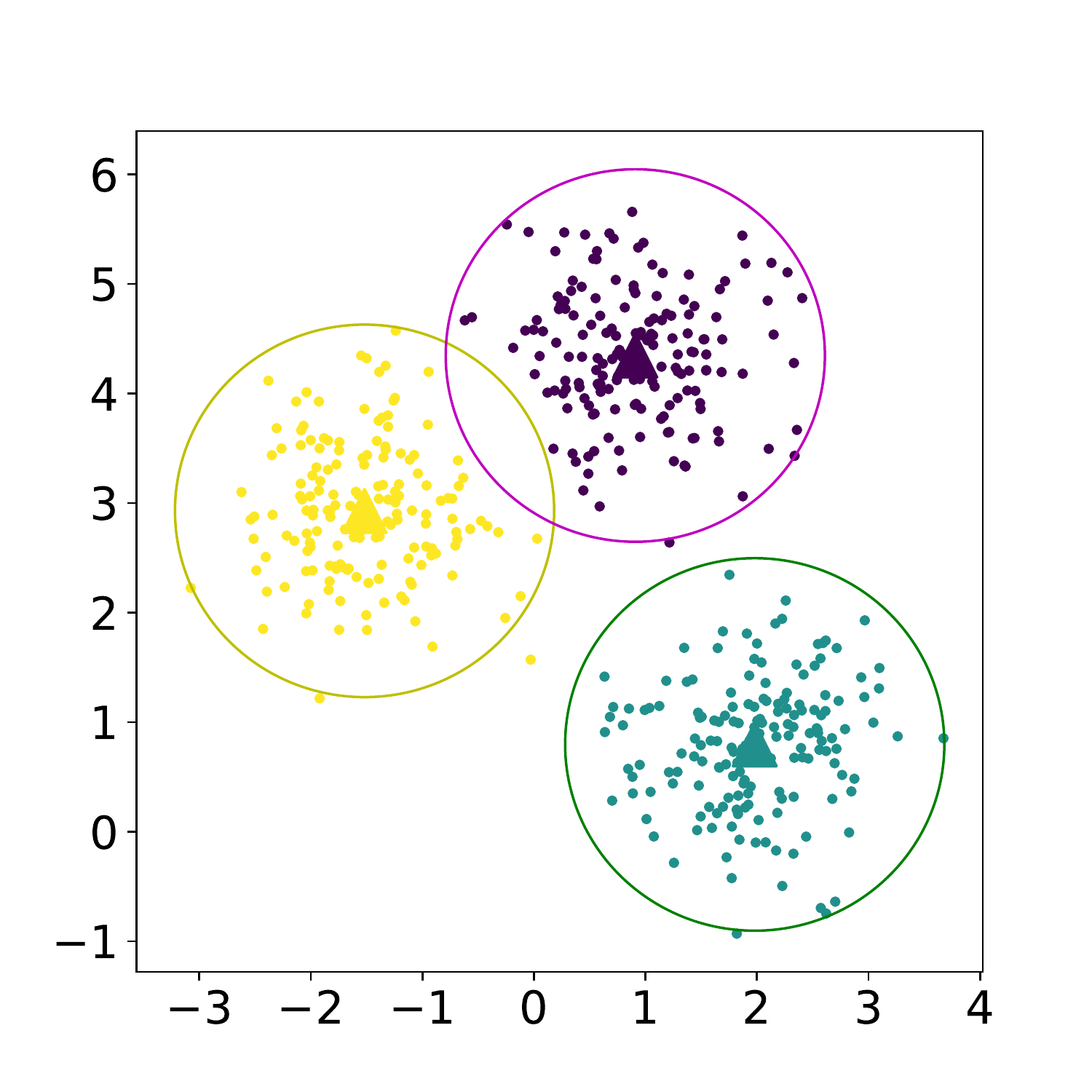}
\caption{Structure of K-means. Yellow, purple and blue are three groups. Triangles present cluster centers of each group.}
\label{fig:kmeans_structure}
\end{figure}

Among partition-based clustering algorithms, K-means and K-mediods are two typical methods; variants such as K-means++, intelligent K-means, genetic K-means, K-modes, and kernel K-means also perform well. The goal of partition-based clustering algorithms is that data in the same group appears more similar than those in different groups. The pseudo-code about partition-based clustering methods is given in Algorithm \ref{Algorithm_1} and Fig. \ref{fig:kmeans_structure} shows the structure of K-means.

\begin{algorithm}  
  \caption{Partition-Based Clustering Algorithm}  
  \label{Algorithm_1}  
  \begin{algorithmic}[1]
    \Require  
    Data;
    K: the number of clusters; 
    \Ensure  
   Data-Labels
     \State Determine initial centers by random or other methods.
    \While{not reach convergence}  
      \State Assign each sample to the most similar cluster center.
      \State Update cluster centers with certain criterion.
    \EndWhile
  \end{algorithmic}  
\end{algorithm}

K-means is a typical partition-based algorithm which has been widely applied to many fields because of its advantages of simple implementation, intuitive design and linear computational cost. It is also robust when tackling high dimensional data and large dataset, making it an attractive tool for star classification, multi-band photometric analysis, supernova light curve surveys, and so on. However, the number of clusters and initial cluster centers selection are two key factors which influence the performance of K-means. The optimal solution to the above problems is still in the continuous exploration stage. Most partition-based methods measure the similarity between samples by Euclidean distance, which is not so reliable when the feature of data is complex, so it is important to choose an appropriate similarity measure for the data under study. Another shortcoming of partition-based methods is that they cannot detect clusters with arbitrary shape, and only round clusters can be recognized.

\subsection{Density-Based Clustering Algorithm}

\begin{table*}
\caption{Investigations of density-based clustering algorithms on astronomical spectra.}
\label{table_DBSCAN_related}
\resizebox{\linewidth}{!}{
\begin{tabular}{lll}
\hline
Merits & Caveats & References \\ \hline
\multirow{3}{*}{\begin{tabular}[c]{@{}l@{}}DBSCAN can be used to classify objects \\ on photometric and spectra data.\\ CFSFDP is efficient on detecting outliers.\end{tabular}} & \multirow{3}{*}{DBSCAN rarely clusters spectra directly.} & {\citealt{2020A&A...633A.154L}},\\
 &  & \citealt{DengT17}, \citealt{9049419},\\
 &  & \citealt{2016MNRAS.463.2939T}\\
 \hline
\multirow{2}{*}{\begin{tabular}[c]{@{}l@{}}DBSCAN can be used to study stellar history \\ by clustering chemical abundances.\end{tabular}} & \multirow{2}{*}{} & \citealt{10.1093/mnras/stz1260},\\
 &  & \citealt{2018ApJ...860...70C}\\
\hline
\multirow{6}{*}{\begin{tabular}[c]{@{}l@{}}DBSCAN can cluster data with \\ arbitrary shape and size.\\ Objects in 3D space can be detected and analysed \\ by density-based clustering algorithms, like DBSCAN \\ and OPTICS.\end{tabular}} & \multirow{6}{*}{\begin{tabular}[c]{@{}l@{}}Determination of lower spatial-density features \\ need be improved.\end{tabular}} & 
{\citealt{2019A&A...628A.123Z}}, \citealt{2013PhRvD..88d3006C},\\
 &  & \citealt{Dehghan_2014},\\
 &  & \citealt{2020ApJ...898...80Y}, \citealt{2021MNRAS.501.4420O},\\

 &  & \citealt{10.1093/mnras/sty1370}, \citealt{2014Membership},\\
 &  & \citealt{2015MNRAS.452.3159A},\\
 &  & {\citealt{2017A&A...599A.143S}}\\
 \hline
\multirow{3}{*}{\begin{tabular}[c]{@{}l@{}}DBSCAN can help to reduce noise.\\ TAD algorithm can analyse the spatial-temporal \\ density of complex or special trajectories efficiently.\end{tabular}} & \multirow{3}{*}{\begin{tabular}[c]{@{}l@{}}The efficiency will decrease when \\ the dataset is too large.\end{tabular}} & \multirow{3}{*}{\begin{tabular}[c]{@{}l@{}}\citealt{seo2020applications}, \citealt{YANG2020112846}\end{tabular}} \\ \\ \\ \hline
\end{tabular}
}
\end{table*}

At present, density-based clustering algorithms are promising for exploring the universe. Table \ref{table_DBSCAN_related} depicts applications of density-based methods.

Density-based clustering methods on spectra data are used to classify celestial objects and they are more frequently used in chemical abundance, photometric data, etc. \citet{2020A&A...633A.154L} classified star/galaxy/quasar from photometric data with hierarchical DBSCAN and the accuracy was high. \citet{2016MNRAS.463.2939T} used DBSCAN and DENCLUE to classify galaxies in the Galaxy Zoo dataset. \citet{Ordonez_2022} combined DBSCAN and nearest neighbor algorithm to analyse quasars; the results showed that gravitational interaction may cause more low redshifts quasars clustering.

Another common application of density-based methods is the detection and analysis of spatial structure of objects, like galaxy clusters, molecular clouds and others \citep{2013PhRvD..88d3006C}. \citet{2021MNRAS.501.4420O} used OPTICS to identify the structure of galaxy/halo and the low-density data points can be identified effectively. HDBSCAN was applied to systematically find members which are up to 50 pc from the cluster centers \citep{Tarricq_2022}, and GMM was used to fit the two dimensional distribution of  cluster members in this work; these two methods helped to find 389 open clusters. \citet{2014Membership} determined membership of open cluster NGC 188 and found that DBSCAN could efficiently select possible member stars in 3D kinematic space without making any assumptions about the distribution of cluster or field stars. \citet{Castro_Ginard_2022} used DBSCAN to identify the statistical overdensities of stellar objects, which help to find more open clusters. The structure, kinematics, and ages of the young stellar populations in the Orion region can also be analysed by DBSCAN \citep{2019A&A...628A.123Z}. Besides this, many researchers committed to multi-space, which is a new perspective of the universe, and software engineering also provides many useful tools for clustering analysis.

Some density-based clustering algorithms can search for outliers and cluster centers, such as Clustering by fast search and find of density peaks (CFSFDP).  \citet{9049419} used CFSFDP to detect special samples in the massive low S/N spectra.

In astronomical tasks, 'data level' and 'algorithm level' are two major approaches to achieve desirable clustering performance. \citet{DengT17} and \citet{2017Optimizing} found optimal parameters and distance measurement methods to obtain satisfactory results from 'algorithm level'. Astronomical spectra can be represented in many forms, such as chemical abundance, line indices, color space and other features. Efficient feature extraction is a particularly interesting way to improve performance from 'data level' \citep{2020MNRAS.496.5101P, YANG2020112846, 10.1093/mnras/stz1260, 2014Membership, 2020A&A...633A.154L, 2015RAA....15.2193G, Traven_2017, 2016MNRAS.463.2939T}. Many new density-based clustering algorithms have also been proposed to improve the efficiency and robustness of clustering \citep{YANG2022414, LIANG2022116410}.

Density-based clustering algorithms group data according to region density. High density areas are defined as classes sharing similar information. Always, they require two input parameters: the density radius (Eps) and the minimum number of data (MinPts) to build a cluster. DBSCAN, DENCLUE, OPTICS, CFSPDP are classic density-based methods. In addition, some optimization approaches have been explored to improve clustering performance, such as manhattan distance density algorithm (MD-DBSCAN) \citep{DengT17}
, normalised euclidean distance (NED-DBSCAN) and so on.  
The procedure of density-based clustering algorithm is presented in Algorithm \ref{Algorithm_2}.

\begin{algorithm}
  \caption{Density-Based Clustering Algorithm}
  \label{Algorithm_2}
  \begin{algorithmic}[1]
    \Require  
    Data;
    Neighbor radius $\epsilon$;
    Radius density $N_{pts}$;
    \Ensure  
    Data-Labels
    \While{$X_{i}$ not none }
     \While{queue not none}  
	  \If{the number of data in $\epsilon > N_{pts}$}
       \State $X_{i} \gets core\,point$
      \Else
       \State $X_{i} \gets outlier$
      \EndIf
      \State Assign core point and it's neighbour points to cluster i.
      \State Add neighbour points to a queue.
     \EndWhile
    \EndWhile
  \end{algorithmic}   
\end{algorithm}
Density-based clustering algorithms have been frequently used for scientific data analysis in various fields. It can identify irregular groups rather than being limited to globular shapes (e.g. K-means.) and the number of clusters need not be determined in advance. In astronomy, density-based clustering algorithms provide crucial information for candidates identification, substructure analysis and category supplement.

\subsection{Model-Based Clustering Algorithm}

\begin{table*}
\caption{Investigations of model-based clustering algorithms on astronomical spectra.}
\label{table_model_related}
\resizebox{\linewidth}{!}{
\begin{tabular}{llll}
\hline
Method & Merits & Caveats & References \\ 
\hline
\multirow{9}{*}{SOM} & \multirow{9}{*}{\begin{tabular}[c]{@{}l@{}}SOM maps high-dimension spectra into 2D graph, \\ and similar spectra are close in graph.\\ The ordering of spectra on the map correspond \\ to the ordering of physical properties.\\ SOM can be used to detect special objects.\\ SOM method with interactive interface can greatly \\ improve the efficiency of spectral analysis.\end{tabular}} & \multirow{9}{*}{\begin{tabular}[c]{@{}l@{}}SOM is not stable on clustering.\\ On the 2D map, there is no clear clustering \\ result.\end{tabular}} & \citealt{8282521} \\
 &  &  & \citealt{Rahmani_2018} \\
 &  &  & \citealt{ORDONEZ2012204} \\
 &  &  & \citealt{article_application} \\
 &  &  & {\citealt{2017A&A...597A.134M}} \\
 &  &  & \citealt{10.1093/mnras/stw1228} \\
 &  &  & \citealt{som.article.outlier} \\
 &  &  & \citealt{7849952} \\
 &  &  & {\citealt{2012A&A...547A.115I}} \\
 \hline
\multirow{6}{*}{GMM} & \multirow{6}{*}{\begin{tabular}[c]{@{}l@{}}GMM can detect object with a Gaussian distribution \\ in space, like galaxy(cluster).\\ GMM distinguish different kinds of gamma-ray bursts.\end{tabular}} & \multirow{6}{*}{\begin{tabular}[c]{@{}l@{}}When the data set is large, GMM will \\ take a long time to run.\end{tabular}} & \citealt{2019MNRAS.488.4106I} \\
 &  &  & \citealt{shin2018detecting} \\
 &  &  & \citealt{2017MNRAS.469.3374C} \\
 &  &  & \citealt{2.gao2020discovery}, \citealt{6.2019MNRAS.486.4823T}\\
 &  &  & \citealt{4.2012MNRAS.426.3435S} \\

 &  &  & \citealt{7.2018MNRAS.475.1708A}\\
 \hline
\end{tabular}
}
\end{table*}

Investigations of model-based clustering algorithm on astronomical spectra are shown in Table \ref{table_model_related}. Model-based methods mainly contain neural network clustering algorithms and probability-based model clustering algorithms. Both of them have already been developed for star, galaxy and peculiar objects clustering \citep{8282521,10.2017arXiv171111101K}. 

Self-organizing map (SOM) is a special type of neural network which maps the data into a two-dimensional grid. It is always used to visualize the distribution of data, and is similar to the dimensionality reduction visualization methods like t-SNE \citep{van2008visualizing,wattenberg2016use} and UMAP \citep{1.umap}. \citet{article_application} used SOM to map 158 spectra into a 13 $\times$ 13 grid and made a classification. The two-dimensional map contains aggregated information of the data and the gaps in the map can be regarded as the separation of different clusters. \citet{ORDONEZ2012204} used SOM to distinguish different objects and divided the grid with Fuzzy C-Means (FCM) which is a fuzzy clustering algorithms based on K-means. \citet{Rahmani_2018} trained SOM using a set of galactic templates which covering the wavelength range from far ultraviolet to near-infrared. Compared with other methods, the spectra of galaxies grouped together by the SOM are more similar. And the order of sample categories on the map corresponded to the order of physical properties.
That is, in addition to representing the type, the order of samples on the map can also represent many other information, like physical parameters distribution. For clustering, researchers divided the 2D map into many clusters \citep{7849952} or mapped the data into 1 $\times$ n where n is the number of categories. For easier use of SOM, some researchers built interactive programs to analyse objects efficiently \citep{2012A&A...547A.115I, 2017A&A...597A.134M}. Many other applications like outlier analysis \citep{som.article.outlier}, exploration of the spectroscopic diversity of Type Ia supernovae \citep{10.1093/mnras/stw1228} can also be handled or assisted by SOM.

Gaussian mixture model (GMM) is one of the most commonly used probability models. It uses multiple multidimensional Gaussian distributions to fit data, and each distribution represents a cluster \citep{reynolds2009gaussian}. In astronomy, GMM can be used in two ways, one is to cluster data based on features, such as 1D spectra, chemical element abundance, etc. The other is to use Gaussian distributions to fit targets directly, such as star clusters and molecular clouds.     \citet{2019MNRAS.488.4106I} reduced X-ray spectra which are obtained from Chandra X-ray Observatory from Tycho’s supernova remnant using a variational autoencoder (VAE) and clustered the low-dimension features with GMM to analyse spatial structures in Tycho’s supernova remnant. \citet{shin2018detecting} used an infinite GMM to detect variability in large amounts of astronomical time series data. GMM was also used to classify different kinds of gamma-ray bursts \citep{2017MNRAS.469.3374C, 6.2019MNRAS.486.4823T, 7.2018MNRAS.475.1708A}. \citet{2.gao2020discovery} used PCA+GMM to identify tidal tail around the old open cluster NGC 2506 based on Gaia-DR2. GMM can also detect and analyze objects which shaped like Gaussian distribution. \citet{4.2012MNRAS.426.3435S} used a mixture model to probe gas motions in the intra-cluster medium and showed that the mixture parameters can be accurately constrained by Astro-H spectra.  \citet{Wagenveld_2022} proposed a probabilistic HzQ selection method which used GMM to obtain likelihoods and used a Bayesian framework for poster probabilities of HzQ and contaminating sources. It is useful to find more complete HzQ samples.

\begin{figure}
\centering
\includegraphics[width=\columnwidth]{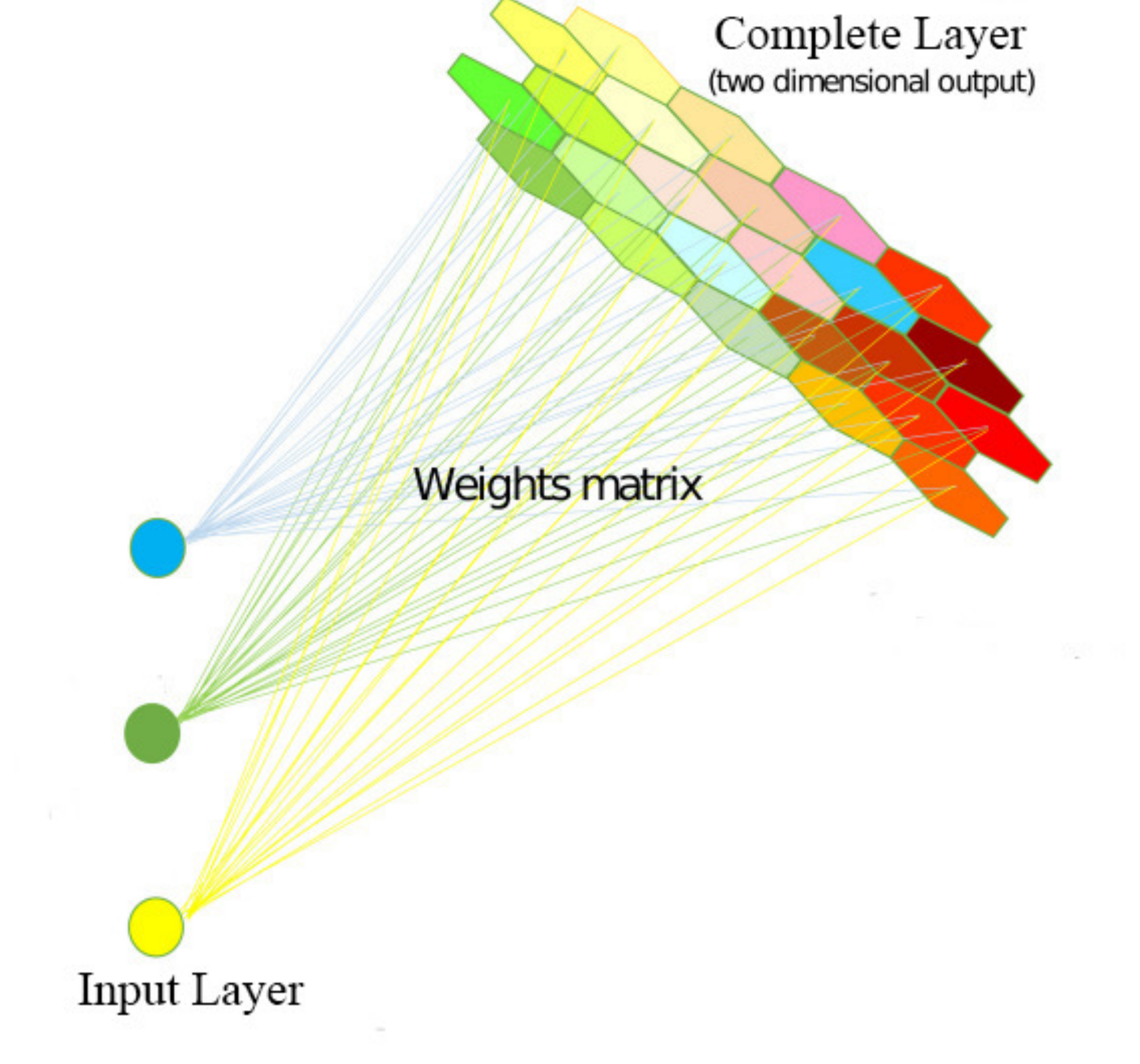}
\caption{Structure of SOM. Three circles are data of input layer. The top right complete layer generates two dimensional output. Weights matrix includes weights between input layer and complete layer.}
\label{fig:som_structure}
\end{figure}

SOM and GMM are the most widely used model-based algorithms. Compared with K-means, they can obtain more valuable information.
The idea of SOM is to establish a mapping between high and low dimensions. An illustration of SOM structure \citep{2018A&A...612A..98G} is given in Fig. \ref{fig:som_structure}. Algorithm \ref{Algorithm_3} is the main idea of SOM.

\begin{algorithm} 
  \caption{Self-Organizing Map}  
  \label{Algorithm_3}  
  \begin{algorithmic}[1]
    \Require  
    Data; Neural units weights; Leaning rate;
    \Ensure  
    Data-The mapping position in the 2D grid   
    \While{neural units weights are not stable}     
      \State Calculate winner neural unit with neural weights and learning rate.
      \State Update the neural units weights.     
    \EndWhile
  \end{algorithmic}   
\end{algorithm}

Neural network is widely used \citep{1385384} because it can learn complex non-linear relationships. But conventional neural network model has many parameters and consumes long time to run. Besides, it is difficult to explain the results. In astronomical research, SOM also has the ability to learn complex non-linear relationships. At the same time, it can visualize the clustering results and alleviate the unexplainability of the model to a certain extent. Numerous studies have shown that SOM can be applied to various astronomical tasks.

\begin{figure}
\centering
\includegraphics[width=\columnwidth]{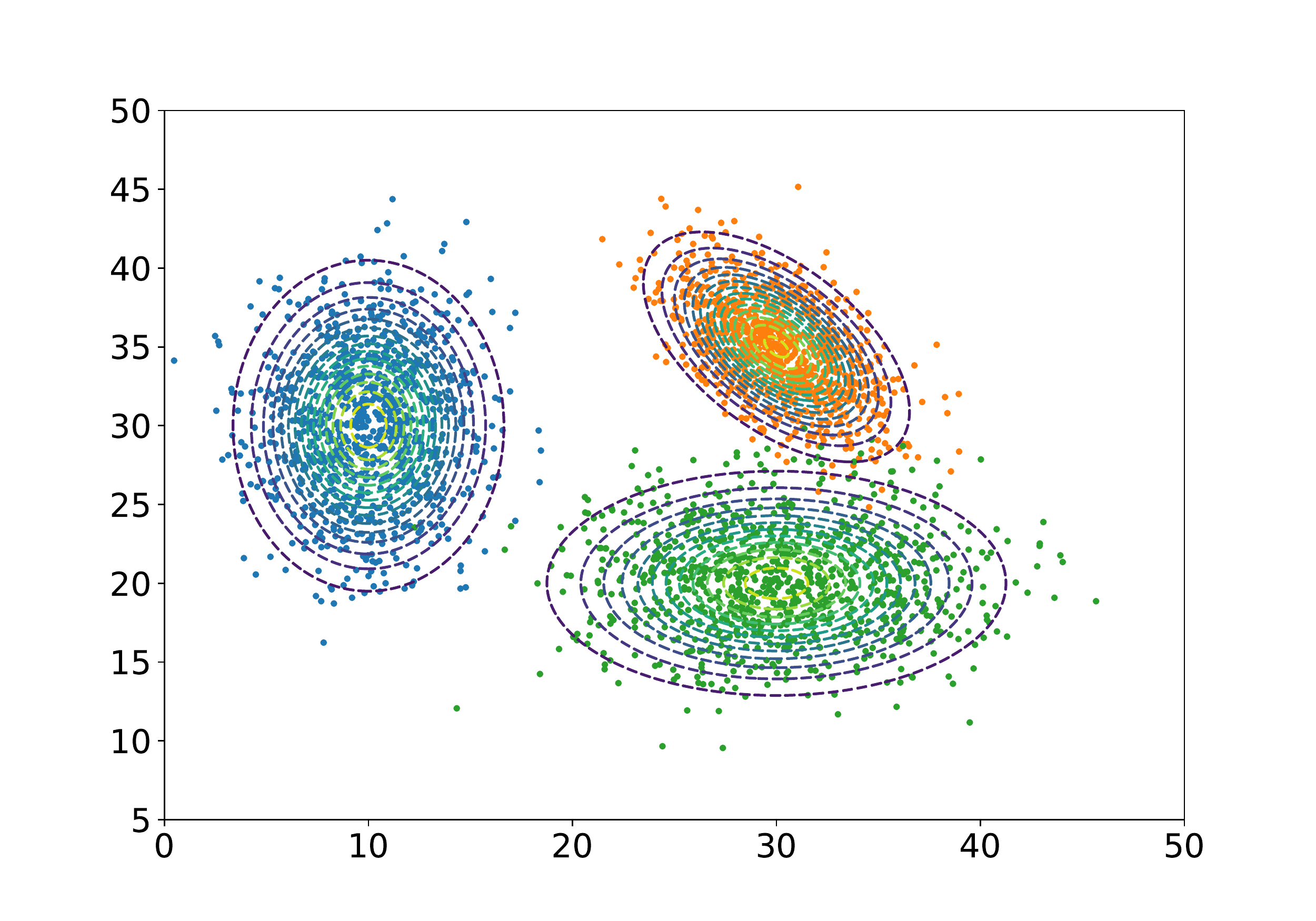}
\caption{Structure of GMM. Yellow, green and blue are three gaussian distributions and each gaussian distribution presents one group.}
\label{fig:gmm_structure}
\end{figure}

GMM is a mixture of several Gaussian distributions. Its advantage is that GMM gets rid of the limitation of measuring similarity based on Euclidean distance and fits each dimension of data with a Gaussian distribution. Data in the same independent Gaussian distribution belong to the same cluster (Fig. \ref{fig:gmm_structure}). It iteratively updates the parameters of the multidimensional distributions through EM (Expectation-Maximum) algorithm \citep{ng2012algorithm}, and Algorithm \ref{Algorithm_4} shows the procedure of GMM.

\begin{algorithm}
  \caption{Gaussian Mixture Model}  
  \label{Algorithm_4}  
  \begin{algorithmic}[1]

    \Require  
    Data; The number of Gaussian model;
    \Ensure  
    Data-Labels
    \State Determine initial parameters of models by some methods.
    \State Calculate the probability that the data belongs to each model.
    \While{probability matrix not stable}     
      \State E-Step: calculate the probability matrix of data.
      \State M-Step: update the parameters again according to probability matrix.     
    \EndWhile
  \end{algorithmic}
\end{algorithm}

Among the model-based clustering algorithms, GMM is a relatively classic algorithm. It gives the probability of classification rather than just the labels. But it needs enough samples to fit a accurate distribution, and needs a long iteration time to find optimal parameters.

\subsection{Hierarchical Clustering Algorithm}
The main idea of hierarchical clustering algorithm \citep{781637} is to classify data into different groups with a hierarchical tree, including hierarchical agglomerative method and hierarchical divisive method. These two types of methods are inverse processes, that is, bottom-up and top-bottom. Balanced Iterative Reducing and Clustering Using Hierarchies (BIRCH) \citep{10.1145/233269.233324}, Clustering Using Representative (CURE) \citep{10.1145/276304.276312} and Chameleon \citep{781637} are commonly used hierarchical clustering methods. \citet{fielding2022classification} used hierarchical clustering (agglomerative clustering) and partition-based clustering (K-means, fuzzy c-means) on extracted features of galaxy for clustering.
Results showed that hierarchical clustering performed best.
Hierarchical agglomerative clustering and hierarchical divisive clustering are presented in Algorithm \ref{Algorithm_5} and Algorithm \ref{Algorithm_6} \citep{2015A}.

\begin{algorithm}  
  \caption{Hierarchical Agglomerative Clustering Algorithm}  
  \label{Algorithm_5}  
  \begin{algorithmic}[1]
    \Require  
    Data;
    \Ensure  
    Data-Labels
    \State Regard each data as an individual cluster.
    \While{more than one cluster left }     
      \State Calculate similarity between every cluster.
      \State Emerge the most similar clusters with minimal distance.     
    \EndWhile
  \end{algorithmic}   
\end{algorithm}

\begin{algorithm} 
  \caption{Hierarchical Divisive Clustering Algorithm}  
  \label{Algorithm_6}  
  \begin{algorithmic}[1]
    \Require  
    Data;
    \Ensure  
    Data-Labels
    \State Suppose all data stands for one cluster.
    \While{not every data is a cluster}     
      \State Calculate similarity between every cluster.
      \State Divide the most dissimilar cluster into two clusters.     
    \EndWhile
  \end{algorithmic}   
\end{algorithm}

Hierarchical clustering can find the hierarchical relationships between classes without determining the number of clusters in advance. But it has high computational complexity and is sensitive to outliers. Therefore, hierarchical clustering is also usually combined with other algorithms to improve the clustering performance. For example, researchers combined SOM and ELM with hierarchical clustering to improve the accuracy of stellar clustering \citep{ORDONEZ2012204, bu2016elm}.

\subsection{Fuzzy Theory-Based Clustering Algorithm}
Clustering algorithms can be divided into two categories, namely hard clustering and soft clustering. Fuzzy theory-based clustering algorithm is a typical soft clustering. The basic idea of this clustering algorithm is that data can be assigned to several clusters with various probabilities \citep{2015A}. Fuzzy C-means (FCM), Fuzzy C-shells (FCS) and Mountain method (MM) are classic clustering algorithms. Main processes of FCM are presented in Algorithm \ref{Algorithm_7}.

\begin{algorithm} 
  \caption{Fuzzy Theory-Based Clustering Algorithm}  
  \label{Algorithm_7}  
  \begin{algorithmic}[1]
    \Require  
    Data; Fuzzy matrix U;
    \Ensure  
    Data-Labels   
    \While{fuzzy matrix U not stable}     
      \State Calculate cluster centers.
      \State Update fuzzy matrix U.    
    \EndWhile
  \end{algorithmic}   
\end{algorithm}

\subsection{Grid-Based Clustering Algorithm}

Grid-based clustering algorithms divide the space into a number of cells. Their processing time is independent of the number of processed objects and the processing speed is fast. Then clustering is based on the grid structure \citep{SAXENA2017664}. Algorithm \ref{Algorithm_8} is the main processes of them.

\begin{algorithm} 
  \caption{Grid-Based Clustering Algorithm}  
  \label{Algorithm_8}  
  \begin{algorithmic}[1]
    \Require  
    Data;
    \Ensure  
    Data-Labels  
    \State Define grid cells. 
    \While{grid cells not stable}     
      \State Partition objects into proper grid cells and compute
the density of each cell.  
      \State Eliminate cells with density below the certain
threshold.  
      \State  Clusters can be generated in contiguous groups of dense cells.
    \EndWhile
  \end{algorithmic}   
\end{algorithm}

\subsection{Graph-Based Clustering Algorithm}
In this section, we introduce a novel approach named graph-based clustering algorithm. Graph-based clustering operates on graphs, nodes are regarded as data points and the edges are regarded as relationships among data points \citep{2015A}. Spectral clustering algorithm and Affinity Propagation (AP) are two classical graph-based clustering algorithms. Graph-based clustering algorithms are intuitive and transform clustering into an optimization problem.

\textbf{Spectral Clustering.} The core idea of spectral clustering is to treat all data as a graph and assign different weights to each edge. It divides the whole graph into many sub-graphs by cutting the edges with small weight, to ensure the weights between different sub-graphs is as small as possible and the sum of weights in the same sub-graph is as large as possible, each sub-graph is considered to be a cluster. \citet{2021MNRAS.500.3027D}  compiled a molecular cloud catalogue with spectral clustering and analysed global properties of some clouds.

\textbf{Affinity Propagation (AP).} The basic idea of AP algorithm is to take all data points as potential clustering centers (called examples) and connect any two data points to form a network (similarity matrix). Then calculate the clustering centers of each sample through the transmission of messages (responsibility and availability) on each side of the network. It has been applied to star/galaxy classification, open clusters clustering and large spectra clustering \citep{10.1109/BigData.2015.7363804, 2015A&A...577A..47B, 7820430}.

\section{Experiment Analysis}
The clustering methods mentioned in Section 2 have been successfully applied to various astronomical data analysis. For astronomical clustering tasks, there are many methods to choose and researchers need to select one which works best. Due to the diversity of clustering tasks and data, it is difficult to evaluate the advantage and disadvantage of these methods from current literatures.

To explore the advantage of each method in various tasks, in this section, we evaluate several commonly used clustering algorithms through building the unified spectral datasets as experimental data, which are observed by LAMOST survey \citep{2015RAA....15.1095L}.

LAMOST (The Large Sky Area Multi-Object Fiber Spectroscopic Telescope, also called the Guo Shou Jing Telescope) is a special reflecting Schmidt telescope with an effective aperture of 3.6 – 4.9 m and a field of view of 5\degree. It is equipped with 4000 fibers, with a spectral resolution of R $\approx$ 1800 and wavelength ranging from 3800 to 9000 $\mathop A\limits^ \circ $
(\url{http://www.lamost.org/}). After seven years of surveys, LAMOST observed tens of millions of low-resolution spectral data, which provides important data for astronomical statistical research. The spectra we used are selected from LAMOST DR8 which released a total of 17.23 million spectra. The number of high-quality spectra of DR8 (S/N > 10) reaches 13.28 million and DR8 includes a catalog of about 7.75 million groups of stellar spectral parameters.

We construct unified datasets from four aspects including feature extraction methods for spectra, quality of spectra, dataset size and outlier detection task, to explore advantages and disadvantages of different clustering algorithms. In the aspect of different feature extraction methods, we organize experiments from stellar classification, star/galaxy/quasar classification, subclasses of A-type star classification and matching sources from LAMOST and SDSS classification. All of these aspects need to be considered in the practical spectral analysis, so comparative experiments are meaningful for researchers. We compare eight clustering algorithms which are widely used in astronomy, including K-means, K-means-dp, SOM, K-mediod, DBSCAN, CFSFDP, Hierarchical Clustering and GMM. And the true labels we use are spectral classes released by LAMOST.

Algorithms are evaluated by clustering accuracy. But we can not know how the samples are divided through average accuracy, and it is not rigorous if the clustering result is undesirable. So t-SNE algorithm and confusion matrix are used to show the clustering errors. Data can be mapped into a 2D graph by t-SNE, and the more similar the samples are, the closer they are in the 2D graph. The t-SNE graphs of data before and after clustering are drawn to observe how data is clustered. The confusion matrix can be used to determine the exact number of false clustering samples.

\subsection{Performance of algorithms on different spectra characteristics }


Low resolution 1D spectra observed by telescope have thousands of dimensions. And in automatic spectral analysis, astronomers always adopt some dimensionality reduction and feature extraction techniques like PCA, extracting spectra line indices, etc \citep{2014IAUS..298..428L, 2004SPIE.5496..756L}. Different clustering methods will achieve different performances on these characteristics, so it is necessary to evaluate these methods on different spectra features. In this section, four types of data including 1D spectra, PCA features, spectra line indices and stellar parameters are constructed to test clustering algorithms. PCA is a feature extraction and dimension reduction method that combines features linearly, spectral features extracted by PCA have no physical meaning. But it is widely used in astronomy and other fields, so we also choose PCA features to compare the clustering performance with other features.

In this section, experiments are carried out from (1) stellar classification, (2) star/galaxy/quasar classification, (3) classification of subclasses of A-type star, (4) classification of matching sources from LAMOST and SDSS. Data selections of the tasks are shown in Table \ref{table:dataset_characteristic}, Table \ref{table:volume_of_dataset_characteristie} and Table \ref{table_lamost_sdss}.

\begin{table}

    \centering
    \caption{Data selection of three tasks.}
    \begin{threeparttable}
    \resizebox{\linewidth}{!}{
    \begin{tabular}{lcccc}
    \hline
        ~ & 1D spectra & PCA & Line Indices & Stellar Parameters \\ \hline
        Task Star & $\surd$ & $\surd$ & $\surd$ & $\surd$ \\ 
        Task SGQ & $\surd$ & ~  & ~ & ~ \\ 
        Task SubA &  $\surd$ & ~ & ~ &  $\surd$ \\ \hline
    \end{tabular}
    }
    \label{table:dataset_characteristic}
    \begin{tablenotes}
    \footnotesize
    \item[1] Task Star: Stellar Classification.
    \item[2] Task SGQ: Classification of Star/Galaxy/Quasar. 1D spectra in this task refer to original spectra and rest wavelength frame spectra.
    \item[3] Task SubA: Classification of Subclass of A-type Star.
    \end{tablenotes}
    \end{threeparttable}

\end{table}

\begin{table}

\centering
\caption{Number of spectra of three tasks.}
\label{table:volume_of_dataset_characteristie}
\resizebox{\linewidth}{!}{
\begin{tabular}{ccc}
\hline
Task & Type & Data volume \\ \hline
\multicolumn{1}{c}{\multirow{4}{*}{Stellar Classification}}
 & A & 5000 \\
\multicolumn{1}{c}{} & F & 5000 \\
\multicolumn{1}{c}{} & G & 5000 \\
\multicolumn{1}{c}{} & K & 5000 \\ \hline
\multicolumn{1}{c}{\multirow{3}{*}{Classification of Star/Galaxy/Quasar}}
 & Star & 5000 \\  
 & Galaxy & 5000 \\ 
 & Quasar & 5000 \\ \hline
\multirow{9}{*}{Classification of Subclasses of A-type Star}
& A0 & 14 \\
 & A1 &  6097 \\
 & A2 & 2329 \\
 & A3 & 514 \\
 & A5 & 5551 \\
 & A6 & 3383 \\
 & A7 & 6046 \\
 & A8 & 305 \\ 
 & A9 & 761 \\ \hline
\end{tabular}
}
\end{table}

\begin{figure*}
\centering
\includegraphics[width=13.42cm,height=5.58cm]{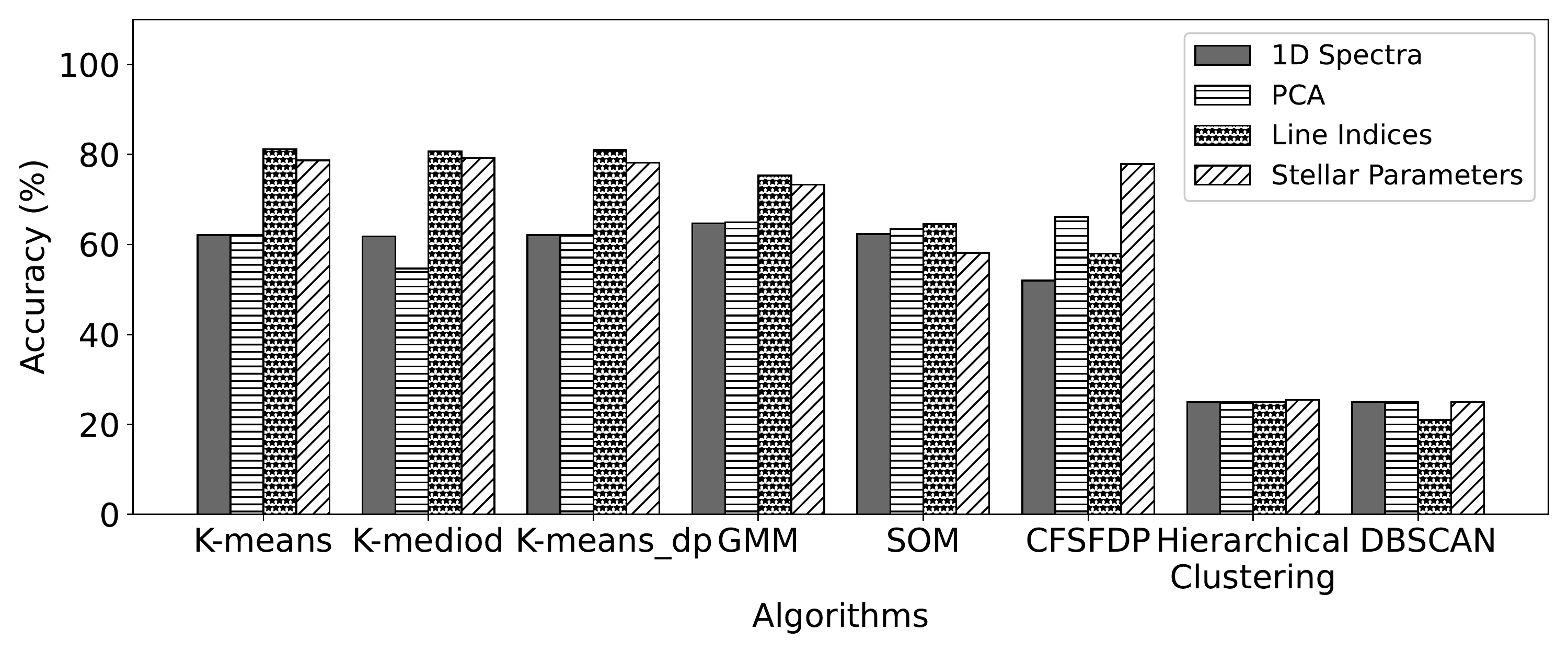}
\caption{Accuracy of Algorithms on Different Characteristics. Four bars represent four data characteristics.}
\label{fig:Algorithms_accuracy_raw_index_PCs_star-parameters_1w_10-_AFGK_bar}
\end{figure*}

\subsubsection{Stellar Classification of Type A/F/G/K}

\begin{figure*}
\centering

\includegraphics[width=15cm,height=3.4cm]{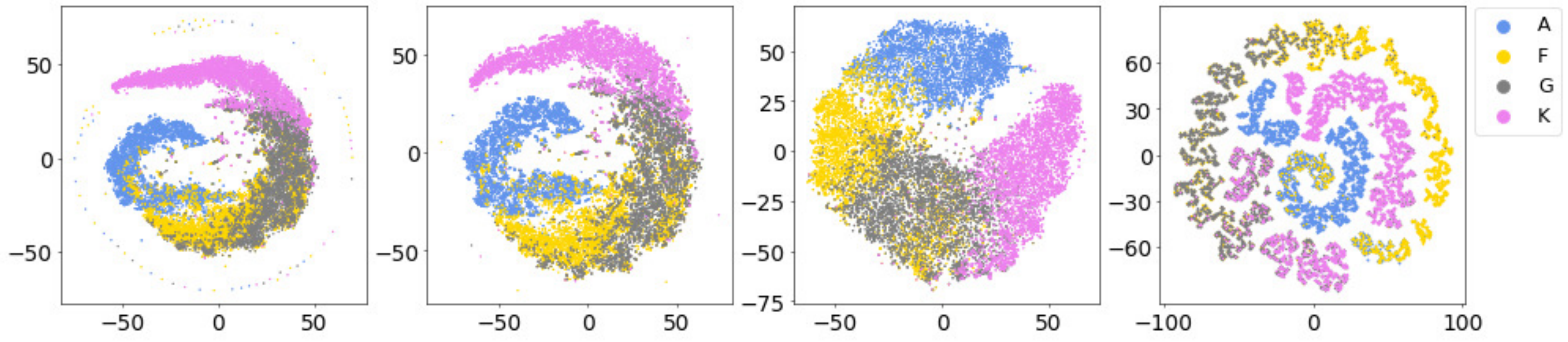}
\caption{t-SNE distribution of true labels for A/F/G/K stars on different data characteristics. From Left to right is 1D spectra, PCA, Line Indices, and Stellar Parameters. Different colors represents various types of stars: Blue-A stars, Yellow-F stars, Gray-G stars, Purple-K stars.}
\label{fig:real_label_component_1w_10-_tsne}
\end{figure*}

\begin{figure*}
\centering
\includegraphics[width=15cm,height=7.2cm]{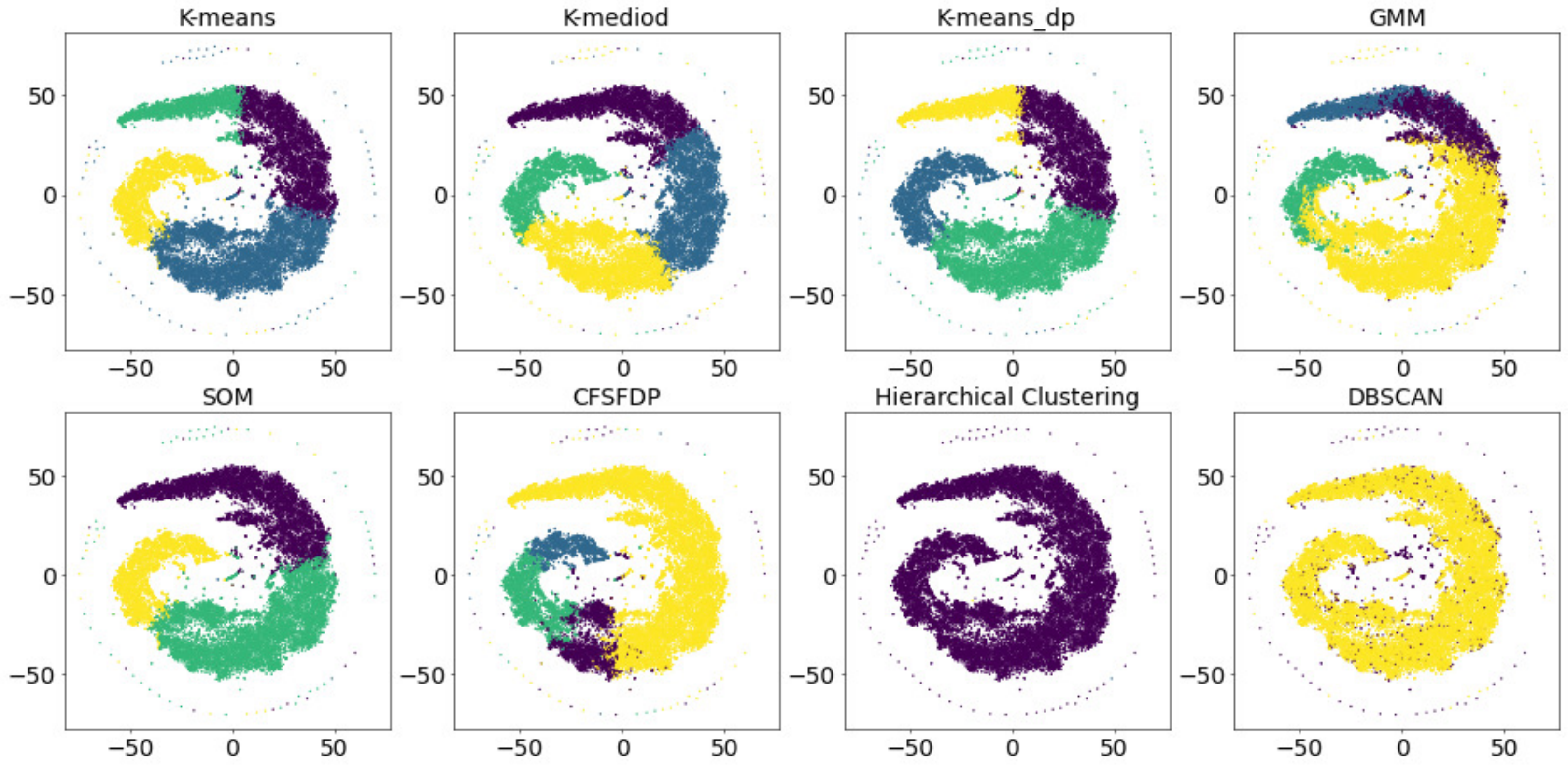}
\caption{t-SNE distribution of eight clustering algorithms results of A/F/G/K stars on 1D spectra. Each subgraph represents the results of one clustering algorithm on the 1D spectra. Different colors in each subgraph represent different classes in the clustering results. The same color in different subplots is not necessarily the same class.}
\label{fig:Algorithms_raw_1w_10-_AFGK_tsne} 
\end{figure*}

\begin{figure*}
\centering
\includegraphics[width=15cm,height=7.2cm]{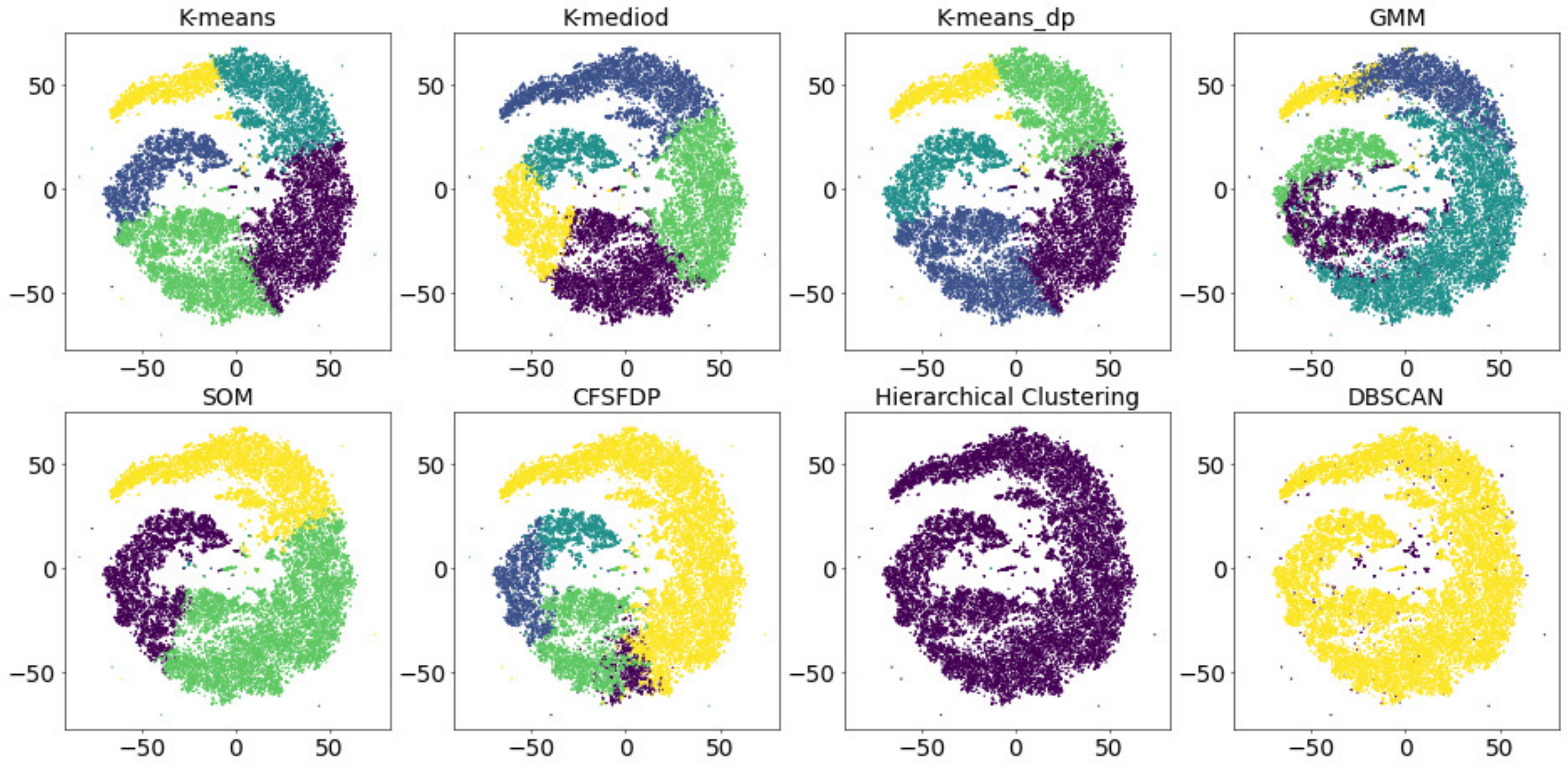}
\caption{t-SNE distribution of eight clustering algorithms results of A/F/G/K stars on PCA. Each subgraph represents the results of one clustering algorithm on the PCA. Different colors in each subgraph represent different classes in the clustering results. The same color in different subplots is not necessarily the same class.}
\label{fig:Algorithms_PCs_1w_10-_AFGK_tsne} 
\end{figure*}

\begin{figure*}
\centering
\includegraphics[width=15cm,height=7.2cm]{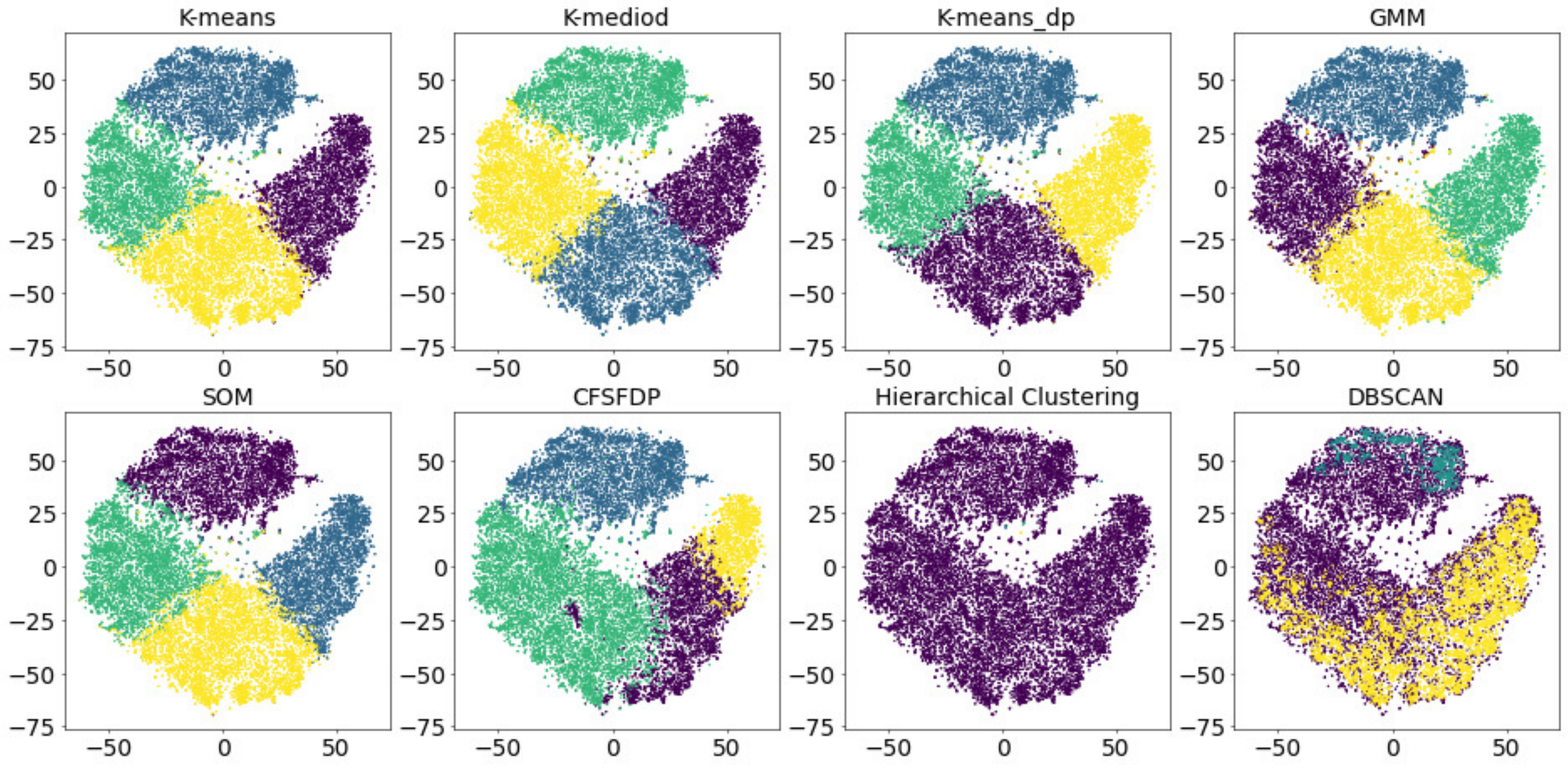}
\caption{t-SNE distribution of eight clustering algorithms results of A/F/G/K stars on line indices. Each subgraph represents the results of one clustering algorithm on the line indices. Different colors in each subgraph represent different classes in the clustering results. The same color in different subplots is not necessarily the same class.}
\label{fig:Algorithms_Index_1w_10-_AFGK_tsne} 
\end{figure*}

\begin{figure*}
\centering
\includegraphics[width=15cm,height=7.2cm]{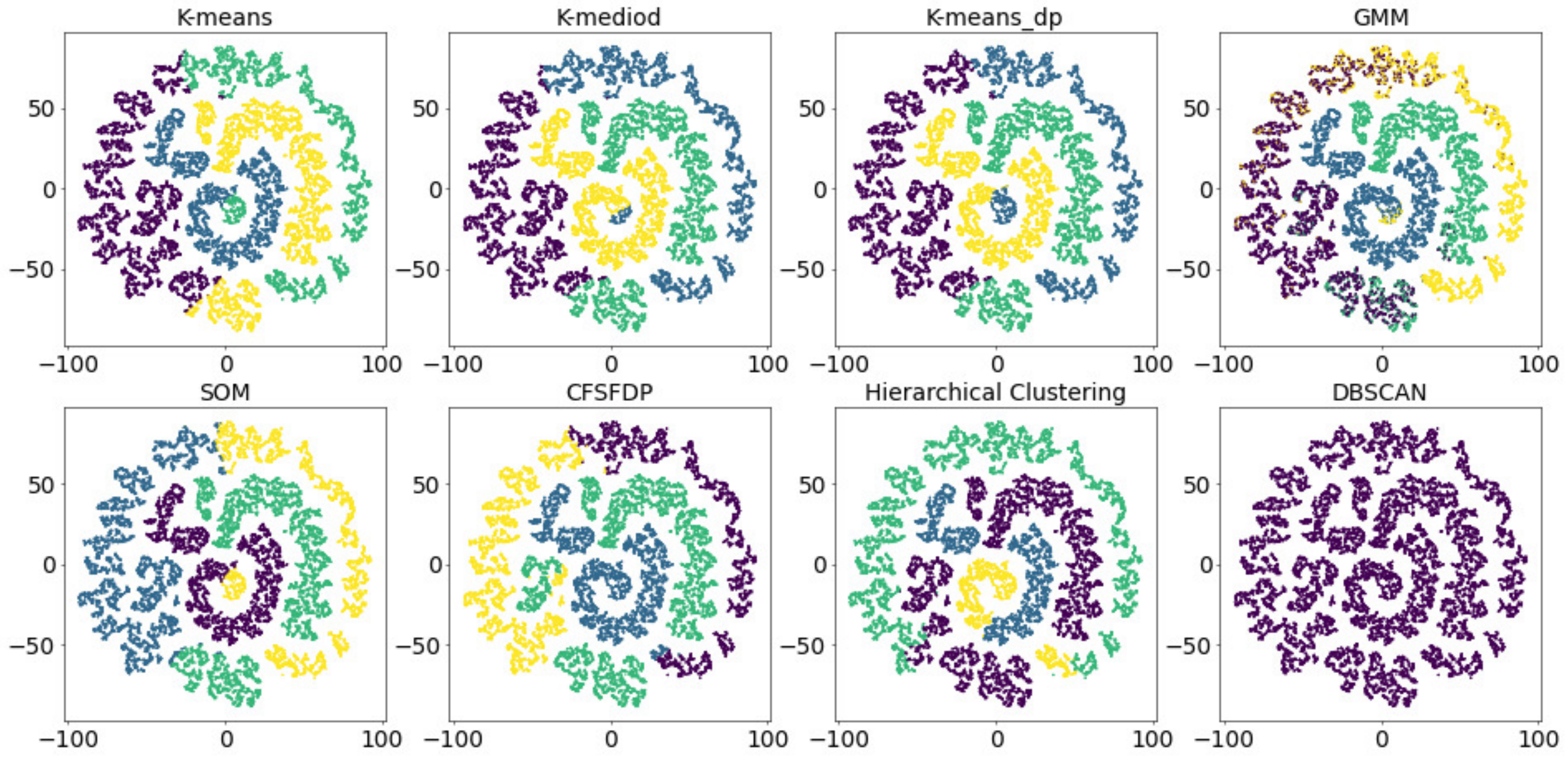}
\caption{t-SNE distribution of eight clustering algorithms results of A/F/G/K stars on stellar parameters. Each subgraph represents the results of one clustering algorithm on the stellar parameters. Different colors in each subgraph represent different classes in the clustering results. The same color in different subplots is not necessarily the same class.}
\label{fig:Algorithms_para_1w_10-_AFGK_tsne} 
\end{figure*}

\begin{figure*}
\centering
\includegraphics[width=15cm,height=5cm]{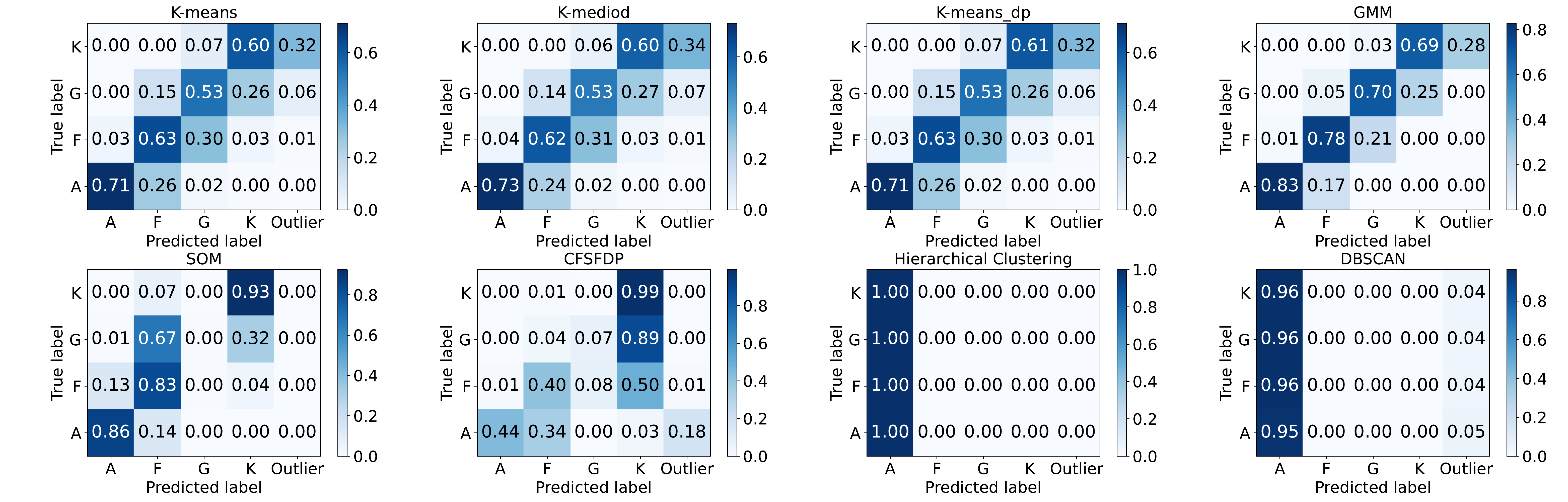}
\caption{Confusion matrix of eight algorithms on 1D spectra. Predicted label: color and digit in each cell are the consistent probability between predicted label and true label. Color is in direct proportion to the figure: bigger numbers and deeper color.}

\label{fig:Algorithms_raw-spectra_1w_10-_confusion_matrix} 
\end{figure*}

\begin{figure*}
\centering
\includegraphics[width=15.36cm,height=5.376cm]{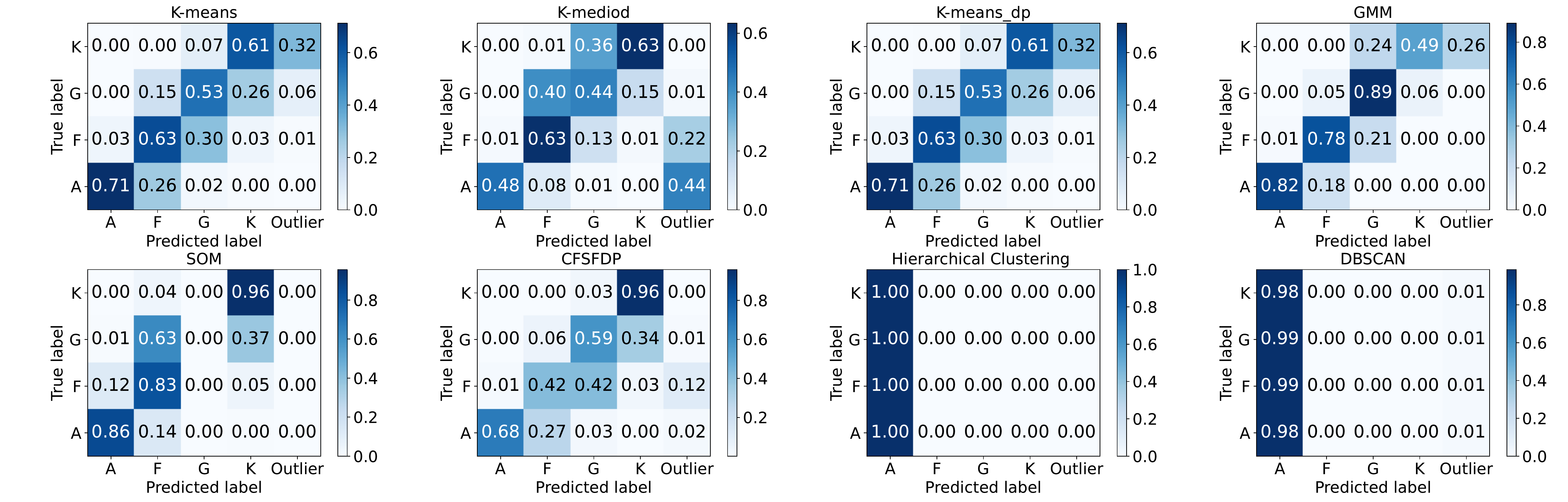}
\caption{Confusion matrix of eight algorithms on PCA. Predicted label: color and digit in each cell are the consistent probability between predicted label and true label. Color is in direct proportion to the figure: bigger numbers and deeper color.}

\label{fig:Algorithms_PCs_1w_10-_confusion_matrix} 
\end{figure*}

\begin{figure*}
\centering
\includegraphics[width=15.36cm,height=5.376cm]{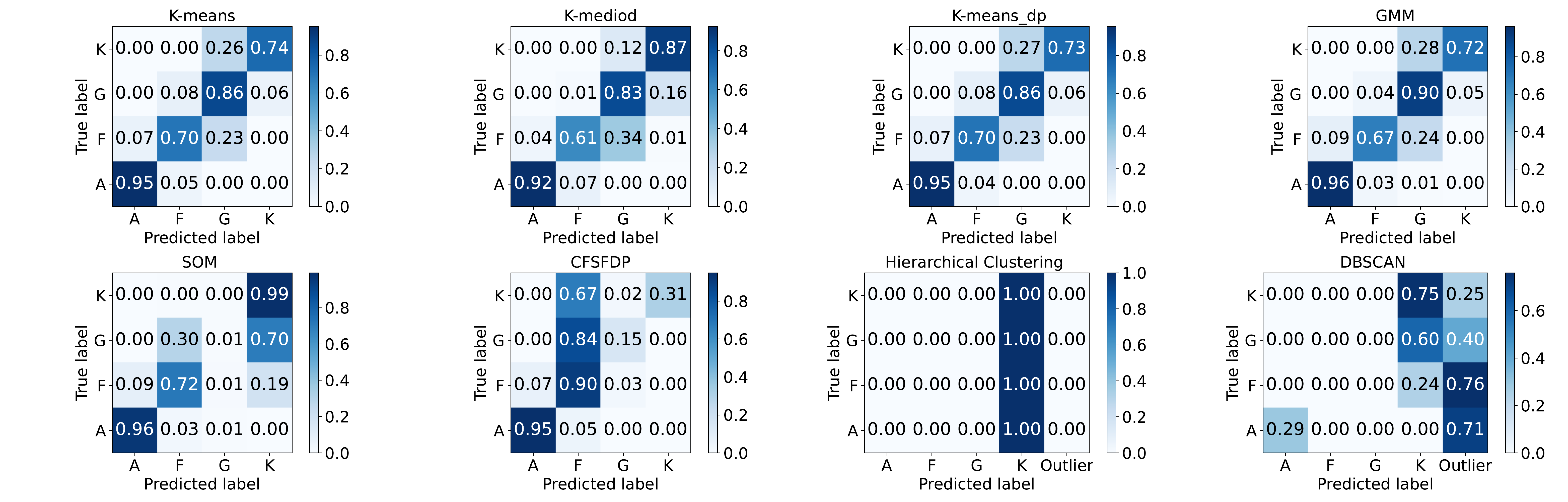}
\caption{Confusion matrix of eight algorithms on line indices. Predicted label: color and digit in each cell are the consistent probability between predicted label and true label. Color is in direct proportion to the figure: bigger numbers and deeper color.}
\label{fig:Algorithms_index_1w_10-_confusion_matrix} 
\end{figure*}

\begin{figure*}
\centering
\includegraphics[width=15.36cm,height=5.376cm]{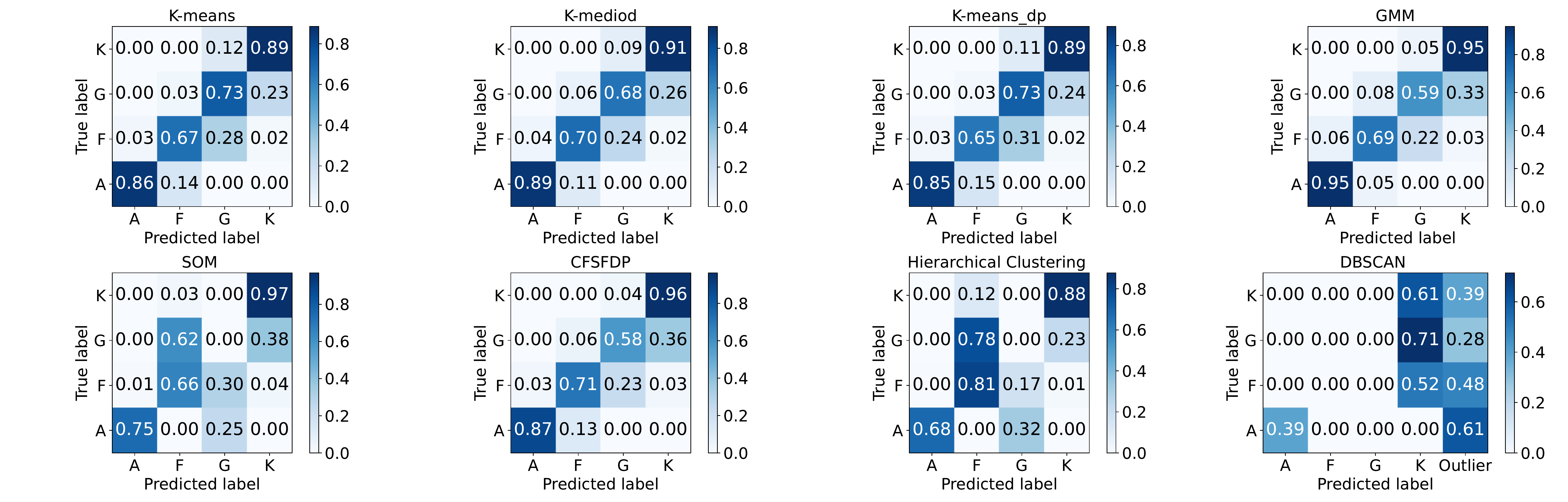}
\caption{Confusion matrix of eight algorithms on stellar parameters. Predicted label: color and digit in each cell are the consistent probability between predicted label and true label. Color is in direct proportion to the figure: bigger numbers and deeper color.}
\label{fig:Algorithms_star-parameters_1w_10-_confusion_matrix} 
\end{figure*}

In the stellar classification, we use four features mentioned above. While, the performance of data analysis using the full spectrum is inevitably affected by reddening, and the LAMOST pipeline corrects it using polynomials during data processing. It is difficult to obtain a consistent extinction model, and at the same time, this paper mainly focuses on providing a performance analysis on astronomical spectra of different techniques. Therefore, we choose the spectra of high galactic latitude ($\mathscr{l}$ > 45°) to construct the datasets. In the preprocessing steps, we cut out the 5700$\mathop A\limits^ \circ $-5900$\mathop A\limits^ \circ $ wavelength range, because LAMOST adopts relative flux calibration and there are lots of noise at the merge between red and blue segment. The redshift of star is small, so we use original 1D spectra released by LAMOST directly. Fig. \ref{fig:Algorithms_accuracy_raw_index_PCs_star-parameters_1w_10-_AFGK_bar} - Fig. \ref{fig:Algorithms_star-parameters_1w_10-_confusion_matrix} are the results of eight clustering algorithms on four spectra features.

From Fig. \ref{fig:Algorithms_accuracy_raw_index_PCs_star-parameters_1w_10-_AFGK_bar}, it is clear that algorithms perform differently on four spectra characteristics. The clustering accuracies are not high and the highest ones are only more than 80\%. Among the four features, line indices work best to separate different types of stars, followed by stellar parameters. And 1D spectra has the same results as PCA features in most methods. But PCA, as typical linear dimensionality-reduction method, can greatly reduce the clustering time compared with 1D spectra. Hierarchical clustering and DBSCAN are the worst and they group most of samples into one cluster because a small number of outlier will be classified into other categories. 
GMM performs well on various data. Compared with partition-based methods, GMM uses multiple multi-dimensional Gaussian distributions to fit the data which is more in line with the real characteristic distribution of the data. The main idea of single-layer SOM is mapping data to two dimensional competitive neural units with topology structures. In clustering, SOM maps data to 1 $\times$ N neural units where N is the number of clusters.  Its accuracy is rarely different from K-means, but presents instability because it maps spectra to fewer data points and needs to set the number of iteration numbers.


In order to intuitively understand how data is divided in the clustering, t-SNE algorithm is used to draw the two-dimensional distribution of datasets before and after clustering (Fig. \ref{fig:real_label_component_1w_10-_tsne}, Fig. \ref{fig:Algorithms_raw_1w_10-_AFGK_tsne} - Fig. \ref{fig:Algorithms_para_1w_10-_AFGK_tsne}). Different colors are used to distinguish different types. But the same color in different graphs after clustering does not represent the same type of stars, because clustering algorithms just divide data into some groups based on the similarity. In the four plots on true labels (Fig. \ref{fig:real_label_component_1w_10-_tsne}), there are overlaps between adjacent types, especially on F stars and G stars, G stars and K stars. t-SNE map of PCA features is basically the same as that of 1D spectra. However, there is a circle of outliers around 1D spectra which does not exist in PCA features. The outliers are some incomplete spectra due to observational conditions and this shows that PCA can reduce the impact of unusual values on the data. The distribution of stars with the same types is more concentrated in the graph of the line indices. In the stellar parameters, the graph shows small aggregations.



From Fig. \ref{fig:Algorithms_raw_1w_10-_AFGK_tsne} and Fig. \ref{fig:Algorithms_PCs_1w_10-_AFGK_tsne}, GMM can tackle complicated data distributions compared with partition-based algorithms. The reason is that, partition-based clustering algorithms use Euclidean distance to measure the similarity between samples which makes it easy to identify clusters with spherical distribution, but they can not consider some special dimensional features. GMM uses Gaussian distribution to fit features in each dimension, so it has better performance on 1D spectra and PCA features. While on the dataset of line indices (Fig. \ref{fig:Algorithms_Index_1w_10-_AFGK_tsne}), which has a more concentrated distribution of each type of star, partition-based methods, GMM and SOM have the same clustering results. DBSCAN and hierarchical algorithms always divide data into one cluster. CFSFDP algorithm sometimes gets results similar to K-means and sometimes gets poor results. Because it divides samples into a cluster according to the local density of samples without considering the global distribution.

Meanwhile, we show confusion matrix of each kind of algorithms to observe misclassified samples straightforward (Fig. \ref{fig:Algorithms_raw-spectra_1w_10-_confusion_matrix} - Fig. \ref{fig:Algorithms_star-parameters_1w_10-_confusion_matrix}). Through experiments, sometimes it is more appropriate to divide four types of spectra into five clusters, for that a small number of samples will be classified as one cluster, and we name them "outliers" in the confusion matrix.

Compared with other methods, performance of partition-based algorithms and GMM are stable and good because they obtain the optimal results from the global perspective of data. Misclassified phenomenon always exists in the adjacent clusters and the clustering accuracy of A stars is highest.

\subsubsection{Classification of Star/Galaxy/Quasar}

\begin{figure*}
\centering
\includegraphics[width=14.5cm,height=5.5cm]{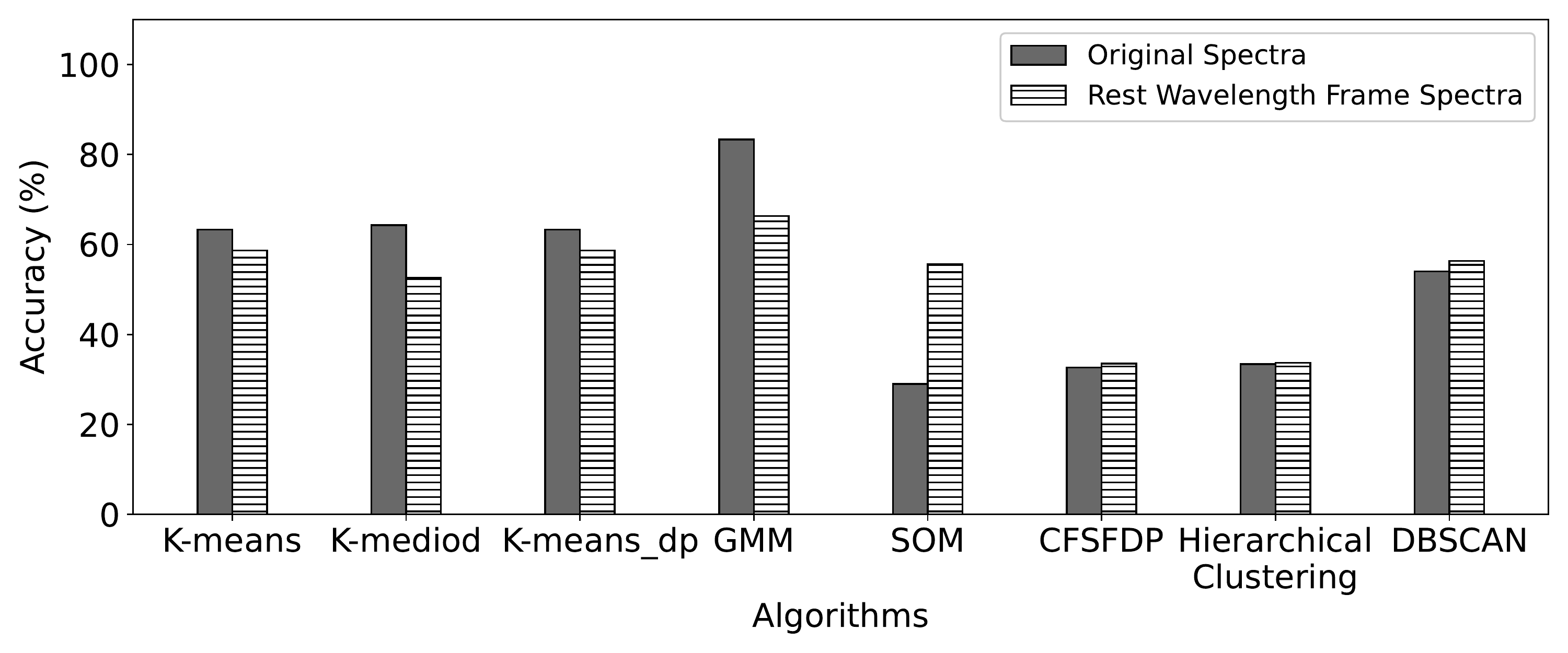}
\caption{Accuracy of eight algorithms for star/galaxy/quasar on original spectra and rest wavelength frame spectra. Two bars represent tow spectra characteristics.}
\label{fig:Algorithms_accuracy_Star_Galaxy_Quasar_1D_rest_bar} 
\end{figure*}

\begin{figure*}
\centering
\includegraphics[width=9cm,height=3.4cm]{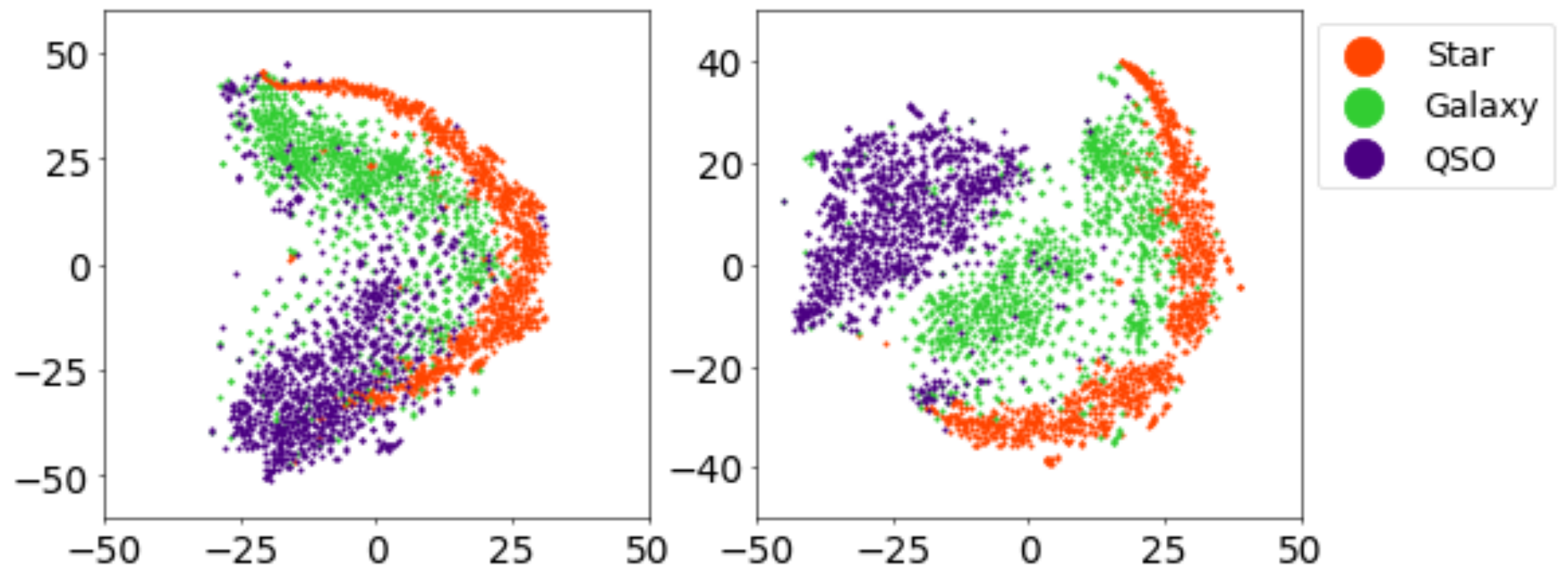}
\caption{t-SNE distribution of star/galaxy/quasar on original spectra and rest wavelength frame spectra. Left is the original spectra, right is the rest wavelength  frame spectra. Red: Star, Green: Galaxy, Purple: QSO.}
\label{fig:SGQ_1D_rest_label_tsne} 
\end{figure*}

\begin{figure*}
\centering
\includegraphics[width=15cm,height=7.2cm]{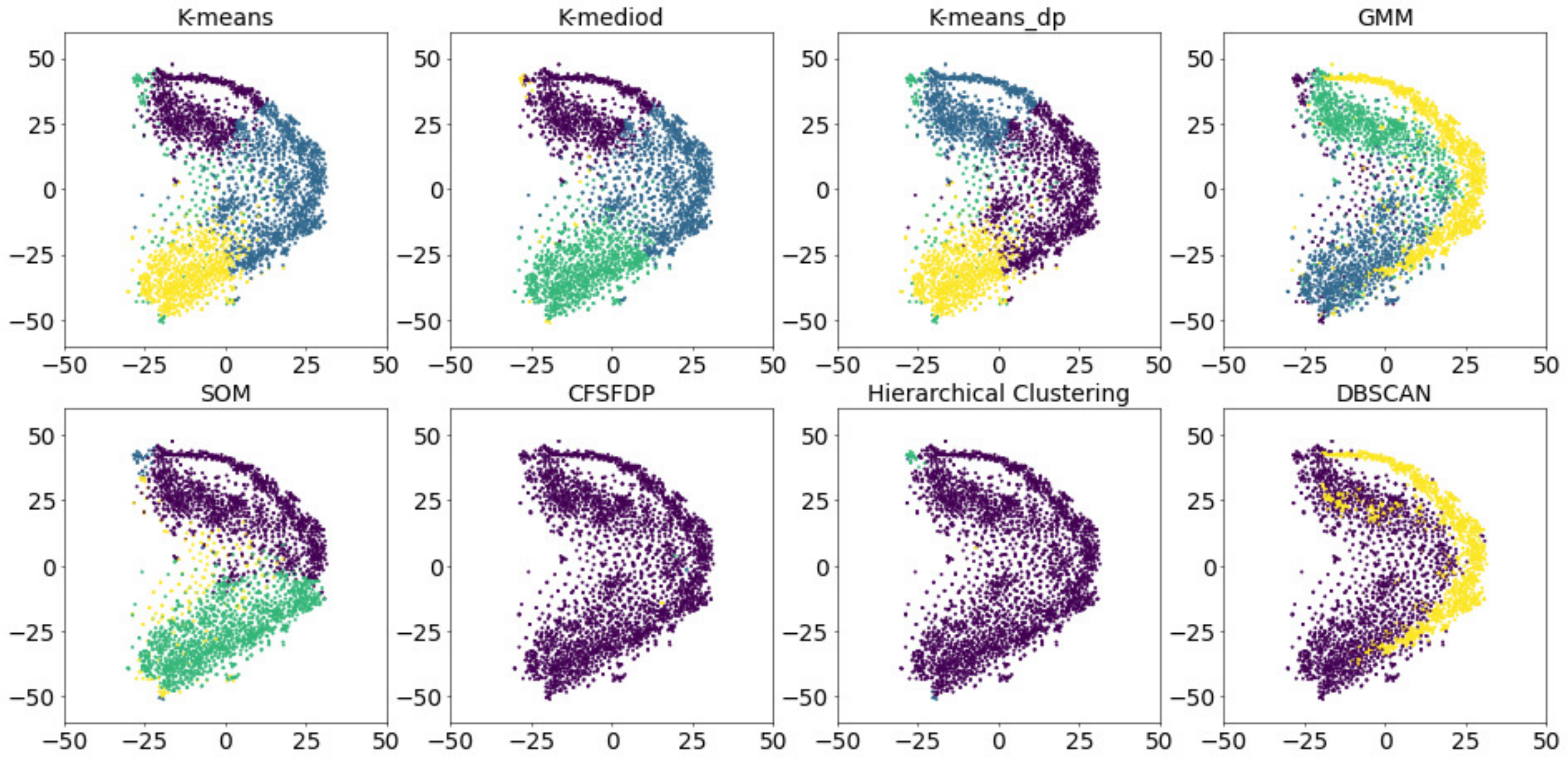}
\caption{t-SNE distribution of the results of algorithms on original spectra.  Each subgraph represents the results of one clustering algorithm on the original spectra. Different colors in each subgraph represent different classes in the clustering results. The same color in different subplots is not necessarily the same class.}
\label{fig:SGQ_1D_10-_1w_tsne} 
\end{figure*}

\begin{figure*}
\centering
\includegraphics[width=15cm,height=7.2cm]{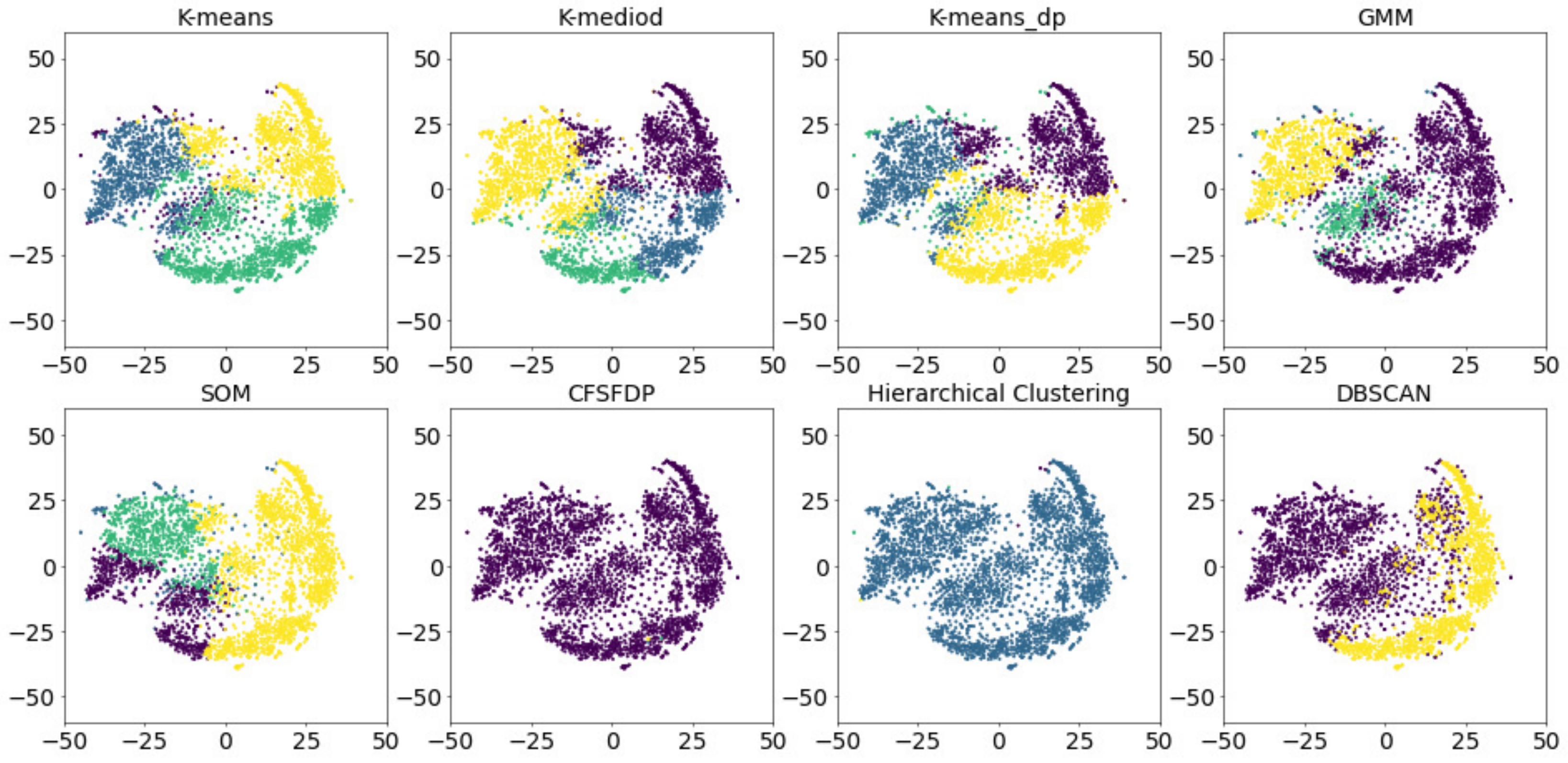}
\caption{t-SNE distribution of the results of algorithms on rest wavelength frame spectra.  Each subgraph represents the results of one clustering algorithm on the rest wavelength frame spectra. Different colors in each subgraph represent different classes in the clustering results. The same color in different subplots is not necessarily the same class.}
\label{fig:SGQ_rest_10-_1w_tsne} 
\end{figure*}

\begin{figure*}
\centering
\includegraphics[width=15.18cm,height=5cm]{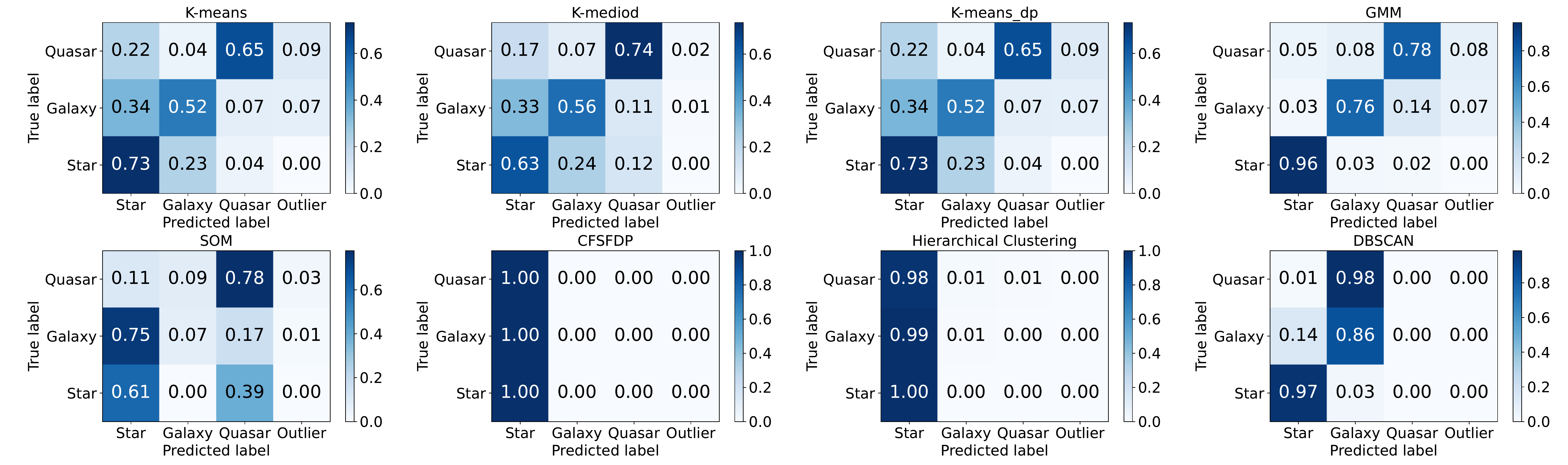}
\caption{Confusion matrix of eight algorithms on original spectra. Predicted label: color and digit in each cell are the consistent probability between predicted label and true label. Color is in direct proportion to the figure: bigger numbers and deeper color.}
\label{fig:Algorithms_star_galaxy_quasar_1D_confusion_matrix} 
\end{figure*}

\begin{figure*}

\centering
\includegraphics[width=15.18cm,height=5cm]{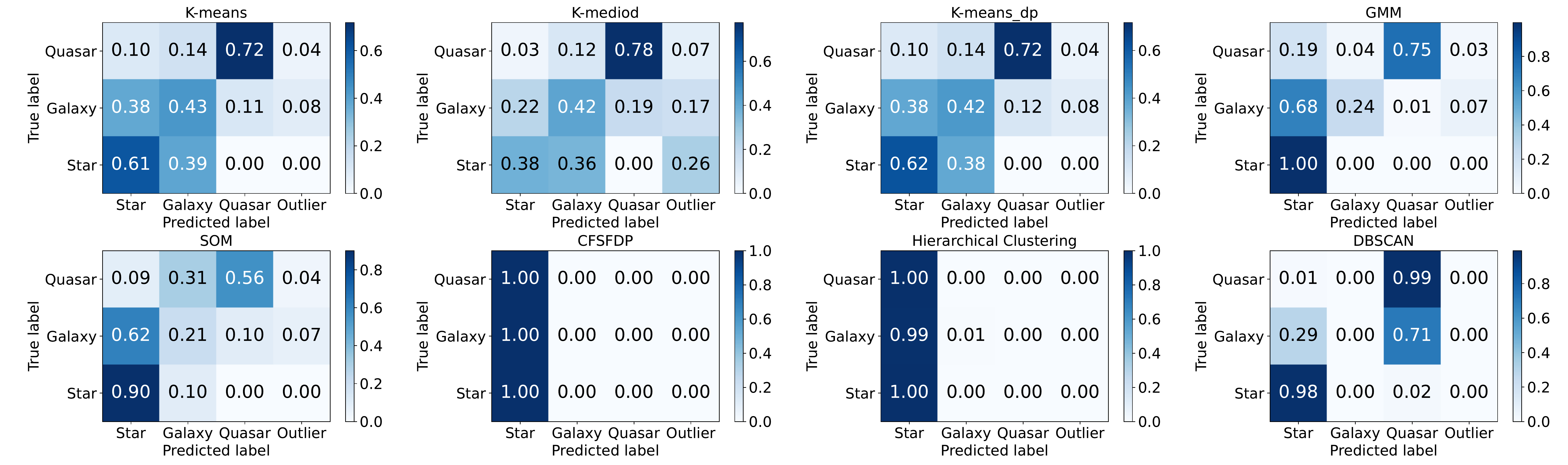}
\caption{Confusion matrix of eight algorithms on rest wavelength frame spectra. Predicted label: color and digit in each cell are the consistent probability between predicted label and true label. Color is in direct proportion to the figure: bigger numbers and deeper color.}
\label{fig:Algorithms_star_galaxy_quasar_rest_confusion_matrix} 
\end{figure*}

Classification of Star/Galaxy/Quasar is an essential part in astronomy. Galaxies and quasars always have large redshift and the positions of line features are shifted variously. So we deredshift the spectra of galaxies and quasars in the dataset. There are many excellent methods to deredshift, but our purpose is to compare clustering algorithms, so we use the redshift values released by LAMOST directly. To have more shared wavelengths of star, galaxy and quasar, we choose galaxies and quasars with redshifts less than 0.3. Meanwhile, we also construct a dataset on original spectra without deredshifting to compare the performance, their results are shown in Fig. \ref{fig:Algorithms_accuracy_Star_Galaxy_Quasar_1D_rest_bar} - Fig. \ref{fig:Algorithms_star_galaxy_quasar_rest_confusion_matrix}.

From Fig. \ref{fig:SGQ_1D_rest_label_tsne}, we can know that the distribution of quasars on rest wavelength frame has less overlaps with stars and galaxies, so the clustering accuracy of quasars is higher than original spectra (Fig. \ref{fig:Algorithms_star_galaxy_quasar_1D_confusion_matrix}, Fig. \ref{fig:Algorithms_star_galaxy_quasar_rest_confusion_matrix}). But the average accuracy of rest wavelength frame spectra is not better than original spectra (Fig. \ref{fig:Algorithms_accuracy_Star_Galaxy_Quasar_1D_rest_bar}). Because galaxies and stars are not separated well. The distribution of stars and galaxies are not spherical (or circular), so partition-based methods cannot classify them, but GMM can fit the distribution of each type very well. Density-based methods (DBSCAN) can distinguish stars from others while cannot separate galaxies and quasars. Because most stars only have emission lines, and galaxies and quasars both have absorption lines. SOM is also a partition-based method when it is used for clustering, but our experiments find that its clustering results are not stable.

\subsubsection{Classification of Subclasses of A-type Star}

\begin{figure*}
\centering
\includegraphics[width=9.48cm,height=3.6cm]{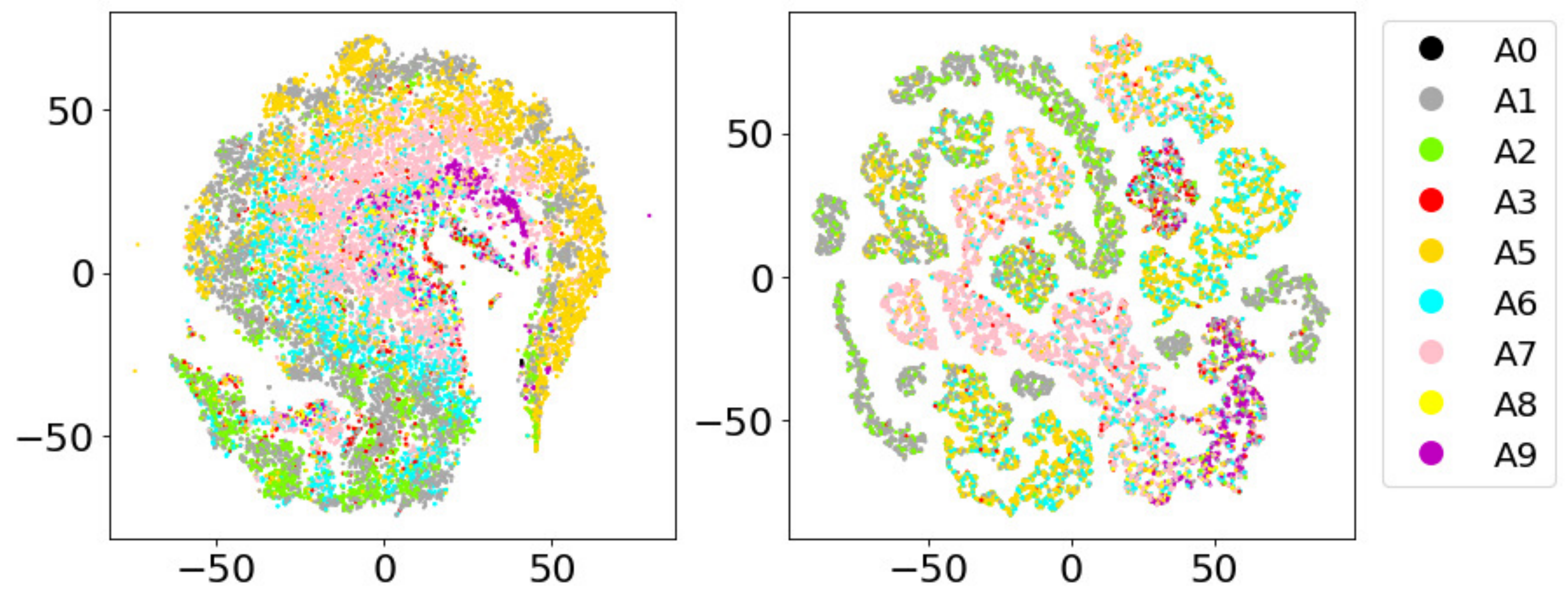}
\caption{t-SNE distribution of subclasses of A-star with true labels. Left to right is 1D spectra and stellar parameters. Different colors represent subclasses of A-star.}
\label{fig:subA_1D_stellar_true_tsne} 
\end{figure*}

\begin{figure*}
\centering
\includegraphics[width=15cm,height=7.2cm]{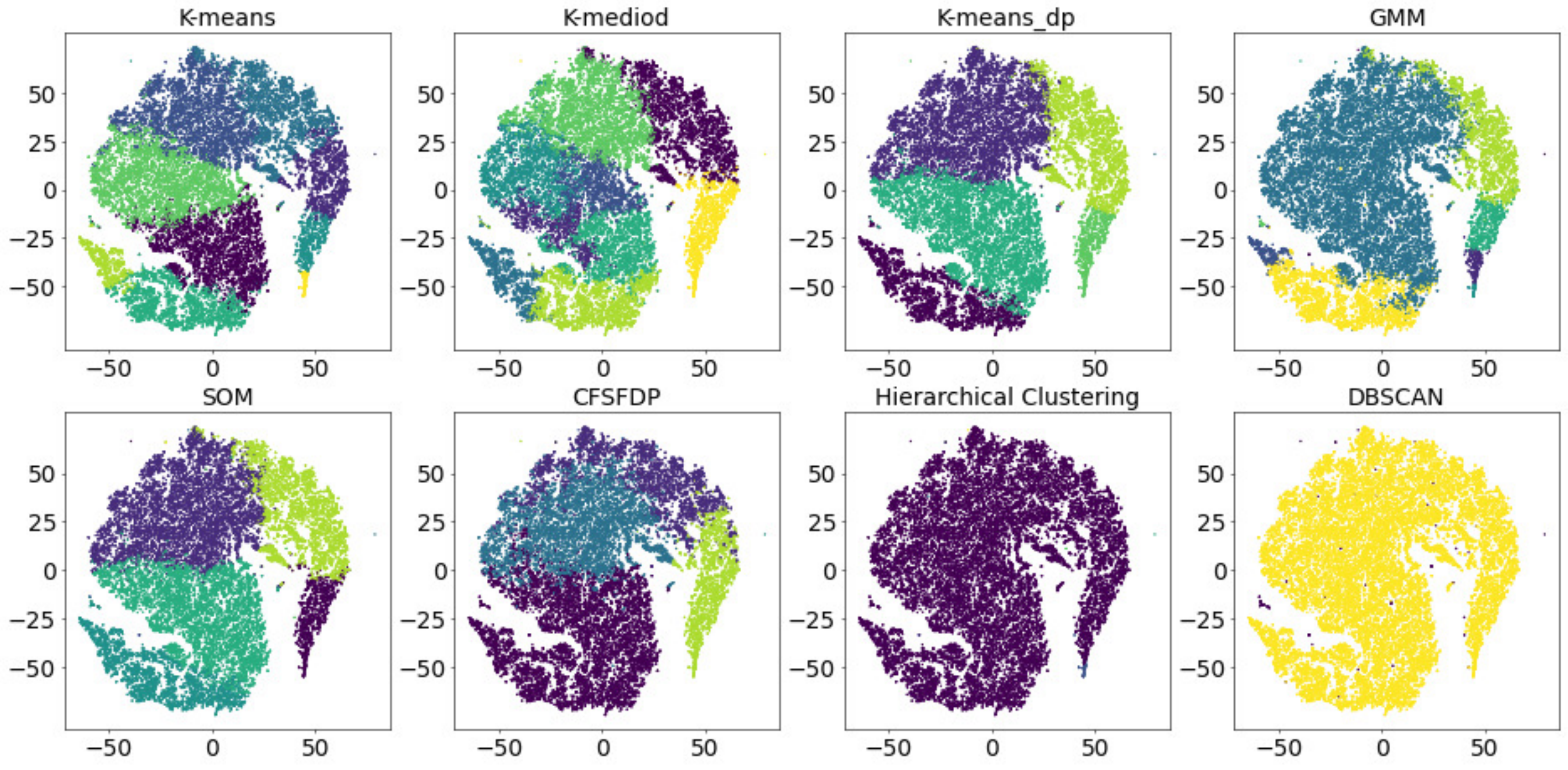}
\caption{t-SNE distribution of eight algorithms on 1D spectra of subclass of A stars.  Each subgraph represents the results of one clustering algorithm on the 1D spectra. Different colors in each subgraph represent different classes in the clustering results. The same color in different subplots is not necessarily the same class.}
\label{fig:subA_1D_tsne} 
\end{figure*}

\begin{figure*}
\centering
\includegraphics[width=15cm,height=7.2cm]{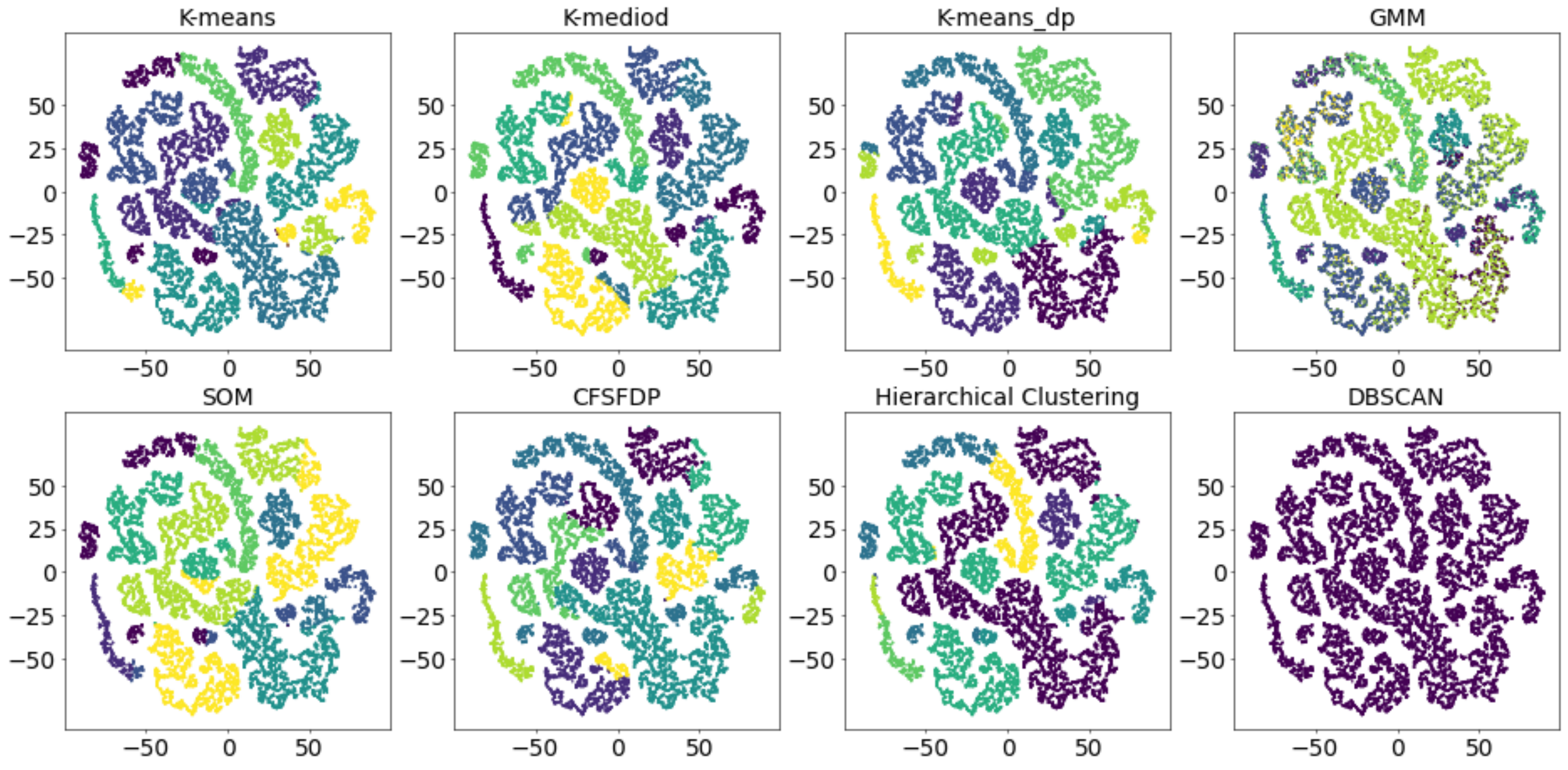}
\caption{t-SNE distribution of eight algorithms on stellar parameters of subclass of A stars.  Each subgraph represents the results of one clustering algorithm on the stellar parameters. Different colors in each subgraph represent different classes in the clustering results. The same color in different subplots is not necessarily the same class.}
\label{fig:subA_stellar_tsne} 
\end{figure*}

1D spectra and stellar parameters are used to cluster subclasses of A-type star.
Considering that there are many subclasses of A-type star, confusion matrices are not plotted in this section, the t-SNE maps and clustering results are shown in Fig. \ref{fig:subA_1D_stellar_true_tsne} - Fig. \ref{fig:subA_stellar_tsne}.

t-SNE map (Fig. \ref{fig:subA_1D_stellar_true_tsne}) shows that there are significant overlaps between different subclasses on 1D spectra. On stellar parameters, there are obvious gaps between the subclasses with large differences, but the overlap between the subclasses with small differences is also serious. And none of the algorithms can separate 1D spectra of subclasses well, even GMM which performs better on other clustering tasks. On stellar parameters, their performances are a little better. Stellar parameters with significant difference can be separated, but very similar ones cannot be separated well. Subclasses with small number are always divided into that subclasses with large number and cannot be separated out individually. So it is not a good idea to classify subclasses of stars by clustering methods.

There are two main difficulties in clustering subclasses of A-type star. One is that  the spectra of some subclasses are very similar, the other is the number of different subclasses varies greatly. This will cause the classifier to make wrong classification, even the pipline of LAMOST which distinguishes subclasses by template matching. Compared to 1D spectra, stellar parameters are more reliable to classify subclasses. The poor classification results of the experiment is not only due to the clustering algorithms but also possibly because the errors exist in the input labels, so clustering accuracies in this task do not make much sense. But the experiment could provide a clear view of distributions of subclasses of A-type star and help researchers to study the principle of clustering on them.

\subsubsection{Classification on Matching Sources of LAMOST and SDSS}

\begin{figure*}
\centering
\includegraphics[width=15.24cm,height=6.3cm]{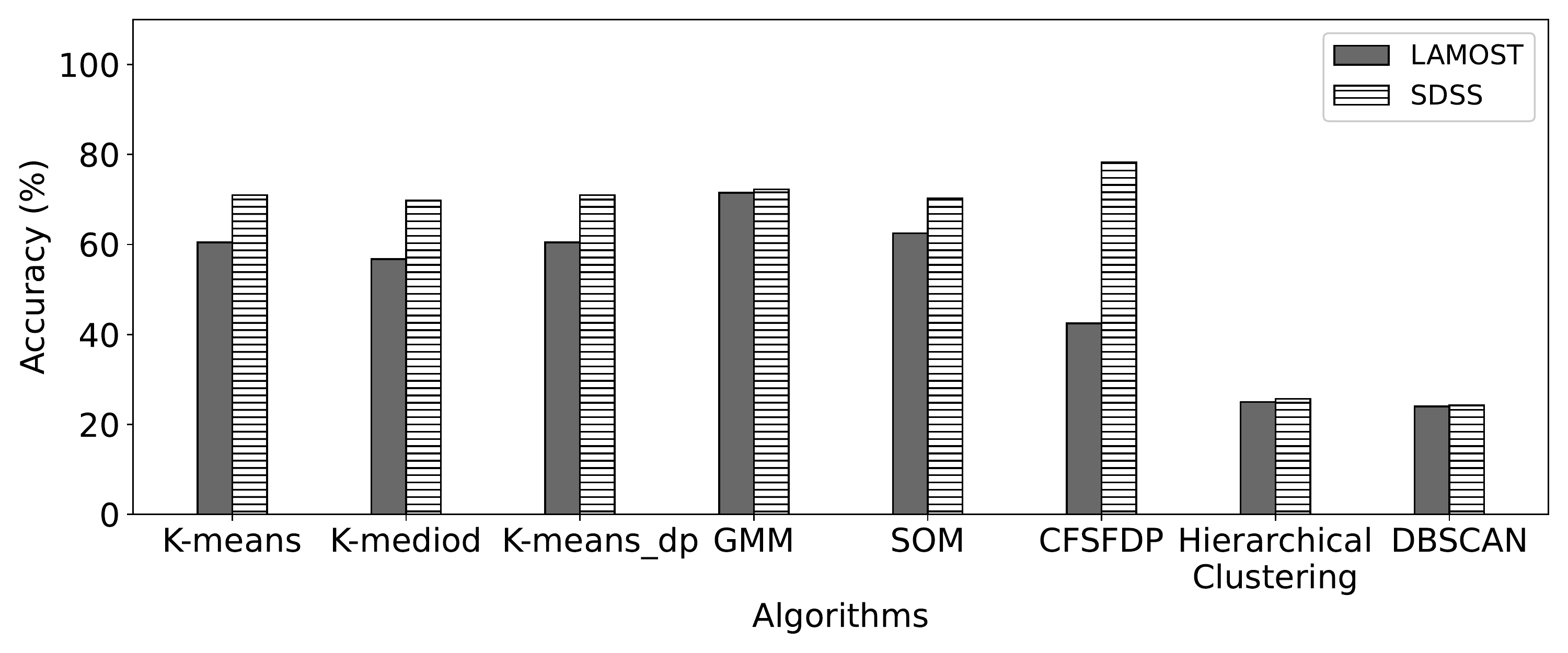}
\caption{Accuracy of eight algorithms on LAMOST and SDSS. Two bars represent spectra from LAMOST and SDSS.}
\label{fig:Algorithms_LAMOST_bar} 
\end{figure*}

\begin{figure*}
\centering
\includegraphics[width=9.39cm,height=3.8cm]{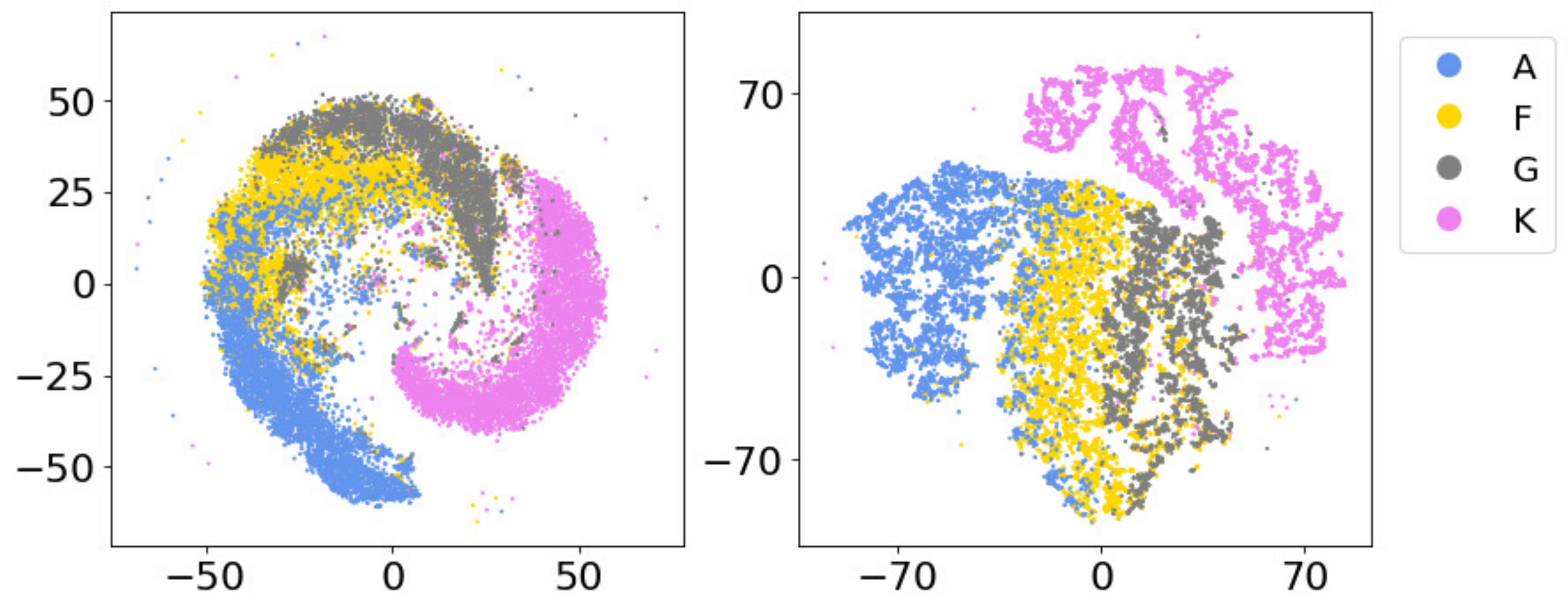}
\caption{t-SNE distribution of true labels from LAMOST and SDSS.  Left to right is LAMOST and SDSS. Different colors represent A, F, G, K stars.}
\label{fig:Algorithms_LAMOST_SDSS_true_tsne} 
\end{figure*}

\begin{figure*}
\centering
\includegraphics[width=15cm,height=7.2cm]{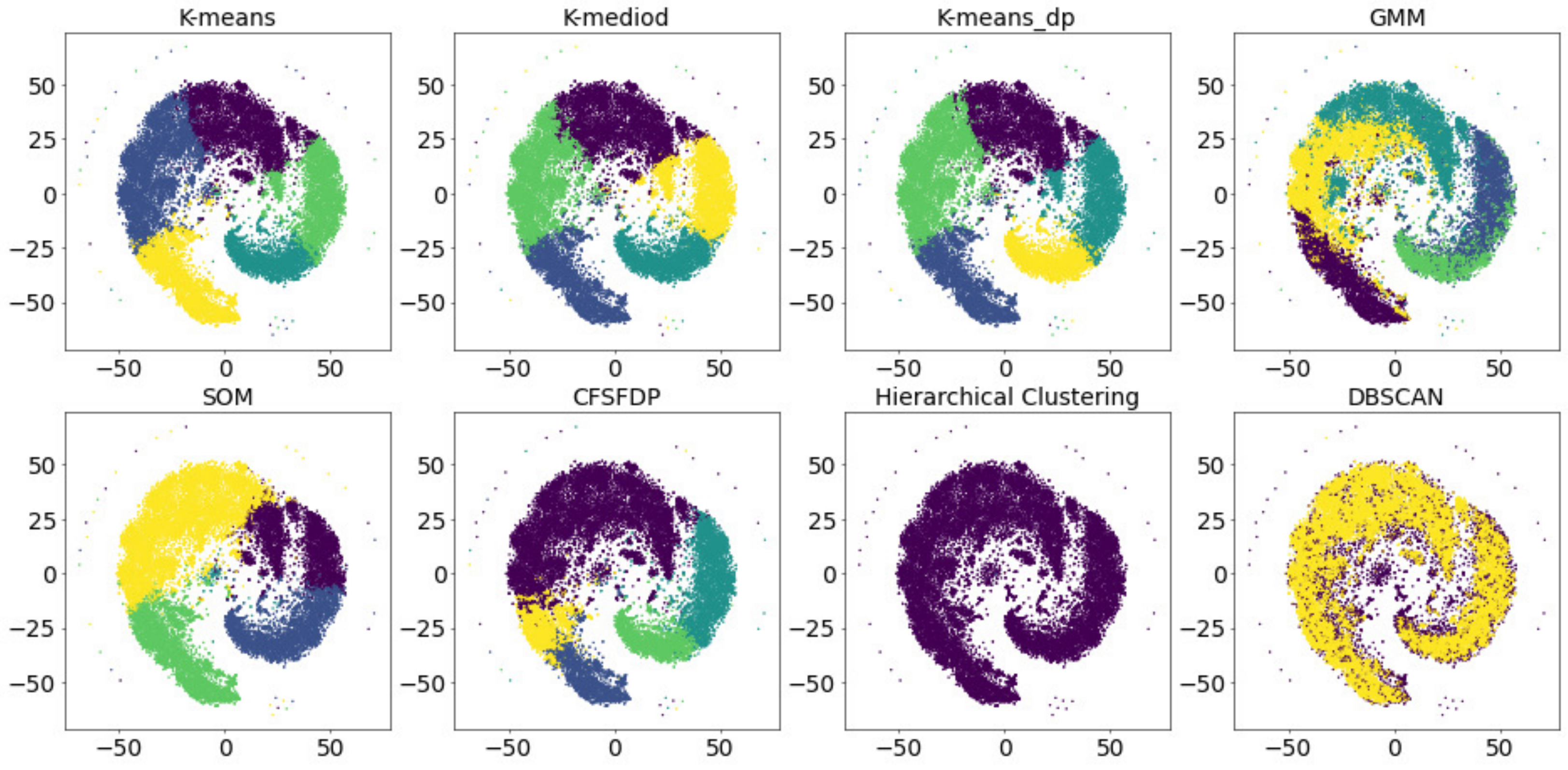}
\caption{t-SNE distribution of eight algorithms on LAMOST spectra.  Each subgraph represents the results of one clustering algorithm on the spectra from LAMOST. Different colors in each subgraph represent different classes in the clustering results. The same color in different subplots is not necessarily the same class.}
\label{fig:Algorithms_LAMOST_tsne} 
\end{figure*}

\begin{figure*}
\centering
\includegraphics[width=15cm,height=7.2cm]{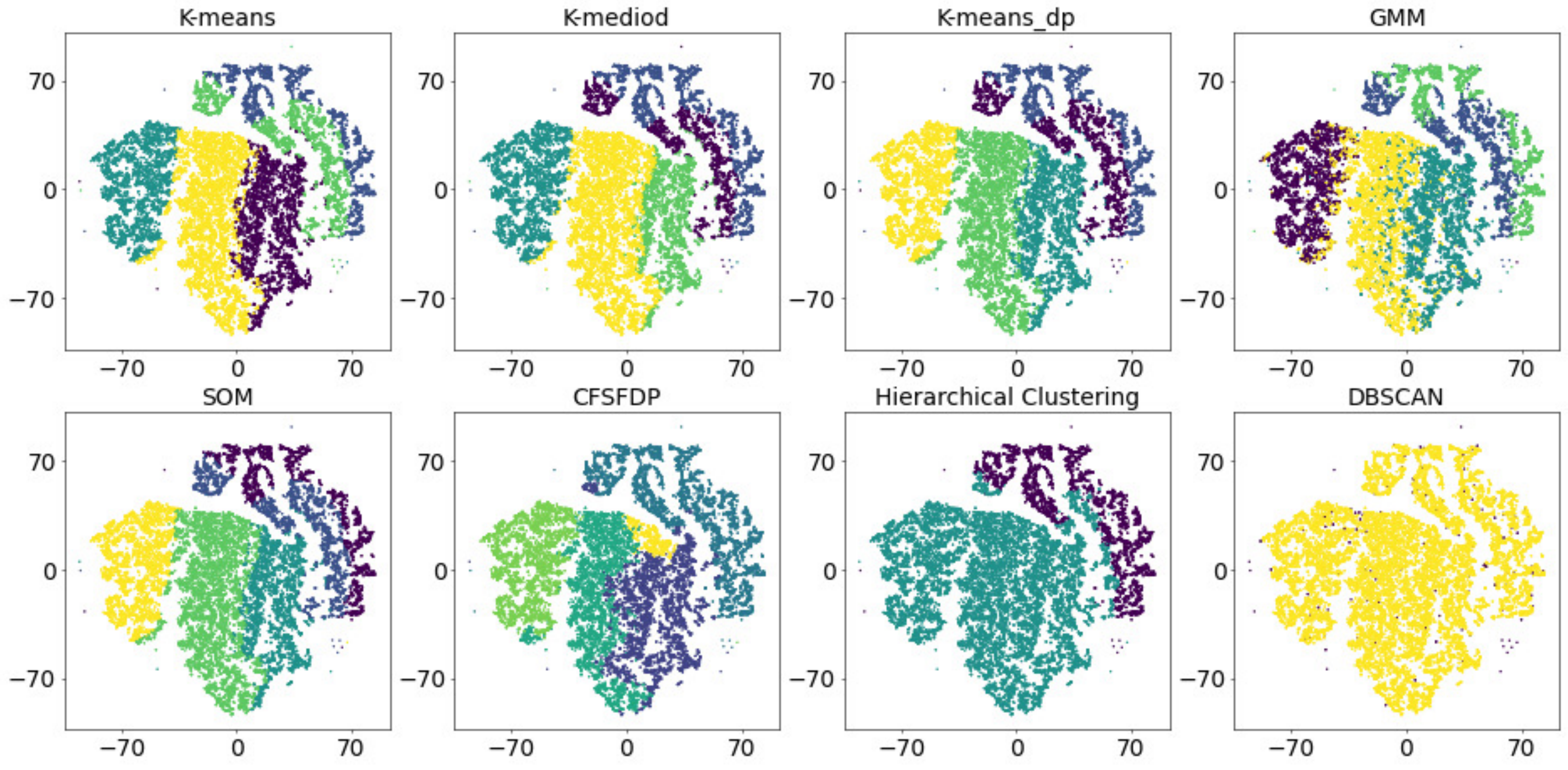}
\caption{t-SNE distribution of eight algorithms on SDSS spectra.  Each subgraph represents the results of one clustering algorithm on spectra from SDSS. Different colors in each subgraph represent different classes in the clustering results. The same color in different subplots is not necessarily the same class.}
\label{fig:Algorithms_SDSS_tsne} 
\end{figure*}

\begin{figure*}
\centering
\includegraphics[width=15.24cm,height=4.872cm]{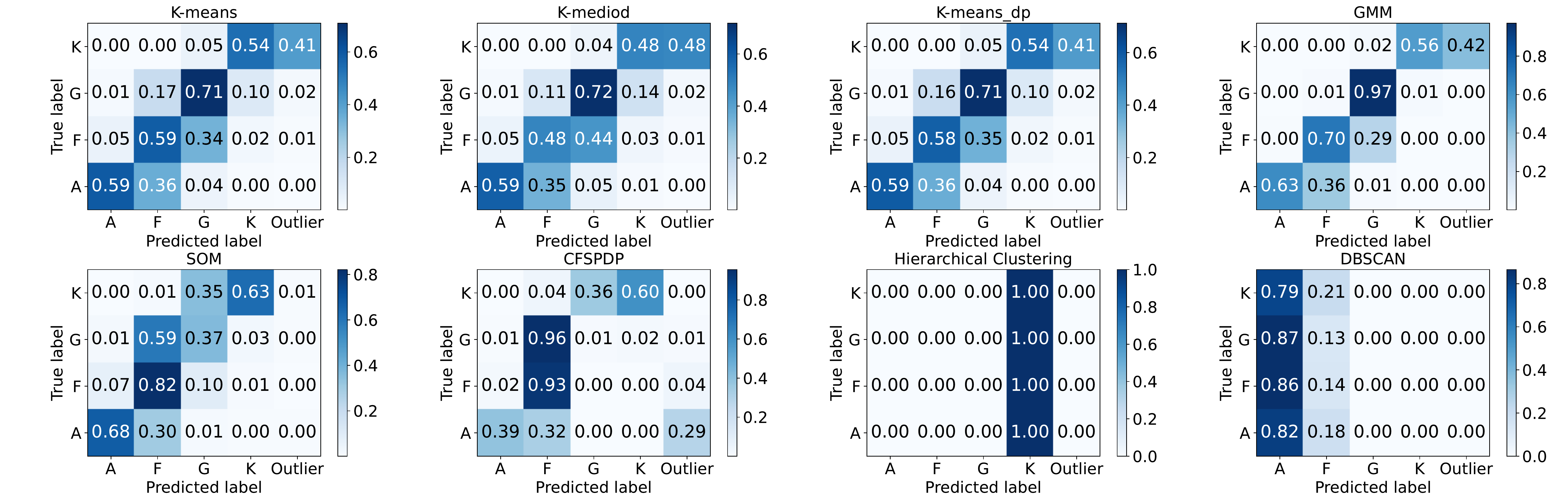}
\caption{Confusion matrix of eight algorithms on LAMOST spectra. Predicted label: color and digit in each cell are the consistent probability between predicted label and true label. Color is in direct proportion to the figure: bigger numbers and deeper color.}
\label{fig:Algorithms_LAMOST_confusion} 
\end{figure*}

\begin{figure*}
\centering
\includegraphics[width=15.24cm,height=4.872cm]{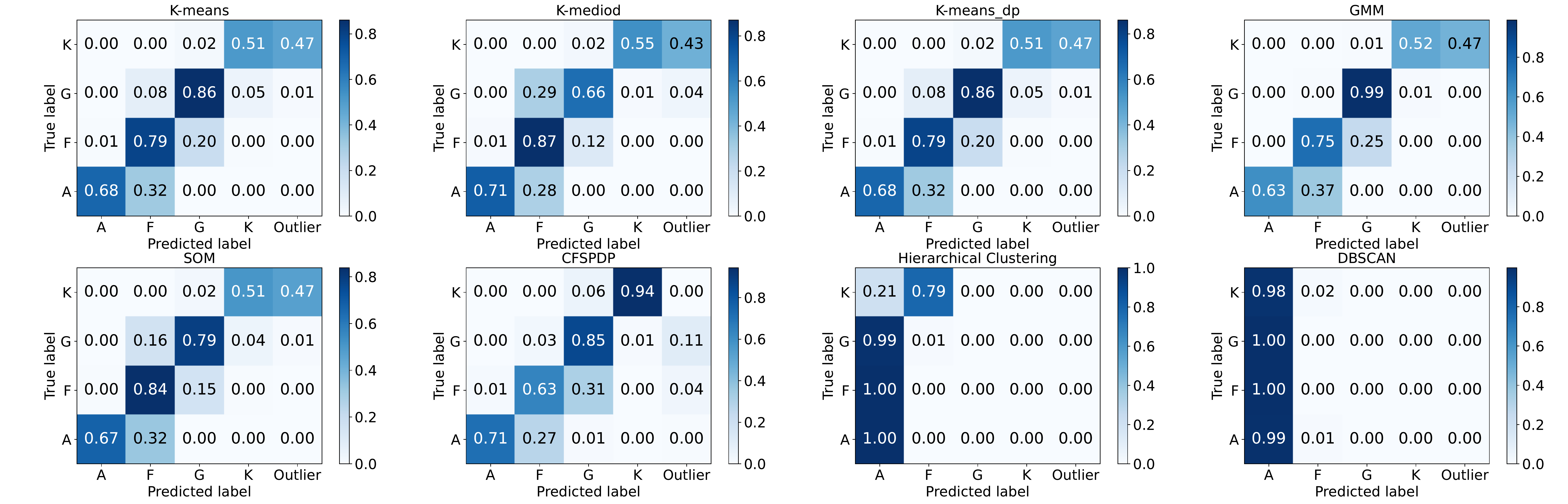}
\caption{Confusion matrix of eight algorithms on SDSS spectra. Predicted label: color and digit in each cell are the consistent probability between predicted label and true label. Color is in direct proportion to the figure: bigger numbers and deeper color.}
\label{fig:Algorithms_SDSS_confusion} 
\end{figure*}

There are lots of overlapping sources of LAMOST and SDSS, we construct two datasets from them to compare clustering performances. Spectra of SDSS are selected from DR16. The task is stellar classification and the datasets are shown in Table \ref{table_lamost_sdss}. The clustering results are shown in Fig. \ref{fig:Algorithms_LAMOST_bar} - Fig. \ref{fig:Algorithms_SDSS_confusion} .

From the t-SNE maps of LAMOST and SDSS (Fig. \ref{fig:Algorithms_LAMOST_SDSS_true_tsne}), it can be seen that K-type star overlap less with other three types. However, A-type and F-type, F-type and G-type have more overlaps. The map of SDSS is clearer than that of LAMOST, it has fewer noise points, mainly because there is always some noise in the splicing at the red and blue segments of LAMOST spectra.

In the process of clustering, we set 5 cluster centers for 4 classes of stars so that clustering algorithms can separate F and G type stellar spectra, but meanwhile, K-type star may be divided into two categories. 

The clustering accuracy of SDSS is slightly better than that of LAMOST (Fig. \ref{fig:Algorithms_LAMOST_bar}). This is mainly due to the calibration quality, compared to SDSS, the fiber-to-fiber sensitivity variations in LAMOST sometimes leads to wrong overall calibration which will affect the classification of spectra.  And the mismatch between red/blue segments in some spectra also introduces problems.
In the clustering results, hierarchical clustering and DBSCAN also cluster most spectra into one class. One thing worth noticing is that GMM can classify G-type stars very well both in LAMOST and SDSS even they have parts of spectra similar to F-type stars. SOM and CFSFDP cannot separate F-type and G-type stars well on LAMOST spectra, but can separate them well on SDSS, this shows that the degree of spectra discernment of SDSS is higher than LAMOST.

\begin{table}
\centering
\caption{Datasets of LAMOST and SDSS}
\label{table_lamost_sdss}
\resizebox{\linewidth}{!}{
\begin{tabular}{lcllcllcl} 
\hline
                  & Survey                  &                   &                   & Type &  &  & Data volume &   \\ 
\hline
\multirow{4}{*}{} & \multirow{4}{*}{LAMOST} & \multirow{4}{*}{} & \multirow{4}{*}{} & A    &  &  & 5824        &   \\
                  &                         &                   &                   & F    &  &  & 5380        &   \\
                  &                         &                   &                   & G    &  &  & 4151        &   \\
                  &                         &                   &                   & K    &  &  & 6240        &   \\ 
\hline
\multirow{4}{*}{} & \multirow{4}{*}{SDSS}   & \multirow{4}{*}{} & \multirow{4}{*}{} & A    &  &  & 5797        &   \\
                  &                         &                   &                   & F    &  &  & 5355        &   \\
                  &                         &                   &                   & G    &  &  & 4144        &   \\
                  &                         &                   &                   & K    &  &  & 6229        &   \\
\hline
\end{tabular}
}
\end{table}

\subsection{Performance of algorithms on different quality of spectra}

The quality of spectra observed by telescope will be affected by the observation conditions, and the signal-to-noise ratio of the spectra will be low if the conditions are not good. Noise in the spectra will increase the difficulty of spectral analysis, so it is important to find out clustering methods which are insensitive to noise. In this section, three spectral datasets of different S/Ns are constructed to study the robustness of clustering algorithms.

\begin{table}

    \centering
    \caption{Three datasets with different S/Ns}
    \begin{threeparttable} 
    \resizebox{\linewidth}{!}{
    \begin{tabular}{lcccc}
    \hline
        ~ & S/N of g-band & S/N of i-band & Size & Data  \\ \hline
        Dataset H & >30 & >30 & 20000 & PCA \\ 
        Dataset M & 10-30 & 10-30 & 20000 & PCA \\ 
        Dataset L & <10 & <10 & 20000 & PCA \\ \hline
    \end{tabular}
    }
    \label{table:datasets_diff_snr}
    \begin{tablenotes}
    \footnotesize
    \item[1] H, M and L represent high S/N, medium S/N and low S/N.
    \item[2] There are equal numbers of four types of stars (A,F,G,K) in each dataset.
    \end{tablenotes}
    \end{threeparttable}
    
\end{table}

Three different S/Ns include high (g band > 30 and i band > 30), middle (g band:10-30 or i band:10-30) and low (g band < 10 and i band < 10). The task of this comparative experiment is to cluster stellar spectra because the number of galaxy and quasar spectra with S/N > 30 is very small. Spectra with S/N > 10 are considered qualified, so we also add them to the accuracy chart. Table \ref{table:datasets_diff_snr} shows the configuration of datasets used in this section.


\begin{figure*}
\centering
\includegraphics[width=12.288cm,height=5.6cm]{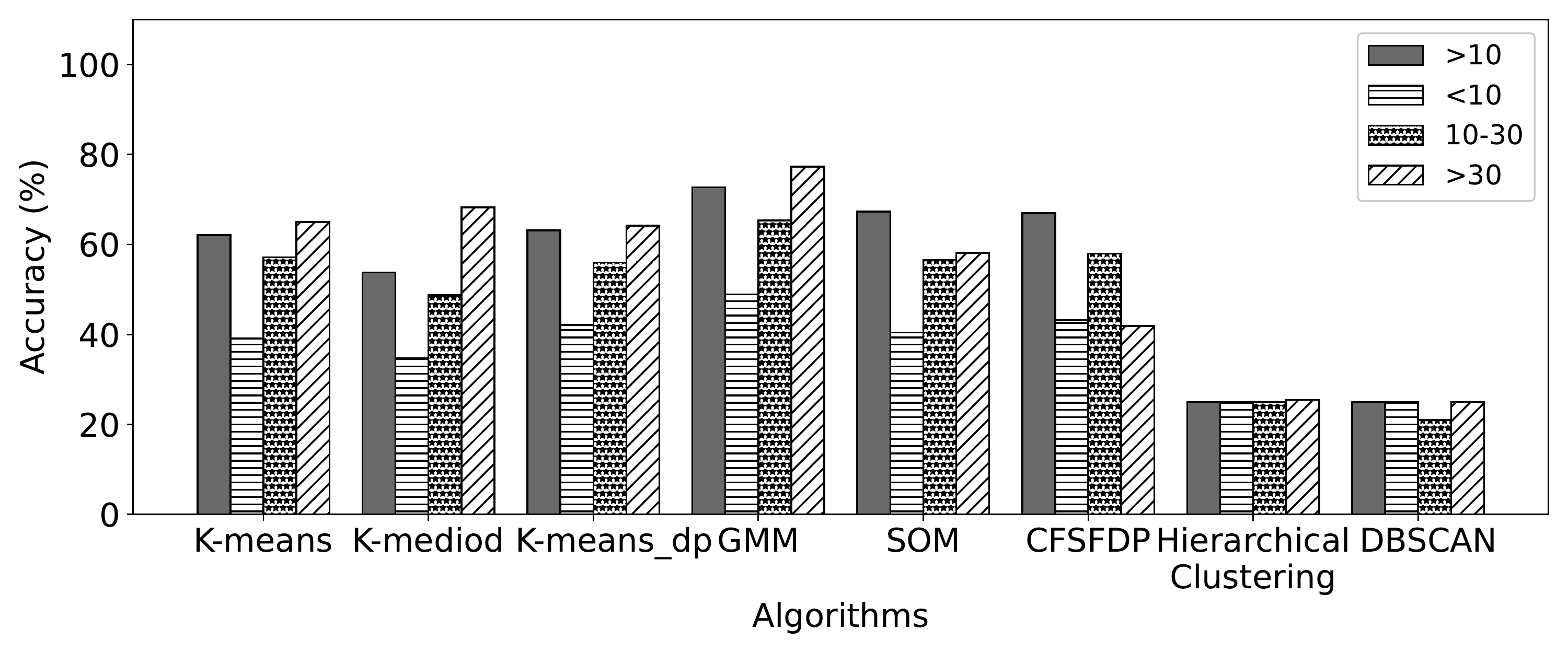}
\caption{Accuracy of eight algorithms on spectra of different S/Ns. Four bars represent four S/Ns.}
\label{fig:Algorithms_accuracy_10-_-10_10-30_30-_1w_PCs_bar}
\end{figure*}

\begin{figure*} 
\centering
\includegraphics[width=10.75cm,height=3.15cm]{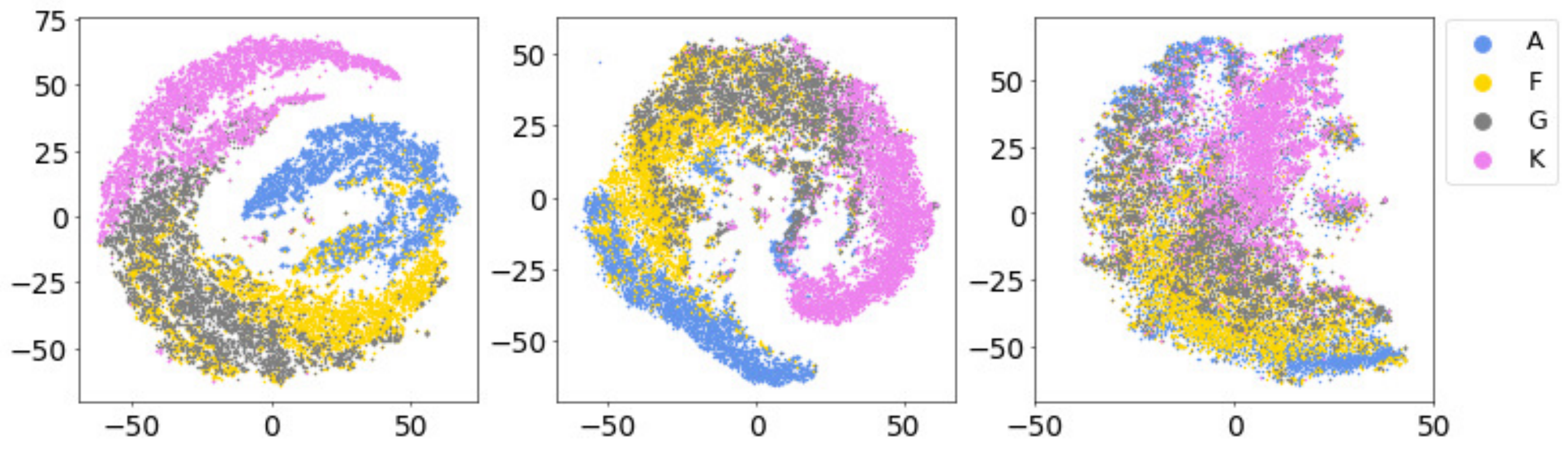}
\caption{t-SNE distribution of true labels of A/F/G/K stars on different S/Ns. Left to right: >30, 10-30, <10. Four colors represent A, F, G, K stars.}
\label{fig:Raw Label_AFGK_SNRs_1w_pca_tsne} 
\end{figure*}

\begin{figure*}
\centering
\includegraphics[width=13.06cm,height=5.95cm]{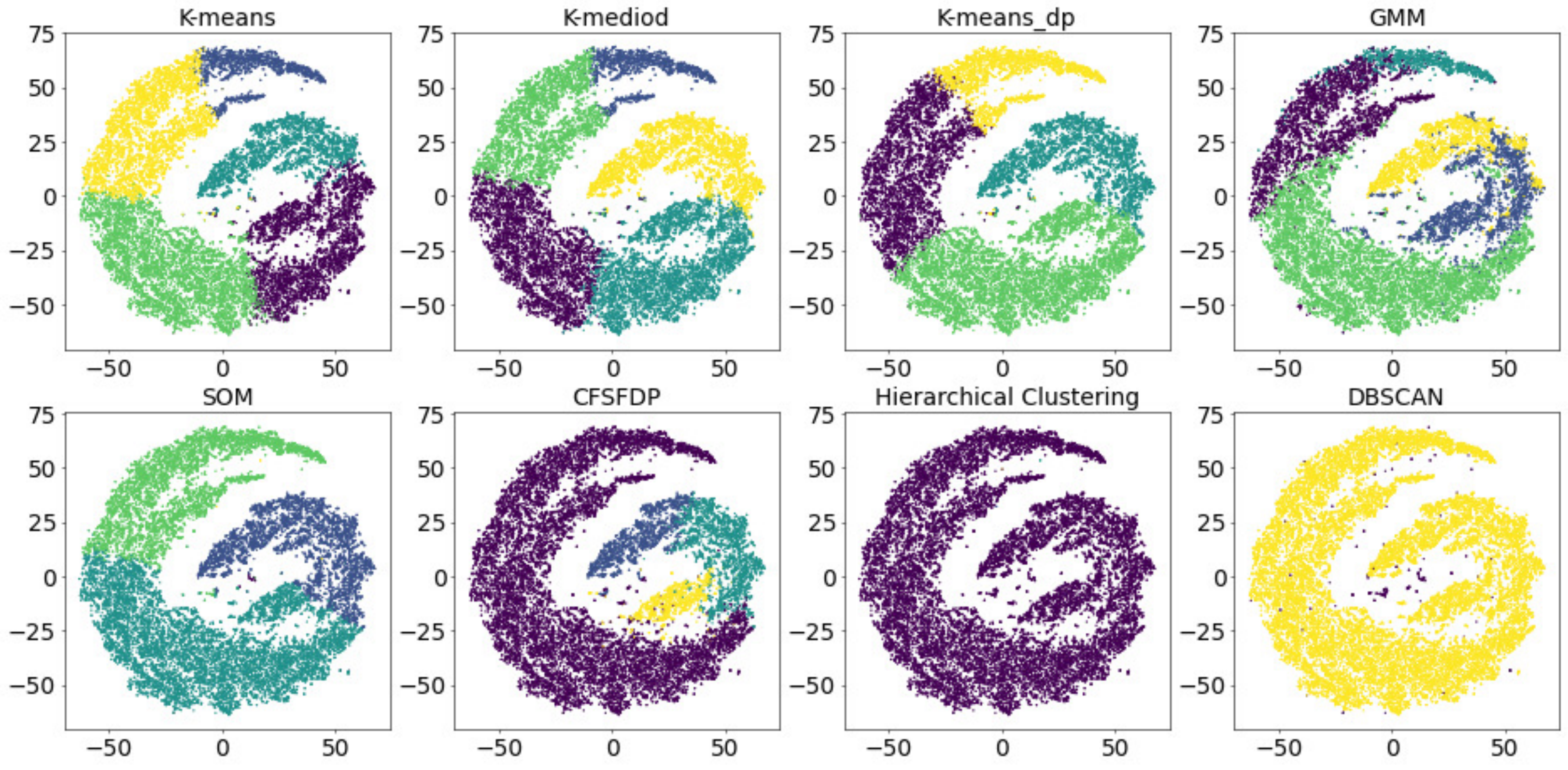}
\caption{t-SNE distribution of eight algorithms on high S/N of spectra.  Each subgraph represents the results of one clustering algorithm on the high S/N of spectra. Different colors in each subgraph represent different classes in the clustering results. The same color in different subplots is not necessarily the same class.}
\label{fig:Algorithm_30-_1w_PCs_tsne_AFGK} 
\end{figure*}

\begin{figure*}
\centering
\includegraphics[width=13.06cm,height=5.95cm]{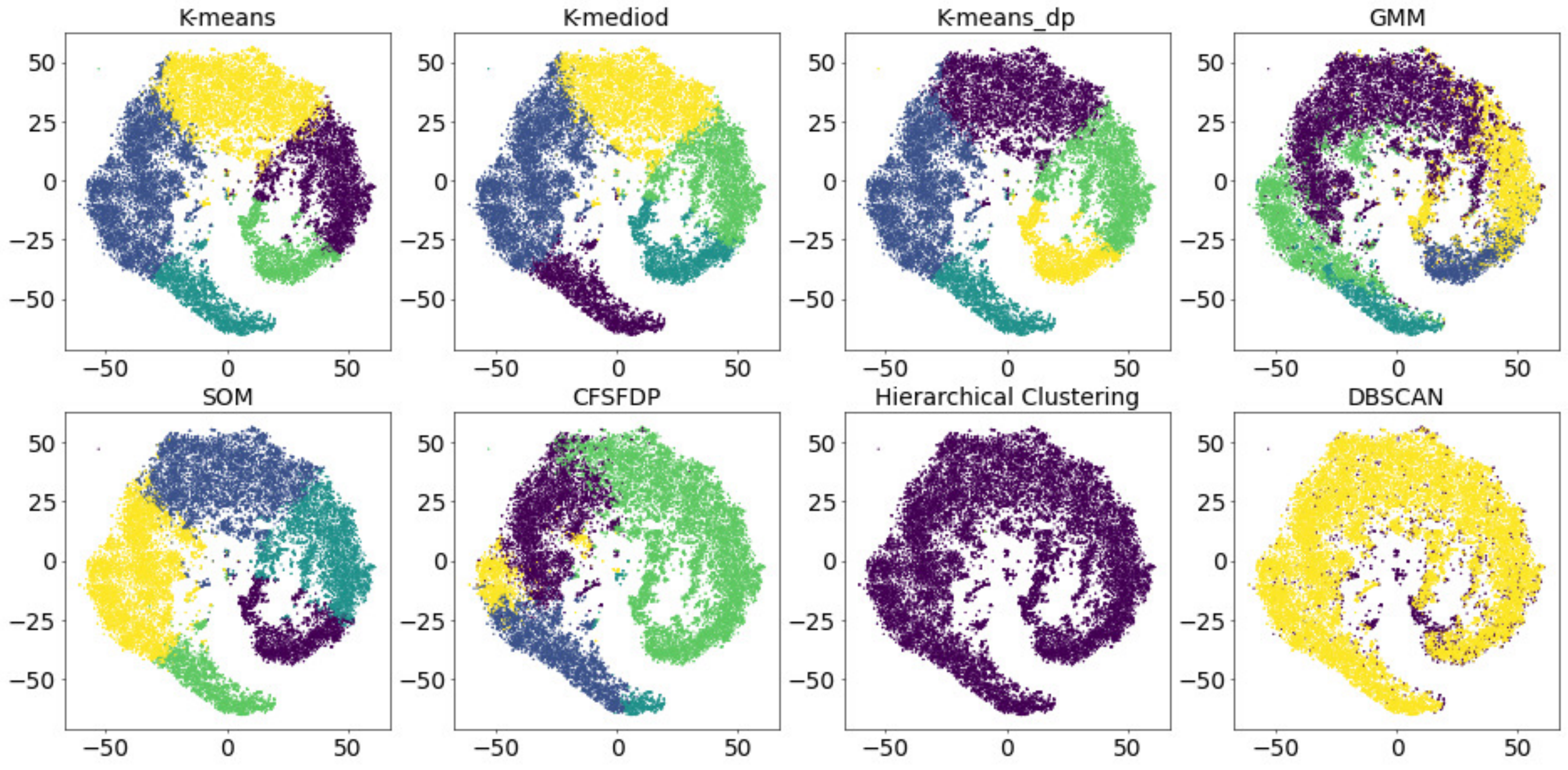}
\caption{t-SNE distribution of eight algorithms on medium S/N of spectra.  Each subgraph represents the results of one clustering algorithm on the medium S/N of spectra. Different colors in each subgraph represent different classes in the clustering results. The same color in different subplots is not necessarily the same class.}
\label{fig:Algorithm_10-30_1w_PCs_tsne_AFGK} 
\end{figure*}

\begin{figure*}
\centering
\includegraphics[width=13.06cm,height=5.95cm]{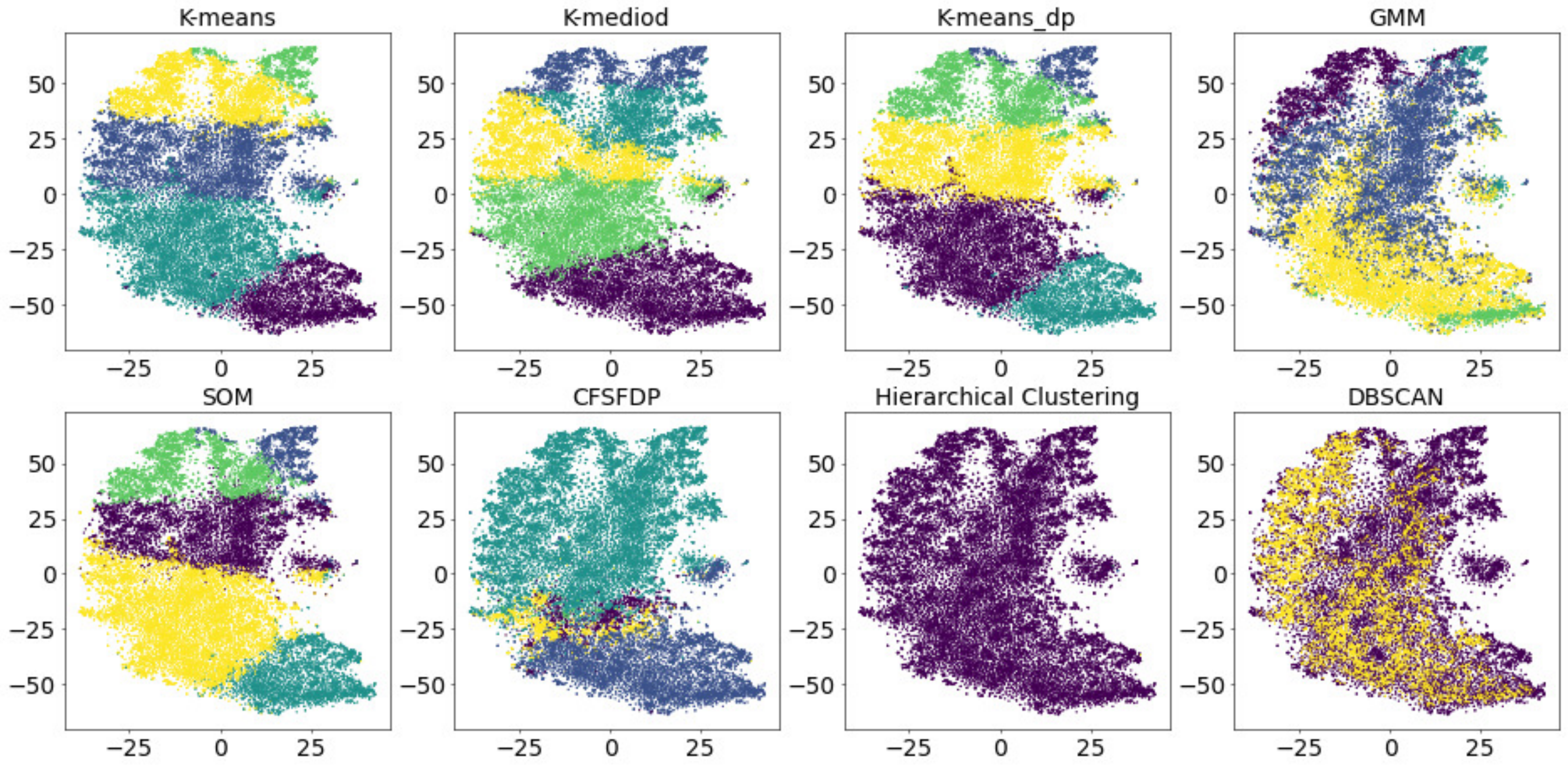}
\caption{t-SNE distribution of eight algorithms on low S/N of spectra.  Each subgraph represents the results of one clustering algorithm on the low S/N of spectra. Different colors in each subgraph represent different classes in the clustering results. The same color in different subplots is not necessarily the same class.}
\label{fig:algorithms_tsne_-10} 
\end{figure*}

\begin{figure*}
\centering
\includegraphics[width=13.97cm,height=4.5cm]{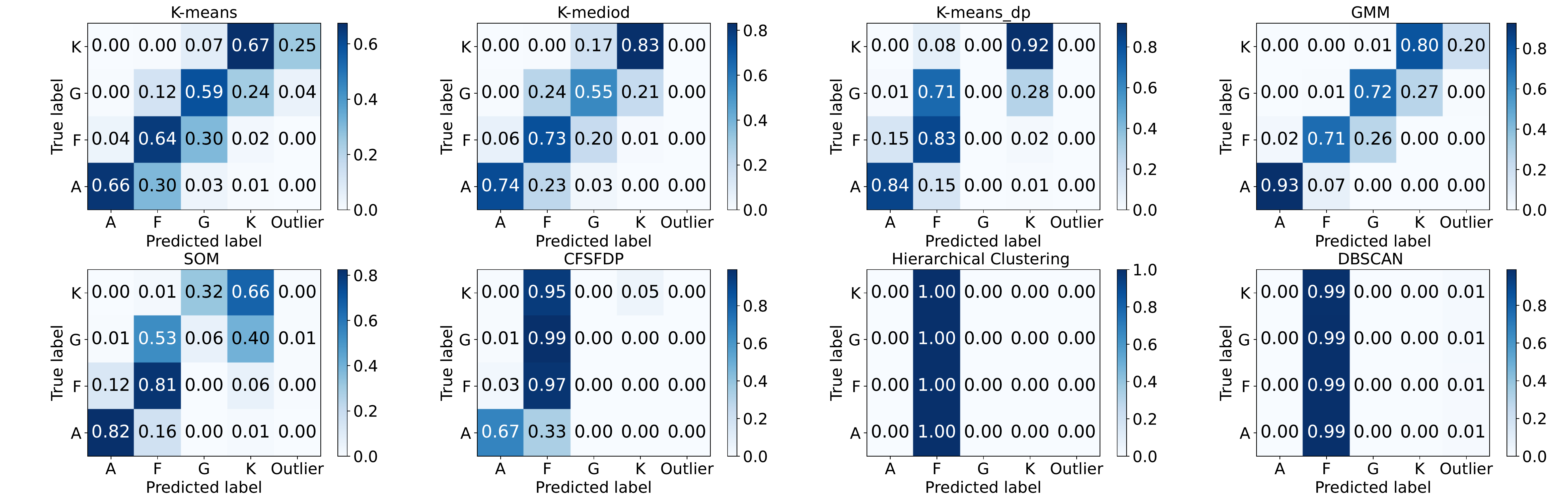}
\caption{Confusion matrix of eight algorithms on high S/N of spectra. Predicted label: color and digit in each cell are the consistent probability between predicted label and true label. Color is in direct proportion to the figure: bigger numbers and deeper color.}
\label{fig:Algorithms_30-_1w_PCs_confusion_matrix} 
\end{figure*}

\begin{figure*}
\centering
\includegraphics[width=13.97cm,height=4.5cm]{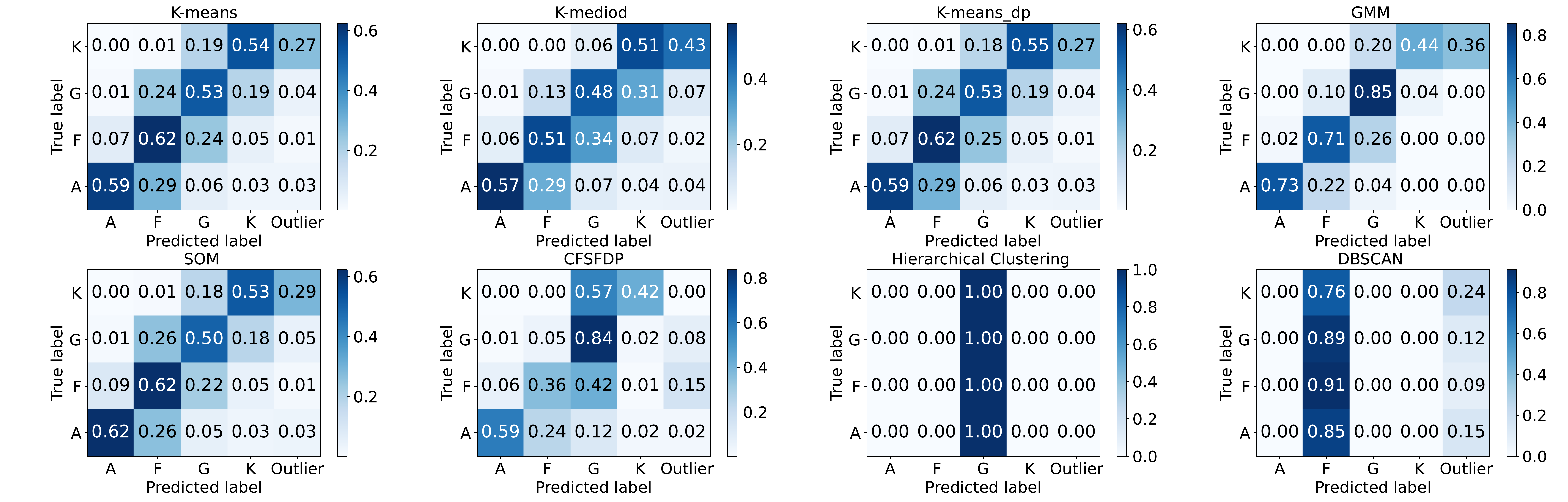}
\caption{Confusion matrix of eight algorithms on medium S/N of spectra. Predicted label: color and digit in each cell are the consistent probability between predicted label and true label. Color is in direct proportion to the figure: bigger numbers and deeper color.}
\label{fig:Algorithms_10-30_1w_PCs_confusion_matrix} 
\end{figure*}

\begin{figure*}
\centering
\includegraphics[width=13.97cm,height=4.5cm]{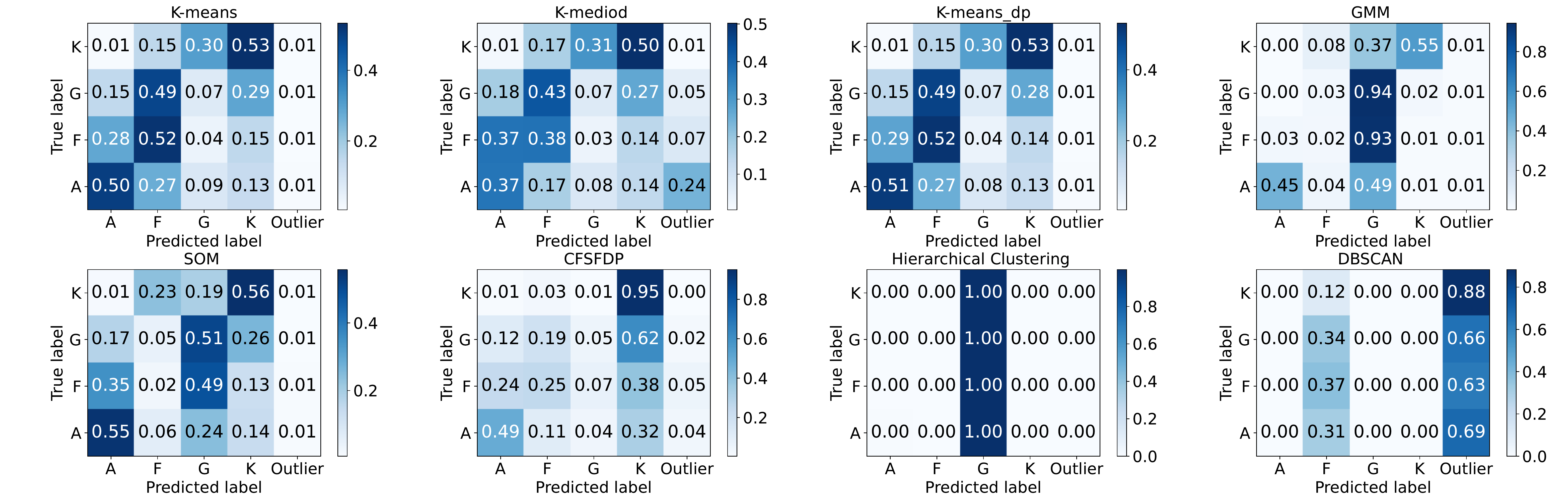}
\caption{Confusion matrix of eight algorithms on low S/N of spectra. Predicted label: color and digit in each cell are the consistent probability between predicted label and true label. Color is in direct proportion to the figure: bigger numbers and deeper color.}
\label{fig:Algorithms_-10_1w_PCs_confusion_matrix} 
\end{figure*}

Fig. \ref{fig:Algorithms_accuracy_10-_-10_10-30_30-_1w_PCs_bar} shows the average accuracy of eight algorithms on different S/Ns datasets and the accuracy is proportional to the signal-to-noise ratio. As before, GMM still has the highest accuracy. Fig. \ref{fig:Raw Label_AFGK_SNRs_1w_pca_tsne} is the t-SNE distribution of datasets and Fig. \ref{fig:Algorithm_30-_1w_PCs_tsne_AFGK} - Fig. \ref{fig:Algorithms_-10_1w_PCs_confusion_matrix} are the clustering results and their confusion matrices. 


Data distributions vary in different S/Ns. There seems to be less overlaps in S/N $>30$, but it is serious in S/N $<10$, and S/N:10-30 is between the above two. Results of partition-based algorithms also tend to be circle. GMM performs very well on the data with S/N $> 10$ and the clustering accuracy of A stars reaches 92.5\%. But there is a phenomenon that misclassification always occurs in backward types but rarely happens in the forward types. For example, in the results of GMM on high S/N, 27 \% of G-type stars are classified as K-type, but only 0.5 \% of K-type stars were classified as G-type and 19.8 \% of K-type stars are grouped separately. Compared with K-means, GMM is more susceptible to signal-to-noise ratio, although GMM has good performance.

Since there is excessive noise in the spectra with S/N < 10, clustering algorithms perform poorly, and the results are of little use for the spectral classification tasks. So for the low S/N spectra, other effective methods should be used for analysis.


\begin{figure*}
\centering
\includegraphics[width=13.97cm,height=4.5cm]{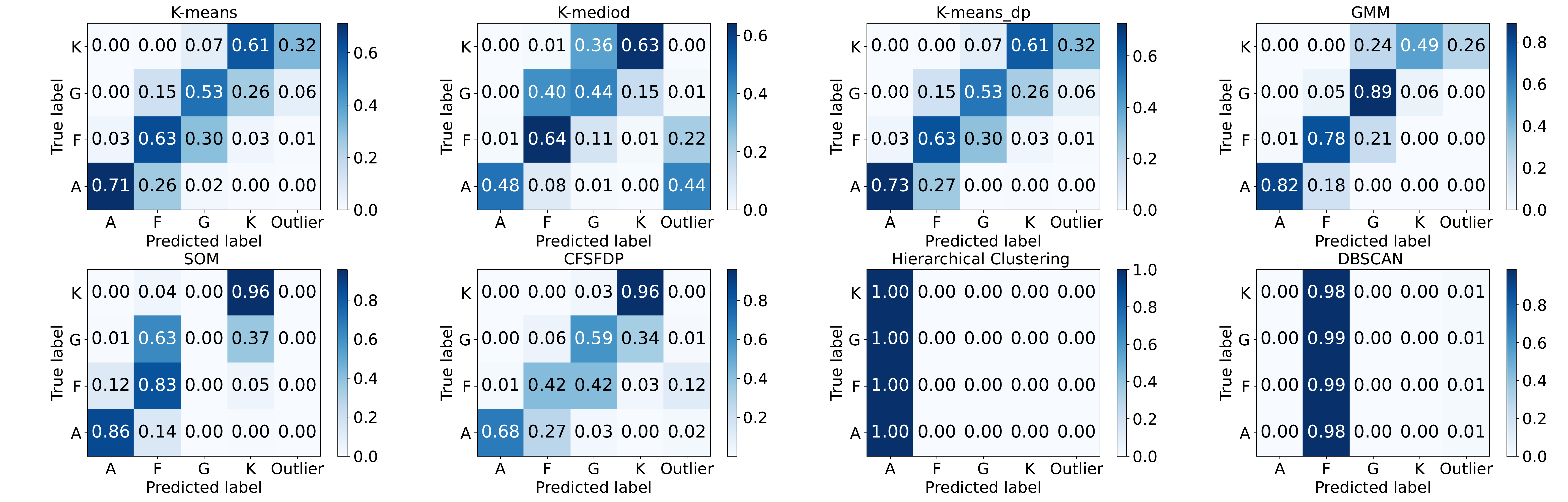}
\caption{Confusion matrix of eight algorithms on S/N $>$10. Predicted label: color and digit in each cell are the consistent probability between predicted label and true label. Color is in direct proportion to the figure: bigger numbers and deeper color.}
\label{fig:Algorithms_10-_1w_PCs_confusion_matrix} 
\end{figure*}



\subsection{Performance of algorithms on different data volumes}

\begin{table}

    \centering
    \caption{Four datasets with different volumes of spectra.}
    \resizebox{\linewidth}{!}{
    \begin{tabular}{lcccccc}
    \hline
        ~ & ~ & Size & ~ & S/N & ~ & Data  \\ \hline
        Dataset s & ~ & 8000 & ~ & >10 & ~ & PCA \\ 
        Dataset m & ~ & 20000 & ~ & >10 & ~ & PCA \\ 
        Dataset l & ~ & 40000 & ~ & >10 & ~ & PCA \\ 
        Dataset e & ~ & 80000 & ~ & >10 & ~ & PCA \\ \hline
    \end{tabular}
    }
    \label{table:datasets_diff_volume}
    
\end{table}

In machine learning tasks, the size of dataset is always a factor that affects the results. When the amount of data is small, overfitting will occur. But if the size is too large, some algorithms with high time complexity will take a long time to run and some algorithms may not be able to run. So, this section performs eight clustering algorithms on spectral datasets size of 8000, 20000, 40000 and 80000 to analyse their effectiveness and efficiency. The clustering task is also stellar classification and the number of four types of stars in the dataset is the same. Table \ref{table:datasets_diff_volume} shows the configuration of four datasets.

\begin{figure*}
\centering
\includegraphics[width=12cm,height=5.2cm]{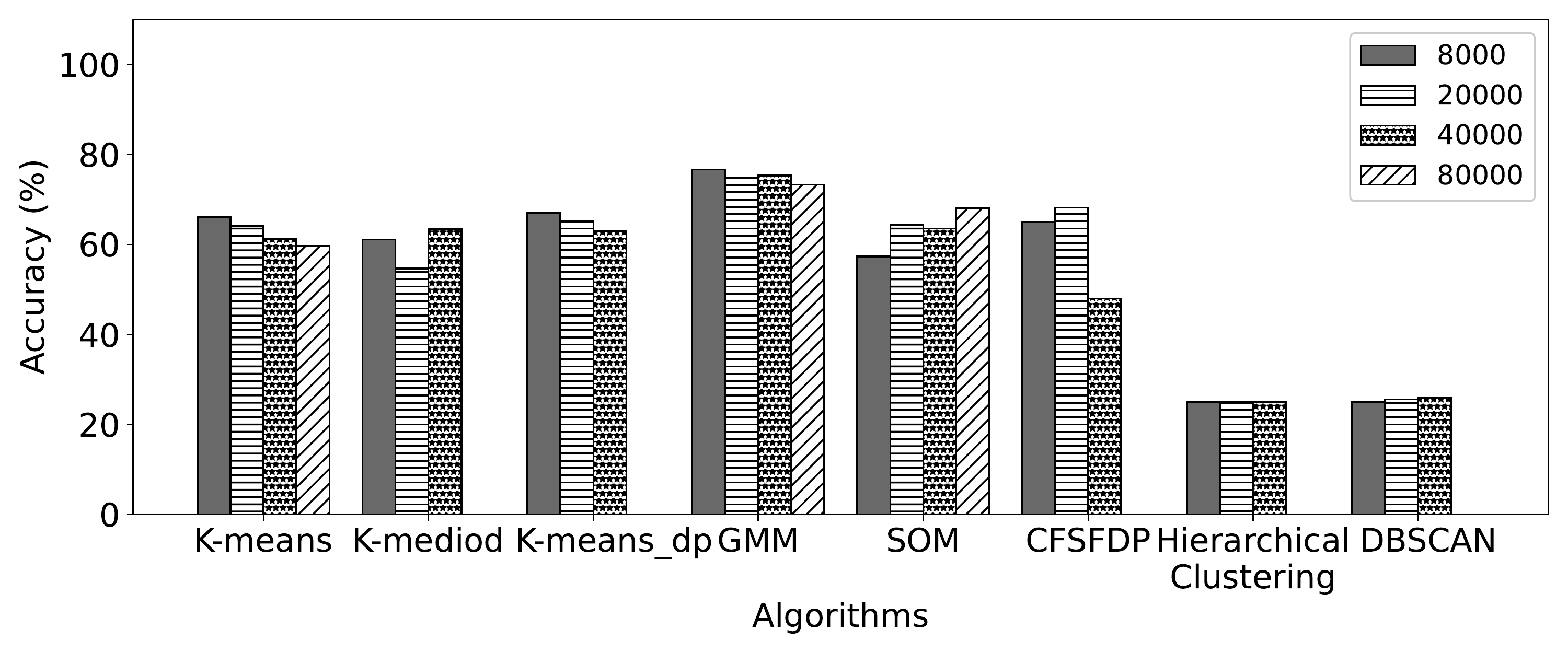}
\caption{Accuracy of algorithms on different data volumes. Four bars represent four data volumes.}
\label{fig:Algorithms_accuracy_2k_5k_1w_5w_raw_10-_bar}
\end{figure*}

\begin{figure*}
\centering
\includegraphics[width=12.25cm,height=2.7cm]{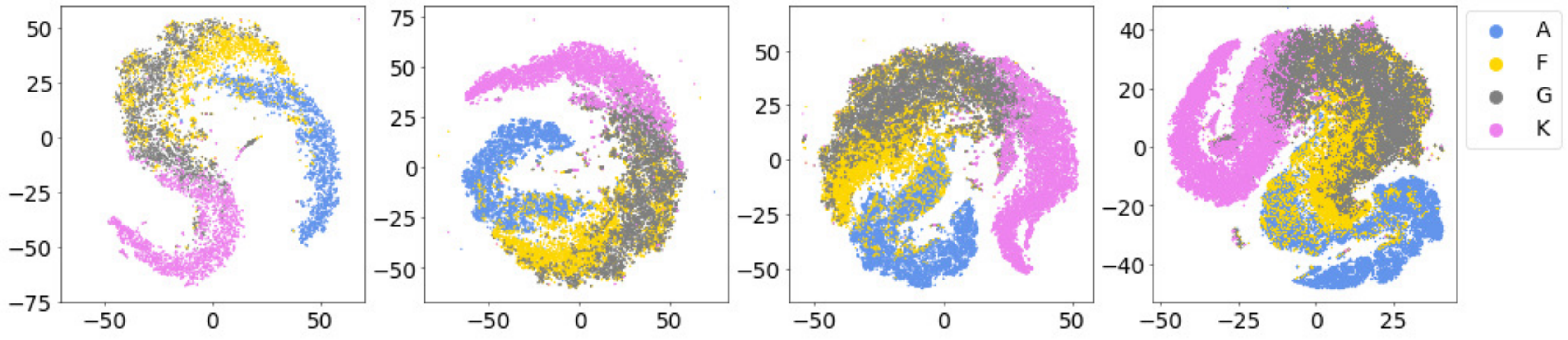}
\caption{t-SNE of eight algorithms on data volumes of 8000, 20000, 40000 and 80000. Four colors represent A, F, G, K stars.}
\label{fig:Algorithms_2k_5k_1w_raw_10-_tsne_raw_label} 
\end{figure*}

\begin{figure*}
\centering
\includegraphics[width=15cm,height=7.2cm]{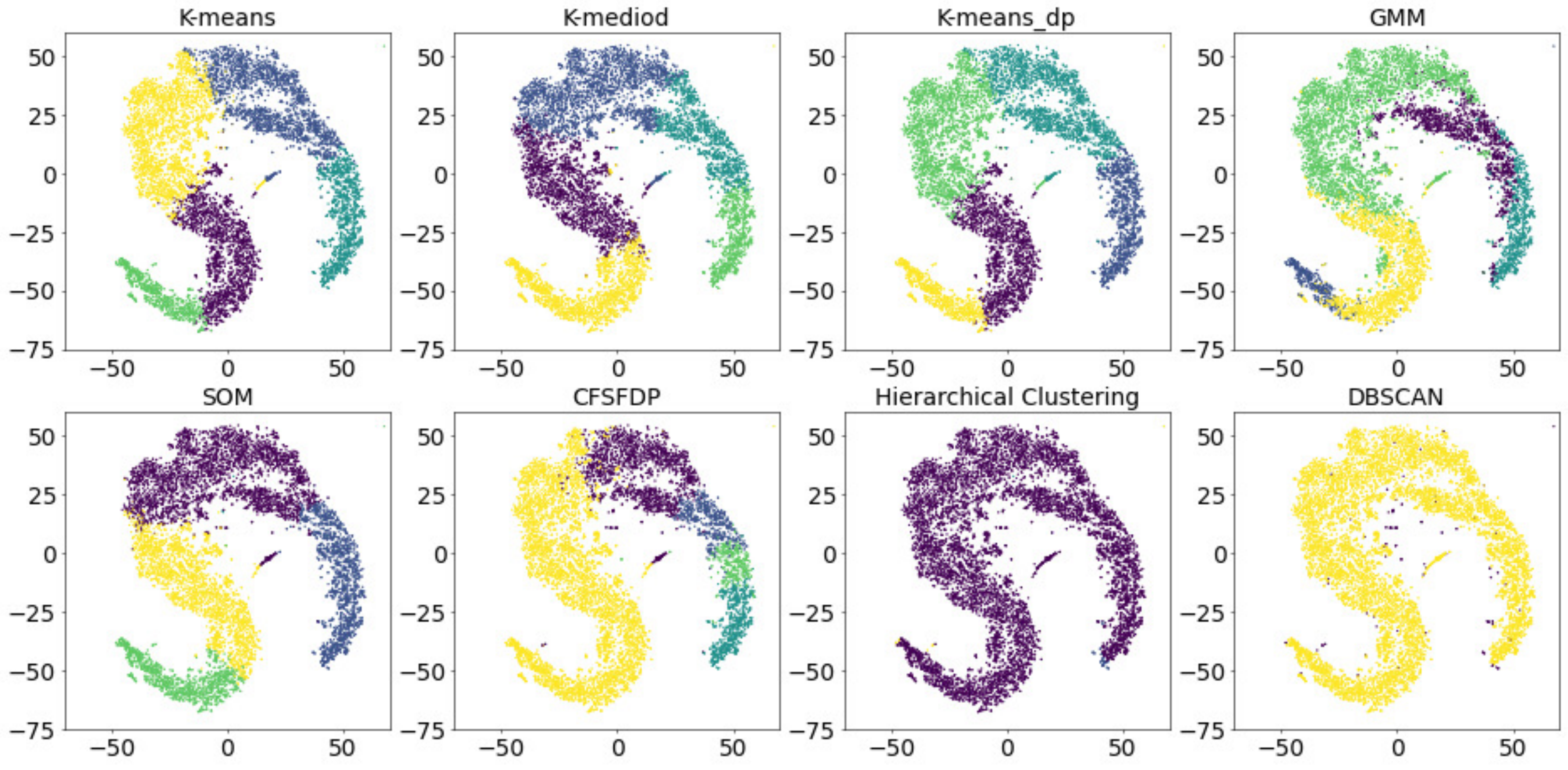}
\caption{t-SNE of eight algorithms on data volume of 8000.  Each subgraph represents the results of one clustering algorithm on the data volume of 8000. Different colors in each subgraph represent different classes in the clustering results. The same color in different subplots is not necessarily the same class.}
\label{fig:Algorithms_2k_raw_10-_tsne} 
\end{figure*}

\begin{figure*}
\centering
\includegraphics[width=15cm,height=7.2cm]{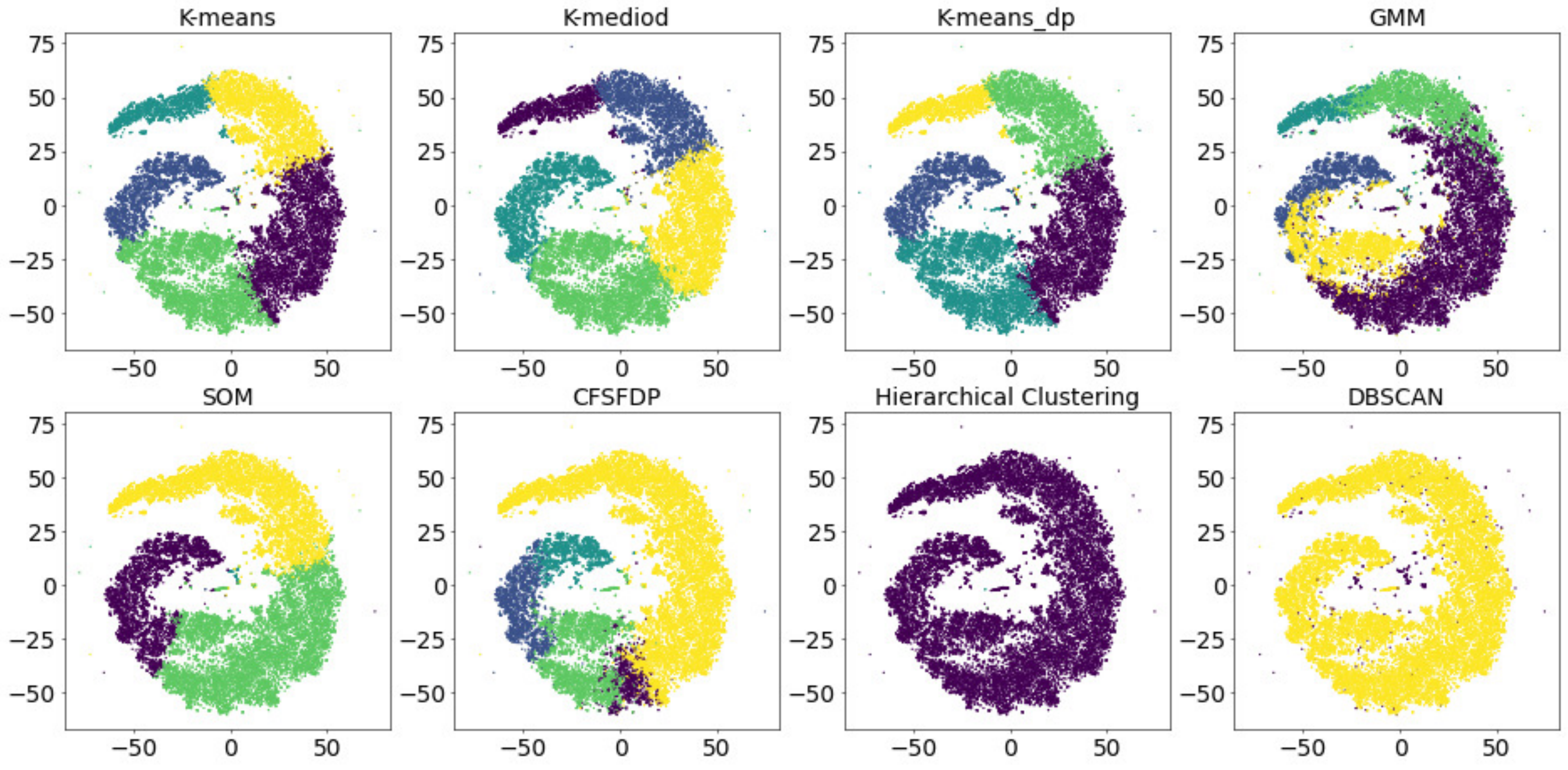}
\caption{t-SNE of eight algorithms on data volume of 20000.  Each subgraph represents the results of one clustering algorithm on data volume of 20000. Different colors in each subgraph represent different classes in the clustering results. The same color in different subplots is not necessarily the same class.}
\label{fig:Algorithms_5k_raw_10-_tsne} 
\end{figure*}

\begin{figure*}
\includegraphics[width=15cm,height=7.2cm]{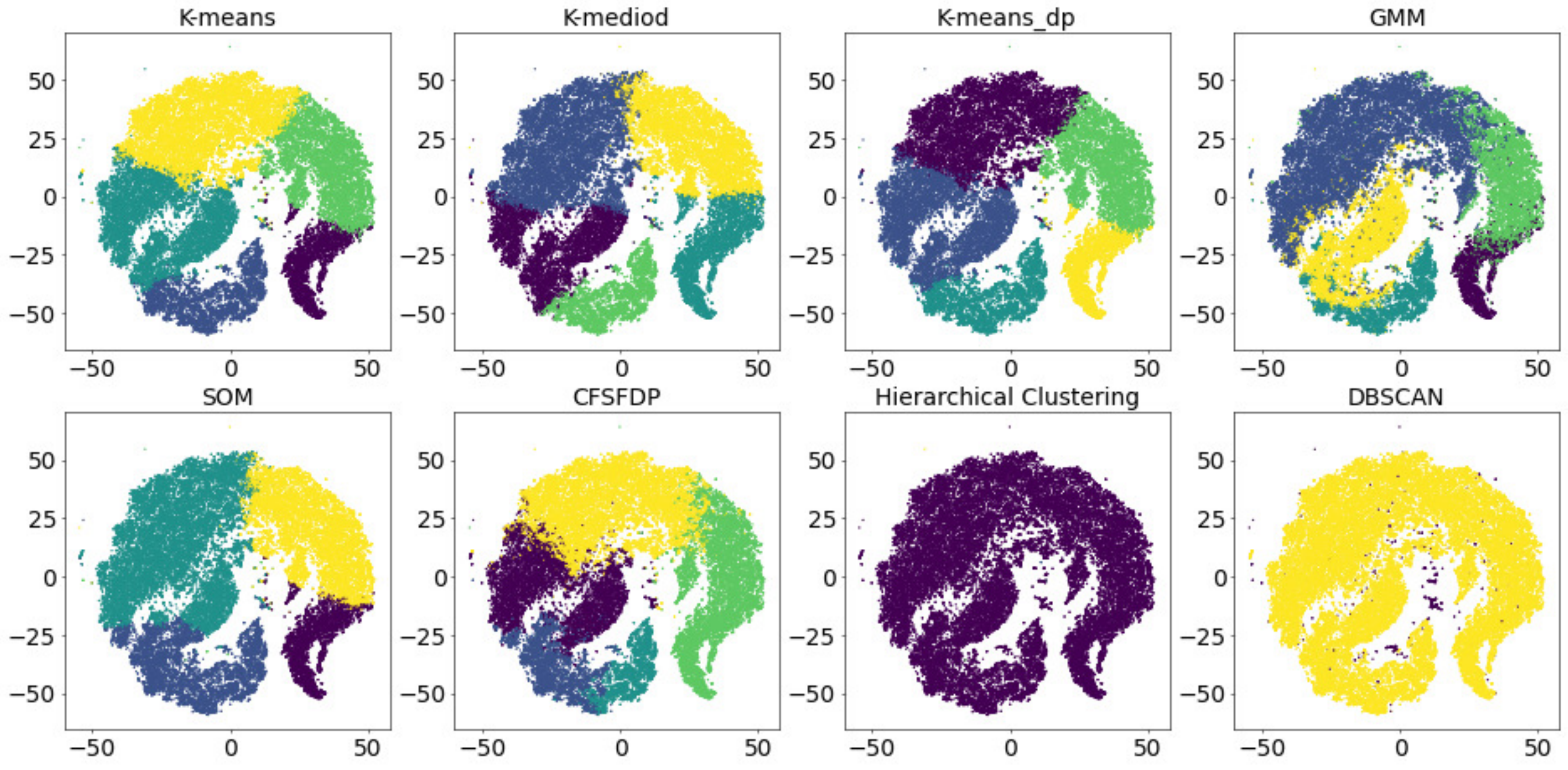}
\caption{t-SNE of eight algorithms on data volume of 40000.  Each subgraph represents the results of one clustering algorithm on data volume of 40000. Different colors in each subgraph represent different classes in the clustering results. The same color in different subplots is not necessarily the same class.}
\label{fig:Algorithms_1w_raw_10-_tsne} 
\end{figure*}

\begin{figure*}
\centering
\includegraphics[width=15.3cm,height=5.1cm]{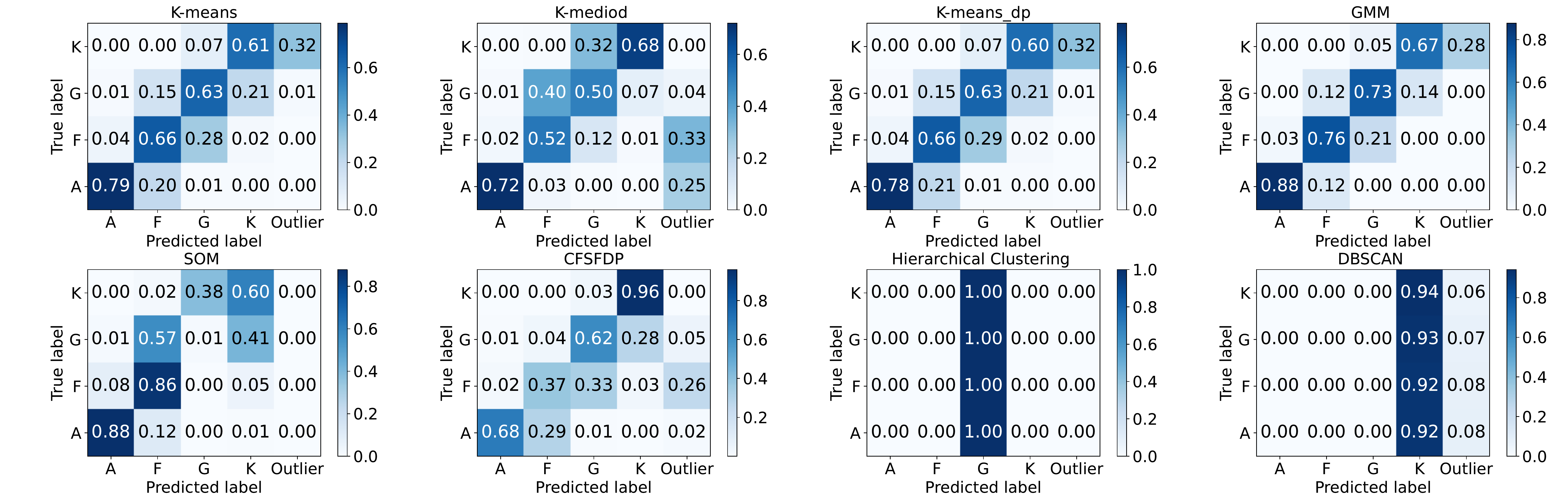}
\caption{Confusion matrix of eight algorithms on data volume of 8000. Predicted label: color and digit in each cell are the consistent probability between predicted label and true label. Color is in direct proportion to the figure: bigger numbers and deeper color.}
\label{fig:Algorithms_2k_raw_10-_confusion_matrix} 
\end{figure*}

\begin{figure*}
\centering
\includegraphics[width=15.3cm,height=5.1cm]{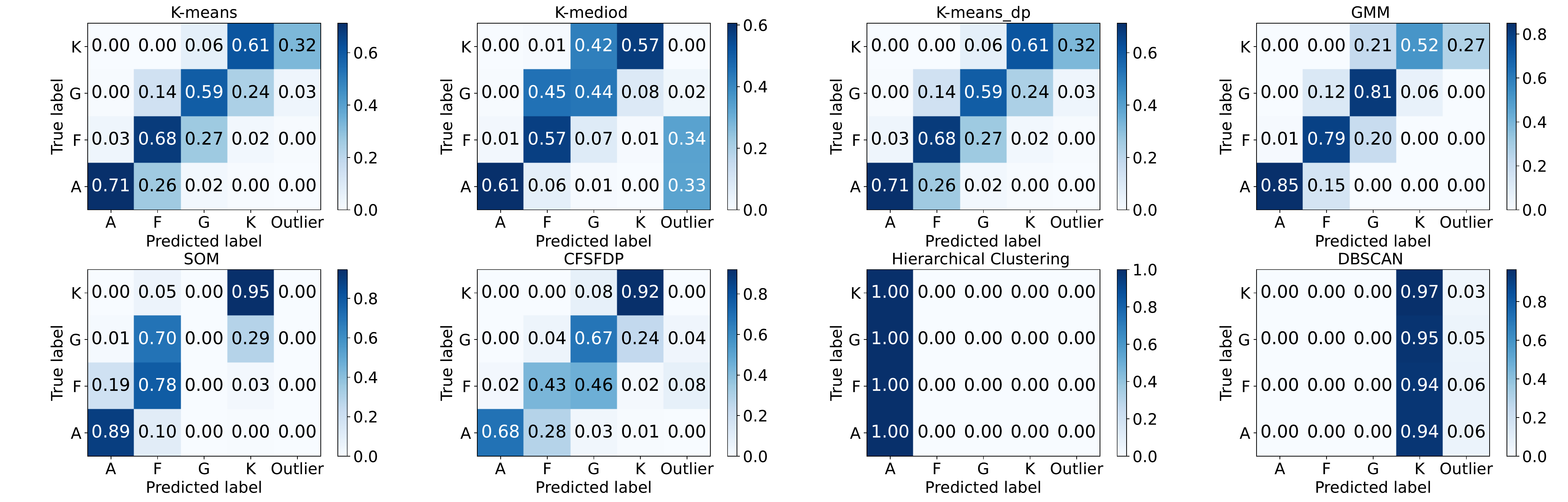}
\caption{Confusion matrix of eight algorithms on data volume of 20000. Predicted label: color and digit in each cell are the consistent probability between predicted label and true label. Color is in direct proportion to the figure: bigger numbers and deeper color.}
\label{fig:Algorithms_5k_raw_10-_confusion_matrix} 
\end{figure*}

\begin{figure*}
\centering
\includegraphics[width=15.3cm,height=5.1cm]{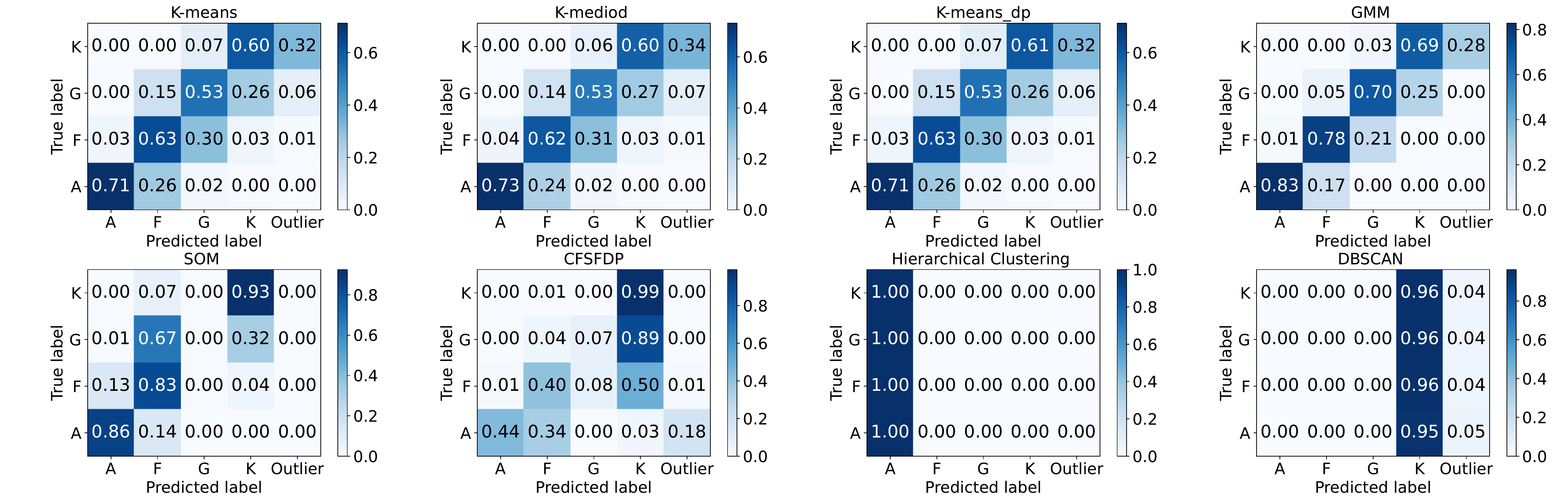}
\caption{Confusion matrix of eight algorithms on data volume of 40000. Predicted label: color and digit in each cell are the consistent probability between predicted label and true label. Color is in direct proportion to the figure: bigger numbers and deeper color.}
\label{fig:Algorithms_1w_raw_10-_confusion_matrix-eps-converted-to.pdf} 
\end{figure*}

\begin{figure*}
\centering
\includegraphics[width=13cm,height=2.89cm]{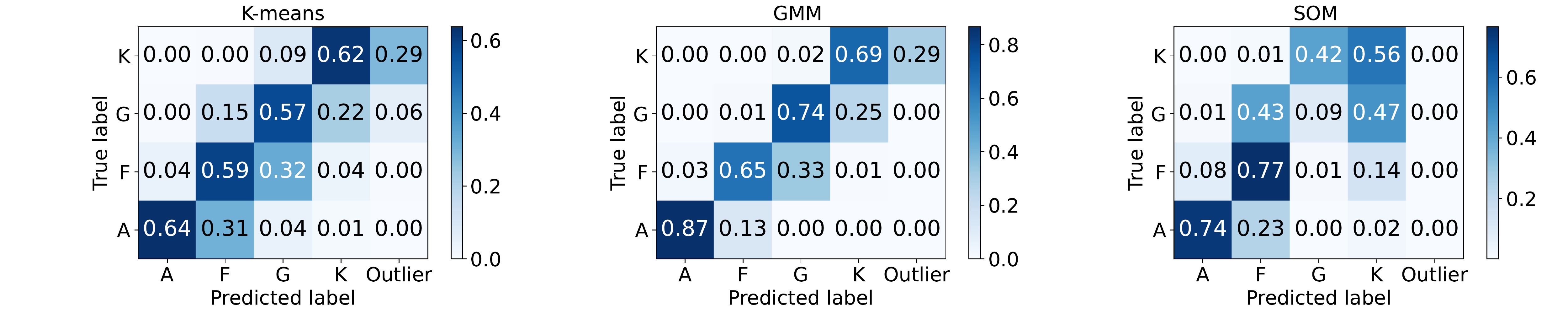}
\caption{K-means, GMM, SOM on data volume of 80000. Predicted label: color and digit in each cell are the consistent probability between predicted label and true label. Color is in direct proportion to the figure: bigger numbers and deeper color.}
\label{fig:Algorithms(kmeans_gmm_som)_2w_raw_10-_confusion_matrix} 
\end{figure*}

Fig. \ref{fig:Algorithms_2k_5k_1w_raw_10-_tsne_raw_label} shows the t-SNE distribution of 8000, 20000, 40000, 80000 of data. Some algorithms do not have the accuracy of size 80000 because they need a lot of memory space and 64GB is not enough for them to run.  Fig. \ref{fig:Algorithms_2k_raw_10-_tsne} - Fig. \ref{fig:Algorithms(kmeans_gmm_som)_2w_raw_10-_confusion_matrix} are the clustering results and confusion matrices and Fig. \ref{fig:Algorithms_accuracy_2k_5k_1w_5w_raw_10-_bar} shows their average accuracies.

Results show that data volume makes a little influence on accuracy of most algorithms in our experiments, and it can also be explained by t-SNE. CFSFDP performs worse with data volume increasing. Because it identifies cluster centers by neighborhood density of adjacent data rather than an iterative method. So data volume has a large impact on CFSFDP.

Although data volume is an important factor affecting the results for supervised classification tasks, it is not the case for clustering. And most clustering algorithms are not sensitive to data volume. Because clustering is to find the optimal division results on all data and it does not need a large amount of data to improve generalization. But some algorithms cannot run when the size of data is too large, because they need to calculate the distance matrix which requires a large memory space.

\subsection{Performance of algorithms on outliers detection}

In the big data era, it is crucial to find outliers. Although outliers sometimes have bad effect on the information we need, they also could provide some useful knowledge to many fields, e.g. credit card fraud, e-commerce crime, medical diagnosis, etc. And rare objects detection is also an attractive research in astronomy.
In addition to typical outlier methods, clustering is also an effective tool to detect outliers. 

In this section, we choose three types of special spectra to verify the ability of clustering methods to find these outliers, (a) physical outliers: carbon star; (b) overlapping sources: double star; (c) artefacts outliers: bad merging of red and blue segments. And we use two types of normal spectra, one containing only stars and the other one containing stars, galaxies and quasars. Put outliers into these normal spectra and observe the effect of detecting them by different algorithms. We also compare the ability of detecting outliers (carbon stars) from datasets on different spectra features, different qualities of spectra and different data volumes. Their data introductions are shown in Table \ref{table:rare_afgk_SGQ} and Table \ref{table:dataset_outlier}.

In this task, K-means, GMM, CFSFDP, DBSCAN, Local Outier Factor (LOF) are selected to detect outliers. LOF is a typical outlier detection algorithm, so it is  also included in our comparison. Various clustering algorithms have different ways to find outliers. In the experiments, K-means regards samples far away from most cluster centers as outliers. Samples with small probabilities of every Gaussian distribution can be outliers in GMM. Data in left-top corner of $\rho$-$\delta$, core graph in CFSFDP, are outliers.
DBSCAN can detect all outliers with small densities and group normal samples into some clusters. 
In LOF we calculate and rank outlier factor of each sample, Samples with large outlier factors are selected as outliers.

\begin{table}
\centering
\caption{Datasets of outlier detection on two types of normal objects}
\label{table:rare_afgk_SGQ}
\resizebox{\linewidth}{!}{
\begin{tabular}{lllll} 
\hline
                                      &  & Normal Spectra                    &  & Outliers          \\ 
\hline
                                      &  &                                    &  & Carbon stars:100  \\
\multicolumn{1}{c}{Dataset outlier 1} &  & A:F:G:K=2500:2500:2500:2500        &  & Double stars:100  \\
                                      &  &                                    &  & Artefacts:100      \\ 
\hline
                                      &  &                                    &  & Carbon stars:200  \\
\multicolumn{1}{c}{Dataset outlier 2} &  & Star:Galaxy:Quasar=10000:6500:3300 &  & Double stars:200  \\
                                      &  &                                    &  & Artefacts:200      \\
\hline
\end{tabular}
}
\end{table}

\begin{table*}

\caption{Nine datasets with three criteria to detect carbon stars.}
\label{table:dataset_outlier}
\resizebox{\linewidth}{!}{
\begin{tabular}{llcccc}
\hline
 & Dataset & Abnormal spectra size & Rare spectra size & S/N & Data type \\ \hline
\multirow{3}{*}{Different Spectral Characteristics} & Dataset ds & 20000 & 100 & \textgreater{}10 & 1D Spectra \\ 
 & Dataset dp & 20000 & 100 & \textgreater{}10 & PCA \\ 
 & Dataset dl & 20000 & 100 & \textgreater{}10 & Line Indices \\ \hline
\multirow{3}{*}{Different S/Ns} & Dataset rh & 20000 & 100 & \textgreater{}30 & PCA \\ 
 & Dataset rm & 20000 & 100 & 10-30 & PCA \\ 
 & Dataset rl & 20000 & 100 & \textless{}10 & PCA \\ \hline
\multirow{3}{*}{Different Data Volumes} & Dataset vs & 8000 & 40 & \textgreater{}10 & PCA \\ 
 & Dataset vm & 20000 & 100 & \textgreater{}10 & PCA \\ 
 & Dataset vl & 40000 & 200 & \textgreater{}10 & PCA \\ \hline
\end{tabular}
}
\end{table*}

We evaluate their performances by ROC curve (receiver operating characteristic curve, or sensitivity curve), which is a widely used index to measure the ability of outliers detection. The horizontal axis of the curve is false positive rate, it is the ratio of normal samples in detected targets to all normal samples. Vertical axis is true positive rate which is the ratio of true outliers in detected targets to all true outliers. The closer the curve is to the upper left corner of the graph, indicates the method has a stronger ability to detect outliers.

\begin{figure}
\centering
\includegraphics[width=\columnwidth]{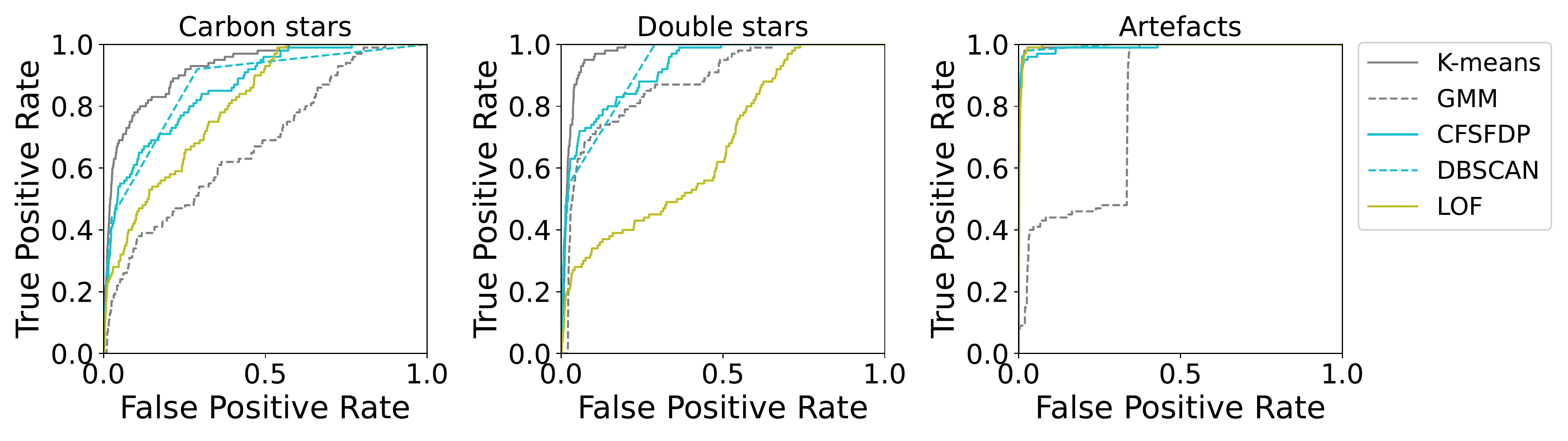}
\caption{ROC curves of detecting three types of outliers from stars. Different colors represents five algorithms. X axis: outlier detection FPR, Y axis: PR. Curves near left-top are prefer to find outliers. }
\label{fig:ROC_star}
\end{figure}

\begin{figure}
\centering
\includegraphics[width=\columnwidth]{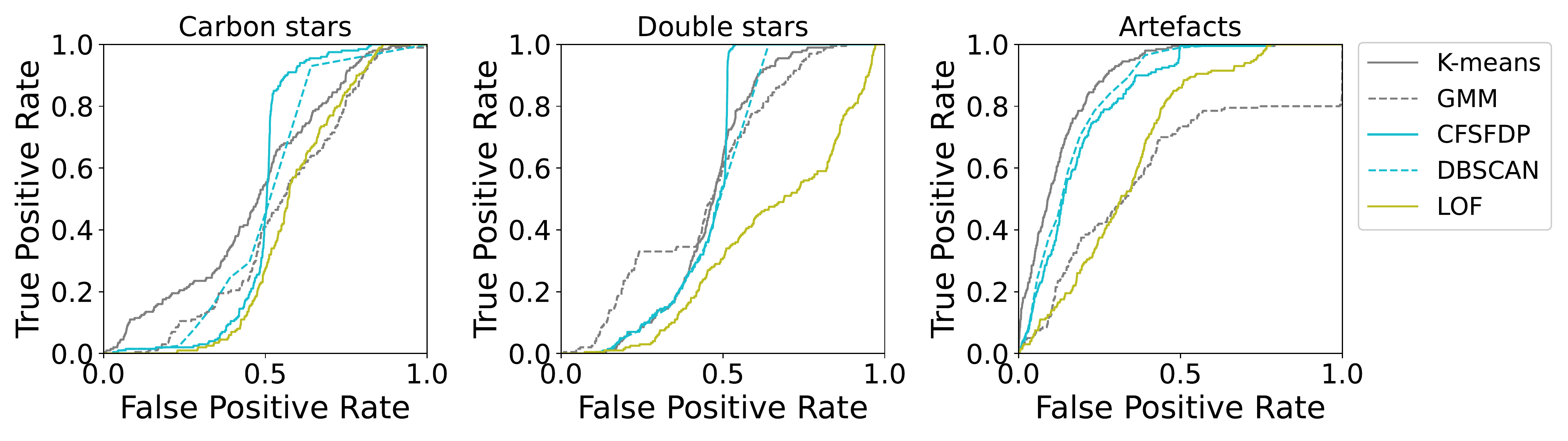}
\caption{ROC curves of detecting three types of outliers from all spectra (stars, galaxies and quasars). Different colors represents five algorithms. X axis: outlier detection FPR, Y axis: PR. Curves near left-top are prefer to find outliers. }
\label{fig:ROC_allgalaxy}
\end{figure}

 From the ROC curves of results (Fig. \ref{fig:ROC_star} and Fig. \ref{fig:ROC_allgalaxy}), we can know that compared with the classic outlier algorithm LOF, clustering algorithms also perform well in outlier detection tasks. It is easier to detect outliers in stellar spectra than in all spectra (including stars, galaxies, and quasars). The reason is that spectra of quasars are also unusual and very few in number, so they are often detected as outliers too. The performance of algorithms on detecting spectra with bad-merge in red and blue segment is better than carbon star and double star only except GMM. Some badly merged spectra have a wavelength range without flux values, and the Euclidean distance between them and normal spectra is far, so they can be detected easily. GMM uses several Gaussian distributions to fit data and does not measure the similarity between spectra, so its ability to detect badly merged spectra is poor.


\begin{figure}
\centering
\includegraphics[width=\columnwidth]{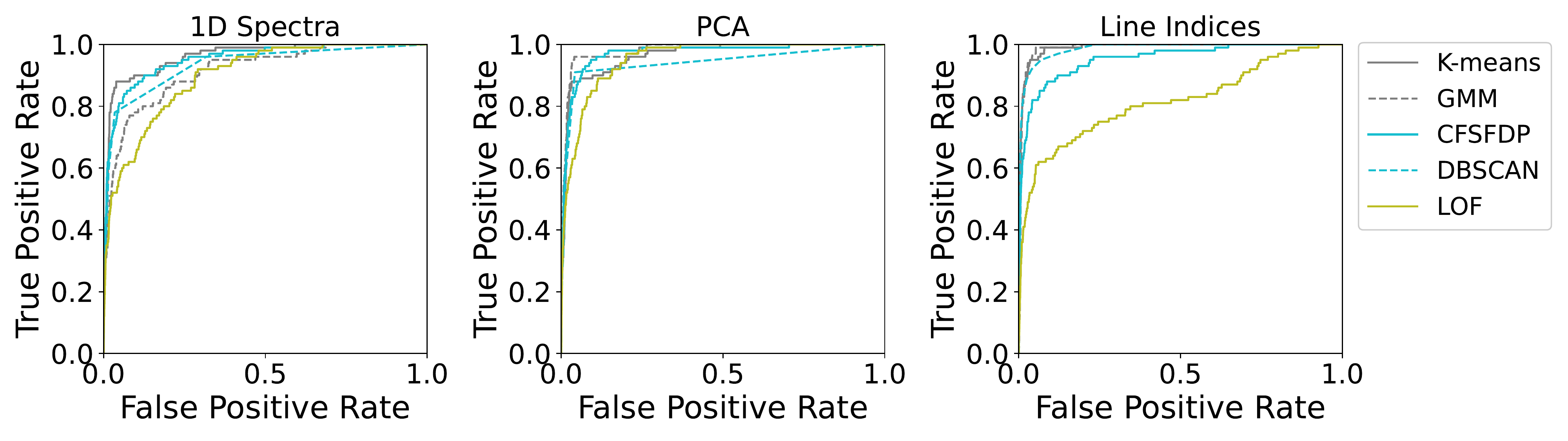}
\caption{ROC curves of detecting carbon stars from stars on different features. Different colors represents five algorithms. X axis: outlier detection FPR, Y axis: PR. Curves near left-top are prefer to find outliers.}
\label{fig:ROC_diff_feature}
\end{figure} 

Fig. \ref{fig:ROC_diff_feature} reveals desirable results of five algorithms on finding outliers on different spectra features. Algorithms perform worse on 1D spectra than on PCA features, because normal 1D spectra sometimes contain few incomplete spectra which are usually regarded as outliers.
In Fig. \ref{fig:ROC_diff_feature}, clustering algorithms can get more satisfactory consequences than LOF algorithm. It indicates that clustering methods also have good ability to outliers detection. Another interesting thing is that though DBSCAN has difficulty on clustering spectra, it is able to detect outliers. K-means and GMM select spectra far away from cluster centers as outliers and they both have good performances. In our experiments, we find that it is not a good idea to regard the samples in clusters with small number in the clustering results as outliers. By clustering methods, clusters or cluster centers are found, and samples farthest from the cluster centers are more likely to be outliers.

\begin{figure}
\centering
\includegraphics[width=\columnwidth]{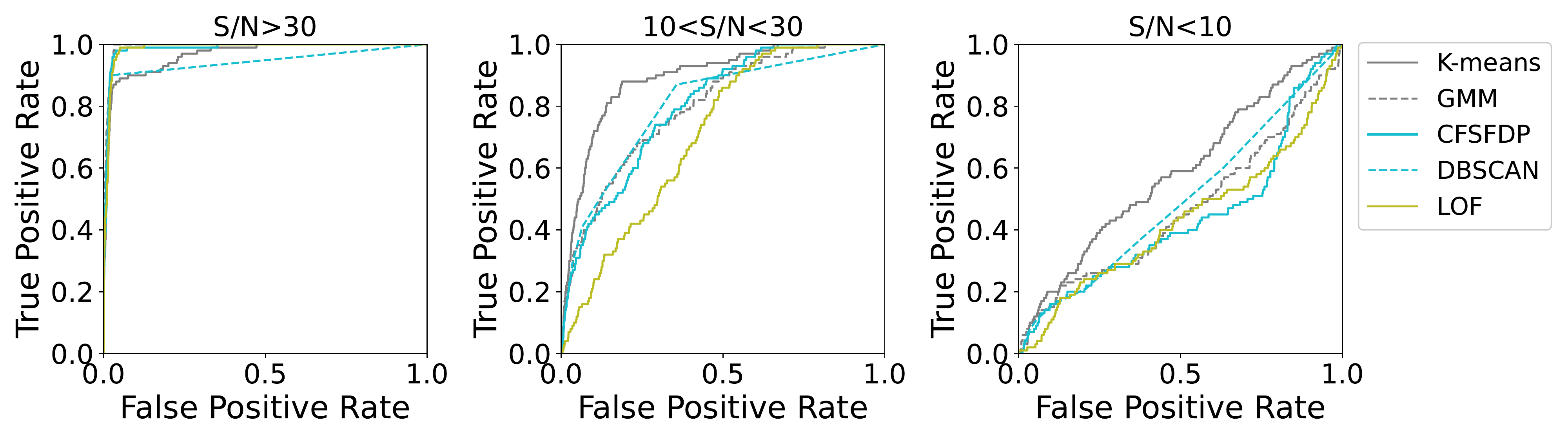}
\caption{ROC curves of detecting carbon stars from stars on different S/Ns. Different colors represents five algorithms. X axis: outlier detection FPR, Y axis: PR. Curves near left-top are prefer to find outliers.}
\label{fig:ROC_diff_snr}
\end{figure} 

 Fig. \ref{fig:ROC_diff_snr} reveals the effect of S/N on outlier detection, and the quality of spectra has a great impact on the results. In the high S/N spectra, each algorithm can find outliers well. But in the low S/N spectra, algorithms basically have no ability to find outliers. K-means and DBSCAN are inferior to others on S/N > 30. While K-means is superior among all algorithms on S/N < 10 and S/N:10-30 that shows a little robustness.

\begin{figure}
\centering
\includegraphics[width=\columnwidth]{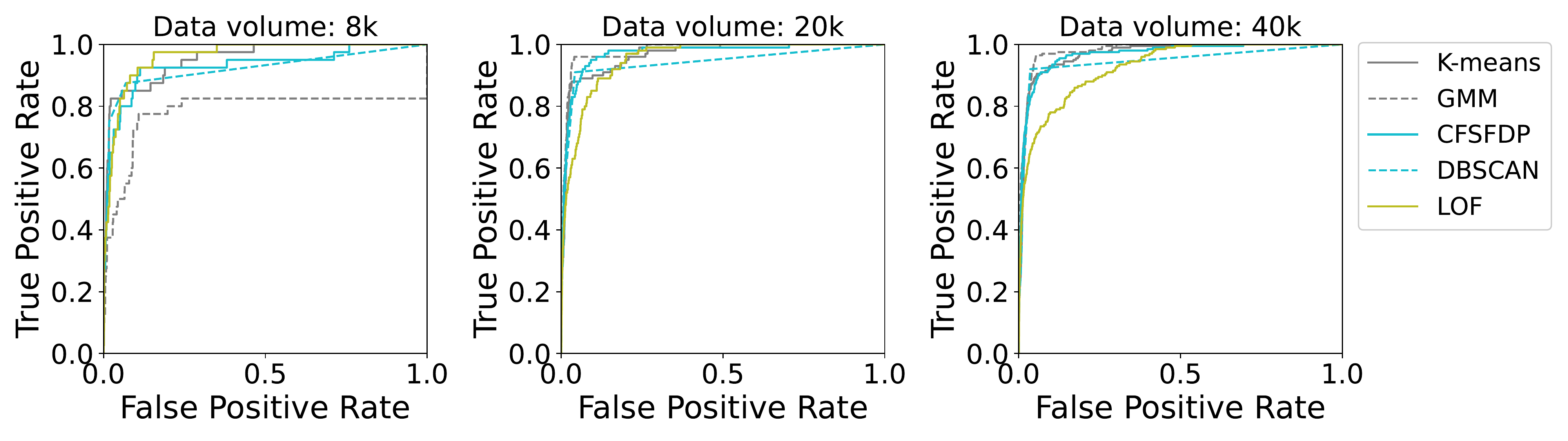}
\caption{ROC curves of detecting carbon stars from stars on different data volumes. Different colors represents five algorithms. X axis: outlier detection FPR, Y axis: PR. Curves near left-top are prefer to find outliers.}
\label{fig:ROC_diff_volume}
\end{figure} 
In the experiment of data volumes (Fig. \ref{fig:ROC_diff_volume}), the proportion of normal spectra and outliers is 1000:5. When the size of dataset is large, clustering algorithms are better than LOF, since there are more abnormal spectra in large dataset, and LOF will treat them as outliers.

\begin{figure*}
\centering
\includegraphics[width=17cm,height=17cm]{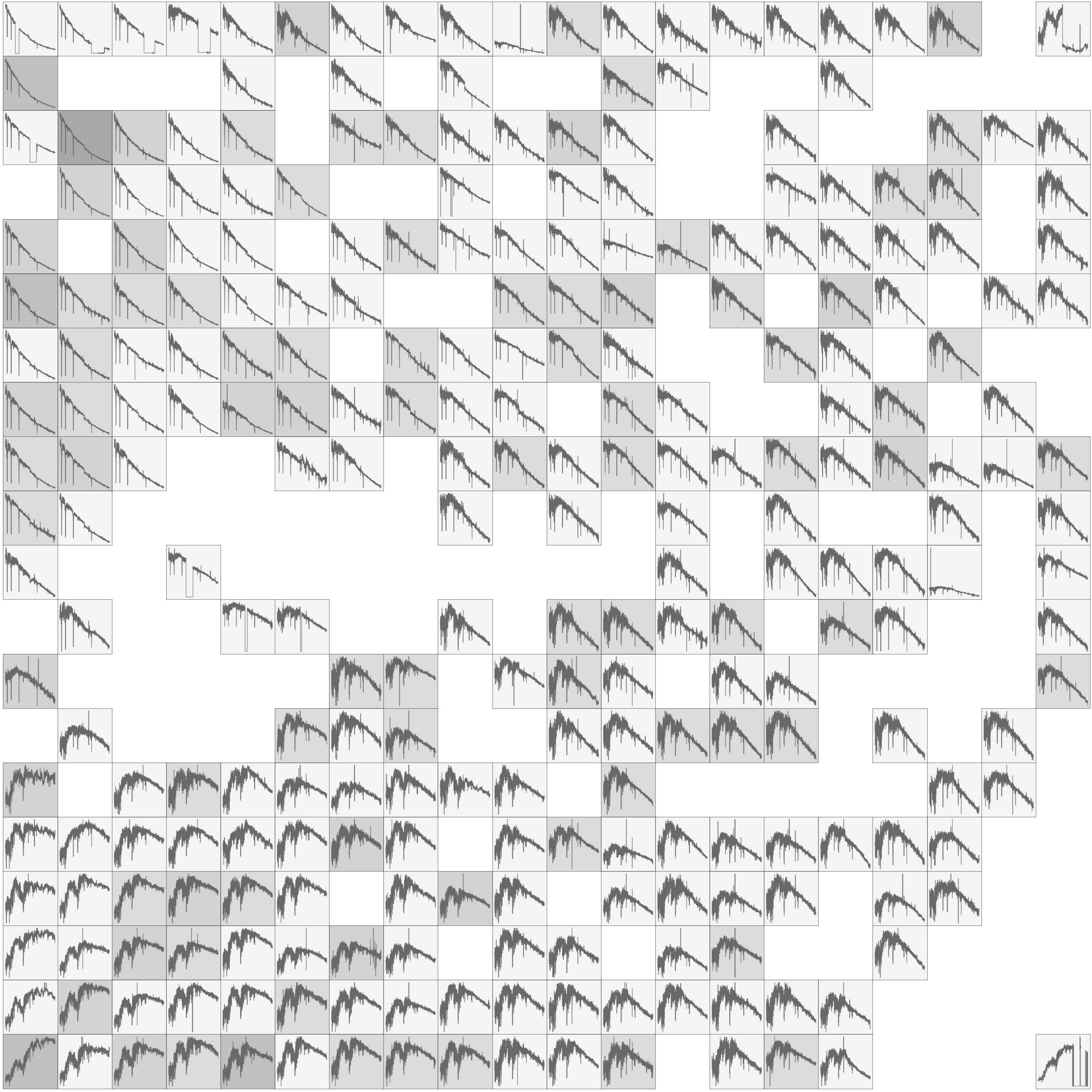}
\caption{Mapping 400 spectra of A, F, G, K stars to a 20 $\times$ 20 grid with  SOM. The darker color means more spectra mapped in this cell and the blank area means no spectra mapped here. Spectra separated by blank space are very different on shape.}
\label{fig:SOM_spectra}
\end{figure*}

\begin{figure}
\flushleft 
\includegraphics[width=9cm,height=8cm]{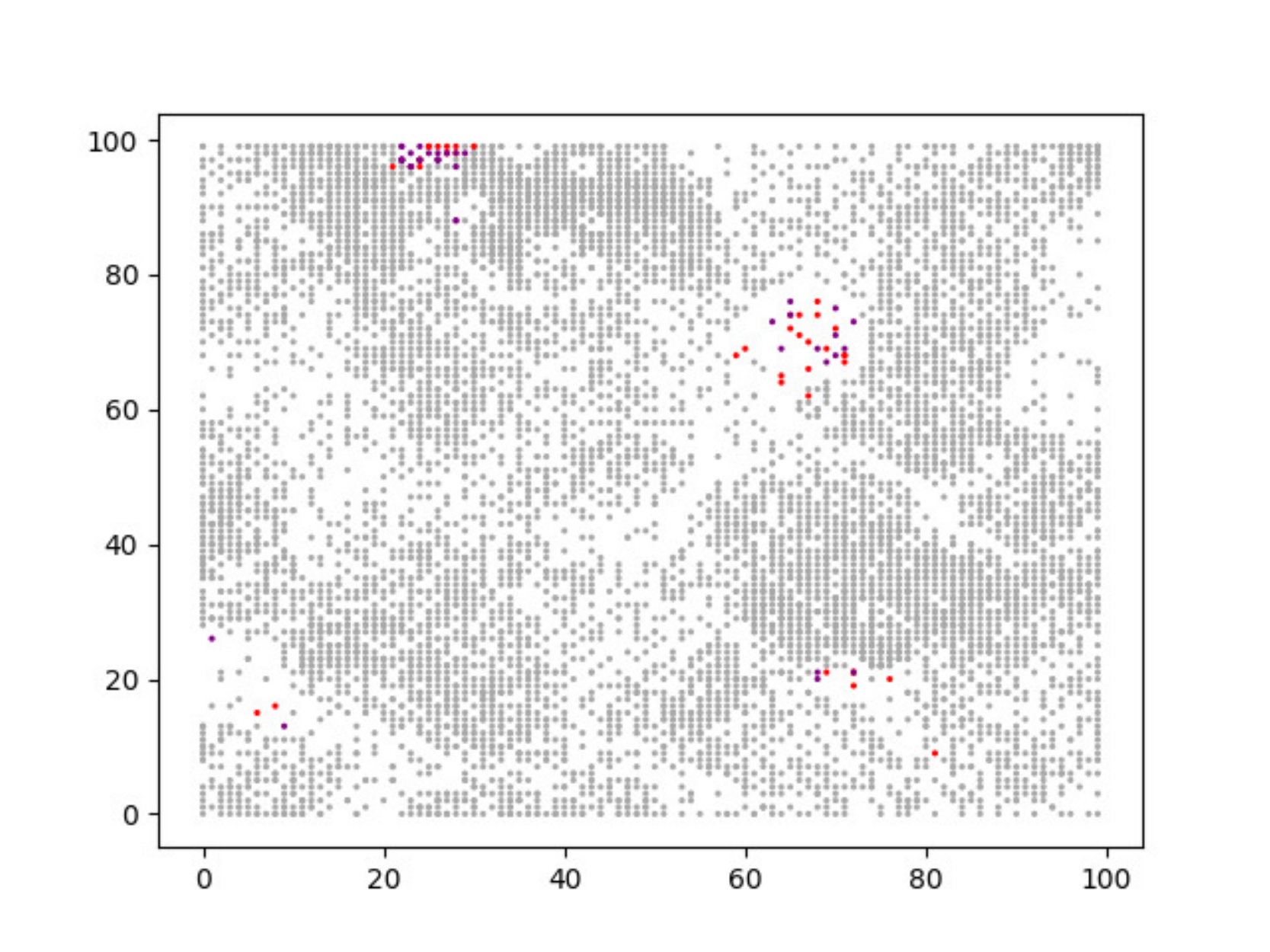}
\caption{Detect rare objects with SOM. Grey points are normal objects. Red points are  known carbon stars. Purple points are unknown carbon stars which are the target of this task. }
\label{fig:SOM_rare}
\end{figure}

Rare objects make significant sense to research in astronomy. To search some kinds of rare objects, we can use supervised classification methods. However, it is difficult to obtain a robust model due to the small number of rare samples. In this situation, clustering methods are good choices to deal with this task. And SOM is also a commonly used method, which can map similar spectra to the same or close locations on two-dimensional competing layer with topology structure. Candidates of rare objects can be adjacent spectra of known rare spectra in the map.

In Fig. \ref{fig:SOM_spectra}, 400 spectra including A, F, G and K stars are mapped on a complete layer of 20 $\times$ 20. Deeper background means more spectra are mapped on this position. Spectra with similar shapes are closer in the map and spectra separated by blank are more different than adjacent spectra. In this graph, A stars, F stars, G stars and K stars are located in different positions respectively. And there are gaps between anomalous spectra and normal spectra, so we also can use SOM to detect outliers.

Fig. \ref{fig:SOM_rare} shows the idea of searching carbon stars by SOM. In the figure, grey points represent normal objects while red points are known carbon stars and purple points are unknown carbon stars. We can recognize adjacent data of red points as rare object candidates and then do further validation manually.

\section{Source Code and Manual}

\begin{table}
\caption{Source codes of algorithms used in this paper.}
\label{table:code}
\centering
\resizebox{\linewidth}{!}{
\begin{tabular}{lllll}
\hline
Code Type &   & & Methods & \\ \hline
\multirow{2}{*}{Spectra processing} &  &  & PCA  & \\
 &  &  & Extract line indices  & \\ \hline
\multirow{8}{*}{Clustering algorithms} &  &  & K-means  & \\
 &  &  & K-mediod  & \\
 &  &  & K-means\_dp  & \\
 &  &  & GMM  & \\
 &  &  & SOM  & \\
 &  &  & CFSFDP  & \\
 &  &  & Hierarchical clustering & \\
 &  &  & DBSCAN & \\ \hline
Others &   & & t-SNE & \\ \hline
\end{tabular}
}
\end{table}

Source codes of clustering algorithms used in this paper is provided (\url{https://github.com/shichenhui/SpectraClustering}) and Table \ref{table:code} presents main codes list. The specific steps and precautions of code usage are also given in the above link. Clustering results of some algorithms will be influenced by the parameters so we adjust parameters and find optimal parameters. 

The code is written in Python which is widely used for machine
learning and data analysis. The dependent packages of the codes are: numpy \citep{harris2020array}, sklearn, matplotlib \citep{hunter2007matplotlib}, pandas, scipy. Each algorithm is organized in the following
steps: 1) load training datasets; 2) plot the t-SNE distribution of the data; 3) cluster data; 4) plot t-SNE distribution of results; 5) evaluate clustering results. 


These codes load data from *.csv files which store tabular data in text and a row of data is a spectrum. Users need to convert their spectra data to such a format or modify the data loading mode. Some of algorithms are implemented directly through the sklearn package. It is efficient to adopt best algorithms for astronomical spectral analysis.

The number of clusters K is the main parameter of partition-based methods and it has a great influence on the clustering results. Sometimes we can set the number of clusters K to be more than the actual number, because some rare objects or outliers will be classified as clusters. In our experiment, one or two more than actual number of clusters can be good results. If we have no idea of the number of clusters, t-SNE or UMAP method could be used to have a rough idea of how the data is aggregated. Euclidean distance is used in K-means to measure the similarity between samples and it can be replaced by other similarity measures, like Manhattan distance, distance in manifold space and others. This should be selected according to the characteristics of the data. K-mediod chooses a sample as cluster center which has the smallest sum of distances from all other samples in the cluster instead of an average point. This can reduce the impact of a small number of outliers on the cluster center. But the time complexity is higher than K-means. As for the selection of initial cluster centers, CFSFDP method is a good choice.

Neighborhood radius and the number of min-samples in a cluster are two main parameters in DBSCAN. Generally, it requires multiple tests to obtain optimal values. To obtain good clustering results, DBSCAN has two requirements for data, one is the degree of sample aggregation of different clusters cannot be very different, otherwise suitable parameter neighborhood radius will not be found. The other is that there needs to be  obvious separation between clusters, so that the clusters are not grouped together. DBSCAN performs bad on spectra because these two conditions are not satisfied. However, in the tasks of analysing spatial structure, DBSCAN can get good results.

SOM needs to set the number of iterations, usually hundreds of times to achieve convergence results. It can map data into 1D or 2D space. 1D space is convenient for analyzing the physical properties and the input data must be a small number of relevant features. But for clustering, the results in 1D space are less stable. 2D space can effectively get the data distribution for clustering or outliers detection. The shape of grids mapped with SOM could be rectangular or hexagonal, and hexagonal grids can represent more detailed distribution information. An important parameter in GMM is covariance-type which describes the type of covariance, four types could be chosen, "spherical", "tied", "diag" and "full". "full" is the most flexible type meaning that each cluster can have its own arbitrary shape. But it may not give the best results in real applications. The choice of covariance-type requires experience or testing to obtain optimal results. Our experimental results found that the "tied" and "diag" were better than "spherical" and "full".

Agglomerative and divisive are two models of hierarchical clustering. In our experiment, hierarchical clustering could not work well but it can be combined with other algorithms. CFSFDP calculates two values ($\delta$ and $\rho$) for each sample to select cluster centers and outliers.  And we can use them to cluster or detect outliers with decision graph method.


\section{Discussion}
\label{sect:discussion}
Clustering methods used in recent astronomical spectroscopic studies are investigated in this paper and they are briefly introduced. Considering the different data used in respective research, it is difficult to find the strength of each method. So we construct some unified spectral datasets to analyse their advantages under different conditions. Some clustering methods can also be used to find outliers, so datasets also include normal spectra and rare spectra to test their performance on searching for outliers.

Through the experiments, we found that GMM performs better than others on 1D spectra and PCA features. On stellar spectra line indices, GMM performs as well as partition-based methods. Spectra line indices can extract stellar spectra features effectively and the clustering results of many methods on line indices are better than 1D spectra. Density-based algorithms and hierarchical clustering perform poorly on spectra related datasets, although they have many advantages on benchmark datasets. The reason is that, in spectra dataset, there is no clear separation between different types of spectra and the density distribution of clusters may be different. So it is impossible to find appropriate parameters for efficient clustering.

Although for supervised classification algorithms, overfitting can be reduced when the dataset is large. But for clustering, the amount of data has little influence on the clustering results. And some algorithms can not run on the large datasets. GMM works well on spectra, but its running time on large datasets is much higher than other methods. K-means is still a good choice if researchers want to make a fast clustering of the data.

The experiments also show that clustering methods are very effective to find abnormal spectra. Multiple cluster centers need to be found first, then the samples far from the cluster centers can be regarded as outliers and these methods are very robust. When researchers want to observe the distribution of spectra data, dimensionality reduction and visualization methods are very intuitive, such as t-SNE, UMAP and SOM. And SOM is also widely used in astronomy to find special spectra.

The purpose of this paper is to provide an analysis of the advantages and disadvantages of different algorithmic ideas. The methods we use are all basic algorithms and the data preprocessing is also general steps. Improved algorithms and better preprocessing will be more effective to improve clustering performance, such as extinction processing, feature extraction, etc.

\section*{Acknowledgements}

We would like to thank the reviewer, Igor Chilingarian, for his valuable comments and suggestions to improve the paper.

The Guo Shou Jing Telescope (the Large Sky Area Multi-Object Fiber Spectroscopic Telescope, LAMOST) is a National Major Scientific Project built by the Chinese Academy of Sciences. Funding for the project has been provided by the National Development and Reform Commission. LAMOST
is operated and managed by National Astronomical Observatories, Chinese Academy of Sciences.

The work is supported by the National Natural Science Foundation of China (Grant Nos. U1931209), Key Research and Development Projects of Shanxi Province (Grant Nos. 201903D121116), and the central government guides local science and technology development funds (Grant No. 20201070). Fundamental Research Program of Shanxi Province(Grant No. 20210302123223, 202103021224275).


\section*{Data Availability}


Experimental data used for this study is obtained from The Guo Shou Jing Telescope (the Large Sky Area Multi-Object Fiber Spectroscopic Telescope, LAMOST) Data Release 8 (\url{http://www.lamost.org/lmusers/}) and Sloan Digital Sky Survey (SDSS) Data Release 16 (\url{https://www.sdss.org/dr16/}). Codes used in this paper is also available online at (\url{https://www.github.com/shichenhui/SpectraClustering}).



\bibliographystyle{mnras}
\bibliography{example} 

\begin{thebibliography}{}
\makeatletter
\relax
\def\mn@urlcharsother{\let\do\@makeother \do\$\do\&\do\#\do\^\do\_\do\%\do\~}
\def\mn@doi{\begingroup\mn@urlcharsother \@ifnextchar [ {\mn@doi@}
  {\mn@doi@[]}}
\def\mn@doi@[#1]#2{\def\@tempa{#1}\ifx\@tempa\@empty \href
  {http://dx.doi.org/#2} {doi:#2}\else \href {http://dx.doi.org/#2} {#1}\fi
  \endgroup}
\def\mn@eprint#1#2{\mn@eprint@#1:#2::\@nil}
\def\mn@eprint@arXiv#1{\href {http://arxiv.org/abs/#1} {{\tt arXiv:#1}}}
\def\mn@eprint@dblp#1{\href {http://dblp.uni-trier.de/rec/bibtex/#1.xml}
  {dblp:#1}}
\def\mn@eprint@#1:#2:#3:#4\@nil{\def\@tempa {#1}\def\@tempb {#2}\def\@tempc
  {#3}\ifx \@tempc \@empty \let \@tempc \@tempb \let \@tempb \@tempa \fi \ifx
  \@tempb \@empty \def\@tempb {arXiv}\fi \@ifundefined
  {mn@eprint@\@tempb}{\@tempb:\@tempc}{\expandafter \expandafter \csname
  mn@eprint@\@tempb\endcsname \expandafter{\@tempc}}}

\bibitem[\protect\citeauthoryear{{Acuner} \& {Ryde}}{{Acuner} \&
  {Ryde}}{2018}]{7.2018MNRAS.475.1708A}
{Acuner} Z.,  {Ryde} F.,  2018, \mn@doi [\mnras] {10.1093/mnras/stx3106}, \href
  {https://ui.adsabs.harvard.edu/abs/2018MNRAS.475.1708A} {475, 1708}

\bibitem[\protect\citeauthoryear{{A.in der Au,}, {Meusinger}, {Schalldach}  \&
  {Newholm}}{{A.in der Au,} et~al.}{2012}]{2012A&A...547A.115I}
{A.in der Au,} {Meusinger} H.,  {Schalldach} P.~F.,   {Newholm} M.,  2012,
  \mn@doi [\aap] {10.1051/0004-6361/201219958}, \href
  {https://ui.adsabs.harvard.edu/abs/2012A&A...547A.115I} {547, A115}

\bibitem[\protect\citeauthoryear{{Armstrong}, {Brown}, {Chadwick}  \&
  {Nolan}}{{Armstrong} et~al.}{2015}]{2015MNRAS.452.3159A}
{Armstrong} T.,  {Brown} A.~M.,  {Chadwick} P.~M.,   {Nolan} S.~J.,  2015,
  \mn@doi [\mnras] {10.1093/mnras/stv1398}, \href
  {https://ui.adsabs.harvard.edu/abs/2015MNRAS.452.3159A} {452, 3159}

\bibitem[\protect\citeauthoryear{Baker et~al.}{Baker et~al.}{2010}]{2010Data}
Baker R.,  et~al., 2010, International encyclopedia of education, 7, 112

\bibitem[\protect\citeauthoryear{{Balazs}, {Garibjanyan}, {Mirzoyan},
  {Hambaryan}, {Kun}, {Fronto}  \& {Kelemen}}{{Balazs}
  et~al.}{1996}]{1996A&A...311..145B}
{Balazs} L.~G.,  {Garibjanyan} A.~T.,  {Mirzoyan} L.~V.,  {Hambaryan} V.~V.,
  {Kun} M.,  {Fronto} A.,   {Kelemen} J.,  1996, \aap, \href
  {https://ui.adsabs.harvard.edu/abs/1996A&A...311..145B} {311, 145}

\bibitem[\protect\citeauthoryear{Bazarghan}{Bazarghan}{2011}]{article_application}
Bazarghan M.,  2011, \mn@doi [Astrophysics and Space Science]
  {10.1007/s10509-011-0822-7}, 337

\bibitem[\protect\citeauthoryear{{Beck}, {Dobos}, {Yip}, {Szalay}  \&
  {Csabai}}{{Beck} et~al.}{2016}]{2016MNRAS.457..362B}
{Beck} R.,  {Dobos} L.,  {Yip} C.-W.,  {Szalay} A.~S.,   {Csabai} I.,  2016,
  \mn@doi [\mnras] {10.1093/mnras/stv2986}, \href
  {https://ui.adsabs.harvard.edu/abs/2016MNRAS.457..362B} {457, 362}

\bibitem[\protect\citeauthoryear{Berry \& Linoff}{Berry \&
  Linoff}{1997}]{1997Data}
Berry M.,  Linoff G.~S.,  1997, Wiley Publishing

\bibitem[\protect\citeauthoryear{{Blanco-Cuaresma} et~al.,}{{Blanco-Cuaresma}
  et~al.}{2015}]{2015A&A...577A..47B}
{Blanco-Cuaresma} S.,  et~al., 2015, \mn@doi [\aap]
  {10.1051/0004-6361/201425232}, \href
  {https://ui.adsabs.harvard.edu/abs/2015A&A...577A..47B} {577, A47}

\bibitem[\protect\citeauthoryear{{Bu}, {Zhao}, {Pan}  \& {Bharat Kumar}}{{Bu}
  et~al.}{2016}]{bu2016elm}
{Bu} Y.,  {Zhao} G.,  {Pan} J.,   {Bharat Kumar} Y.,  2016, \mn@doi [\apj]
  {10.3847/0004-637X/817/1/78}, \href
  {https://ui.adsabs.harvard.edu/abs/2016ApJ...817...78B} {817, 78}

\bibitem[\protect\citeauthoryear{Cai-Xia, Hai-feng, Jiang-hui  \&
  Ya-ling}{Cai-Xia et~al.}{2020}]{Li.18.P-Cygni}
Cai-Xia Q.,  Hai-feng Y.,  Jiang-hui C.,   Ya-ling X.,  2020, Spectroscopy and
  Spectral Analysis, 40, 1304

\bibitem[\protect\citeauthoryear{{Cai}, {Li}  \& {Yang}}{{Cai}
  et~al.}{2020}]{2020JApA...41...15C}
{Cai} J.,  {Li} Y.,   {Yang} H.,  2020, \mn@doi [Journal of Astrophysics and
  Astronomy] {10.1007/s12036-020-09634-x}, \href
  {https://ui.adsabs.harvard.edu/abs/2020JApA...41...15C} {41, 15}

\bibitem[\protect\citeauthoryear{Cai, Yang, Yang, Zhao  \& Hao}{Cai
  et~al.}{2022}]{10.1145/3522592}
Cai J.,  Yang Y.,  Yang H.,  Zhao X.,   Hao J.,  2022, \mn@doi [ACM Trans.
  Knowl. Discov. Data] {10.1145/3522592}, 16, 1

\bibitem[\protect\citeauthoryear{{Carlson}, {Linden}, {Profumo}  \&
  {Weniger}}{{Carlson} et~al.}{2013}]{2013PhRvD..88d3006C}
{Carlson} E.,  {Linden} T.,  {Profumo} S.,   {Weniger} C.,  2013, \mn@doi
  [\prd] {10.1103/PhysRevD.88.043006}, \href
  {https://ui.adsabs.harvard.edu/abs/2013PhRvD..88d3006C} {88, 043006}

\bibitem[\protect\citeauthoryear{{Castro-Ginard} et~al.,}{{Castro-Ginard}
  et~al.}{2022}]{Castro_Ginard_2022}
{Castro-Ginard} A.,  et~al., 2022, \mn@doi [\aap]
  {10.1051/0004-6361/202142568}, \href
  {https://ui.adsabs.harvard.edu/abs/2022A&A...661A.118C} {661, A118}

\bibitem[\protect\citeauthoryear{{Chattopadhyay} \& {Maitra}}{{Chattopadhyay}
  \& {Maitra}}{2017}]{2017MNRAS.469.3374C}
{Chattopadhyay} S.,  {Maitra} R.,  2017, \mn@doi [\mnras]
  {10.1093/mnras/stx1024}, \href
  {https://ui.adsabs.harvard.edu/abs/2017MNRAS.469.3374C} {469, 3374}

\bibitem[\protect\citeauthoryear{{Chattopadhyay}, {Sharina}, {Davoust}, {De}
  \& {Chattopadhyay}}{{Chattopadhyay} et~al.}{2012}]{2012ApJ...750...91C}
{Chattopadhyay} T.,  {Sharina} M.,  {Davoust} E.,  {De} T.,   {Chattopadhyay}
  A.~K.,  2012, \mn@doi [\apj] {10.1088/0004-637X/750/2/91}, \href
  {https://ui.adsabs.harvard.edu/abs/2012ApJ...750...91C} {750, 91}

\bibitem[\protect\citeauthoryear{{Chen}, {Sun}  \& {Yan}}{{Chen}
  et~al.}{2018a}]{2018RAA....18...73C}
{Chen} S.-X.,  {Sun} W.-M.,   {Yan} Q.,  2018a, \mn@doi [Research in Astronomy
  and Astrophysics] {10.1088/1674-4527/18/6/73}, \href
  {https://ui.adsabs.harvard.edu/abs/2018RAA....18...73C} {18, 073}

\bibitem[\protect\citeauthoryear{{Chen}, {D'Onghia}, {Pardy}, {Pasquali},
  {Bertelli Motta}, {Hanlon}  \& {Grebel}}{{Chen}
  et~al.}{2018b}]{2018ApJ...860...70C}
{Chen} B.,  {D'Onghia} E.,  {Pardy} S.~A.,  {Pasquali} A.,  {Bertelli Motta}
  C.,  {Hanlon} B.,   {Grebel} E.~K.,  2018b, \mn@doi [\apj]
  {10.3847/1538-4357/aac325}, \href
  {https://ui.adsabs.harvard.edu/abs/2018ApJ...860...70C} {860, 70}

\bibitem[\protect\citeauthoryear{Connell \& Jain}{Connell \&
  Jain}{1998}]{Connell98learningprototypes}
Connell S.~D.,  Jain A.~K.,  1998, in In Proceedings of the 14th International
  Conference on Pattern Recognition. pp 182--184

\bibitem[\protect\citeauthoryear{Couillet \& Benaych-Georges}{Couillet \&
  Benaych-Georges}{2016}]{8.2015arXiv151003547C}
Couillet R.,  Benaych-Georges F.,  2016, \mn@doi [Electronic Journal of
  Statistics] {10.1214/16-EJS1144}, 10, 1393

\bibitem[\protect\citeauthoryear{Dehghan \& Johnston-Hollitt}{Dehghan \&
  Johnston-Hollitt}{2014}]{Dehghan_2014}
Dehghan S.,  Johnston-Hollitt M.,  2014, \mn@doi [\aj]
  {10.1088/0004-6256/147/3/52}, 147, 52

\bibitem[\protect\citeauthoryear{Deng \& Tu}{Deng \& Tu}{2017}]{DengT17}
Deng S.,  Tu L.,  2017, in 2017 10th International Congress on Image and Signal
  Processing, BioMedical Engineering and Informatics (CISP-BMEI). pp~1--6,
  \mn@doi{10.1109/CISP-BMEI.2017.8302164}

\bibitem[\protect\citeauthoryear{Dorai \& Jain}{Dorai \&
  Jain}{1995}]{10.5555/839284.841388}
Dorai C.,  Jain A.,  1995, in Proceedings., International Conference on Image
  Processing. pp 340--343 vol.3, \mn@doi{10.1109/ICIP.1995.538548}

\bibitem[\protect\citeauthoryear{Du, Wang, Ren, Zhong  \& Gao}{Du
  et~al.}{2016}]{7820430}
Du H.,  Wang Y.,  Ren C.,  Zhong J.,   Gao X.,  2016, in 2016 12th
  International Conference on Computational Intelligence and Security (CIS). pp
  134--137, \mn@doi{10.1109/CIS.2016.0039}

\bibitem[\protect\citeauthoryear{{Duarte-Cabral} et~al.,}{{Duarte-Cabral}
  et~al.}{2021}]{2021MNRAS.500.3027D}
{Duarte-Cabral} A.,  et~al., 2021, \mn@doi [\mnras] {10.1093/mnras/staa2480},
  \href {https://ui.adsabs.harvard.edu/abs/2021MNRAS.500.3027D} {500, 3027}

\bibitem[\protect\citeauthoryear{Durbin, Eddy, Krogh  \& Mitchison}{Durbin
  et~al.}{1998}]{3ac78af0492d473c8e1492890c612e8e}
Durbin R.,  Eddy S.,  Krogh A.,   Mitchison G.,  1998, Biological sequence
  analysis: Probabilistic models of proteins and nucleic acids.
Cambridge University Press, United Kingdom

\bibitem[\protect\citeauthoryear{Everton}{Everton}{2012}]{everton_2012}
Everton S.~F.,  2012, Social Network Analysis: An Introduction.
Cambridge University Press, p. 3–31, \mn@doi{10.1017/CBO9781139136877.003}

\bibitem[\protect\citeauthoryear{Fielding, Nyirenda  \& Vaccari}{Fielding
  et~al.}{2022}]{fielding2022classification}
Fielding E.,  Nyirenda C.~N.,   Vaccari M.,  2022, arXiv preprint
  arXiv:2206.06165

\bibitem[\protect\citeauthoryear{Forsyth \& Ponce}{Forsyth \&
  Ponce}{2011}]{forsyth:hal-01063327}
Forsyth D.,  Ponce J.,  2011, {Computer Vision: A Modern Approach. (Second
  edition)}.
{Prentice Hall}, \url {https://hal.inria.fr/hal-01063327}

\bibitem[\protect\citeauthoryear{Fotheringham, Charlton  \&
  Brunsdon}{Fotheringham et~al.}{1998}]{fotheringham1998geographically}
Fotheringham A.~S.,  Charlton M.,   Brunsdon C.,  1998, Environment and
  Planning A, 30, 1905

\bibitem[\protect\citeauthoryear{{Fraix-Burnet}, {Chattopadhyay},
  {Chattopadhyay}, {Davoust}  \& {Thuillard}}{{Fraix-Burnet}
  et~al.}{2012}]{2012A&A...545A..80F}
{Fraix-Burnet} D.,  {Chattopadhyay} T.,  {Chattopadhyay} A.~K.,  {Davoust} E.,
   {Thuillard} M.,  2012, \mn@doi [\aap] {10.1051/0004-6361/201218769}, \href
  {https://ui.adsabs.harvard.edu/abs/2012A&A...545A..80F} {545, A80}

\bibitem[\protect\citeauthoryear{Fustes, Dafonte, Varela, Manteiga, Smith,
  Vallenari  \& Luri}{Fustes et~al.}{2013}]{som.article.outlier}
Fustes D.,  Dafonte C.,  Varela B.,  Manteiga M.,  Smith K.,  Vallenari A.,
  Luri X.,  2013, \mn@doi [Expert Syst. Appl.] {10.1016/j.eswa.2012.08.069},
  40, 1530

\bibitem[\protect\citeauthoryear{{Gao}}{{Gao}}{2014}]{2014Membership}
{Gao} X.-H.,  2014, \mn@doi [Research in Astronomy and Astrophysics]
  {10.1088/1674-4527/14/2/004}, \href
  {https://ui.adsabs.harvard.edu/abs/2014RAA....14..159G} {14, 159}

\bibitem[\protect\citeauthoryear{{Gao}}{{Gao}}{2020}]{2.gao2020discovery}
{Gao} X.,  2020, \mn@doi [\apj] {10.3847/1538-4357/ab8560}, \href
  {https://ui.adsabs.harvard.edu/abs/2020ApJ...894...48G} {894, 48}

\bibitem[\protect\citeauthoryear{{Gao}, {Xu}  \& {Chen}}{{Gao}
  et~al.}{2015}]{2015RAA....15.2193G}
{Gao} X.-H.,  {Xu} S.-K.,   {Chen} L.,  2015, \mn@doi [Research in Astronomy
  and Astrophysics] {10.1088/1674-4527/15/12/007}, \href
  {https://ui.adsabs.harvard.edu/abs/2015RAA....15.2193G} {15, 2193}

\bibitem[\protect\citeauthoryear{{Garcia-Dias}, {Allende Prieto}, {S{\'a}nchez
  Almeida}  \& {Ordov{\'a}s-Pascual}}{{Garcia-Dias}
  et~al.}{2018}]{2018A&A...612A..98G}
{Garcia-Dias} R.,  {Allende Prieto} C.,  {S{\'a}nchez Almeida} J.,
  {Ordov{\'a}s-Pascual} I.,  2018, \mn@doi [\aap]
  {10.1051/0004-6361/201732134}, \href
  {https://ui.adsabs.harvard.edu/abs/2018A&A...612A..98G} {612, A98}

\bibitem[\protect\citeauthoryear{{Garcia-Dias}, {Allende Prieto}, {S{\'a}nchez
  Almeida}  \& {Alonso Palicio}}{{Garcia-Dias}
  et~al.}{2019}]{chemical_abundance}
{Garcia-Dias} R.,  {Allende Prieto} C.,  {S{\'a}nchez Almeida} J.,   {Alonso
  Palicio} P.,  2019, \mn@doi [\aap] {10.1051/0004-6361/201935223}, \href
  {https://ui.adsabs.harvard.edu/abs/2019A&A...629A..34G} {629, A34}

\bibitem[\protect\citeauthoryear{Gowanlock, Blair  \& Pankratius}{Gowanlock
  et~al.}{2017}]{2017Optimizing}
Gowanlock M.,  Blair D.,   Pankratius V.,  2017, IEEE Transactions on Parallel
  \& Distributed Systems, pp 2595--2607

\bibitem[\protect\citeauthoryear{Guha, Rastogi  \& Shim}{Guha
  et~al.}{1998}]{10.1145/276304.276312}
Guha S.,  Rastogi R.,   Shim K.,  1998, in Proceedings of the 1998 ACM SIGMOD
  International Conference on Management of Data. SIGMOD '98.
Association for Computing Machinery, New York, NY, USA, p. 73–84,
  \mn@doi{10.1145/276304.276312}

\bibitem[\protect\citeauthoryear{Harris et~al.,}{Harris
  et~al.}{2020}]{harris2020array}
Harris C.~R.,  et~al., 2020, Nature, 585, 357

\bibitem[\protect\citeauthoryear{Hayes et~al.,}{Hayes
  et~al.}{2020}]{10.1093/mnras/staa978}
Hayes J. J.~C.,  et~al., 2020, \mn@doi [\mnras] {10.1093/mnras/staa978}, 494,
  4492

\bibitem[\protect\citeauthoryear{Hogg et~al.,}{Hogg et~al.}{2016}]{Hogg_2016}
Hogg D.~W.,  et~al., 2016, \mn@doi [\apj] {10.3847/1538-4357/833/2/262}, 833,
  262

\bibitem[\protect\citeauthoryear{Hunter}{Hunter}{2007}]{hunter2007matplotlib}
Hunter J.~D.,  2007, Computing in science \& engineering, 9, 90

\bibitem[\protect\citeauthoryear{{Iwasaki}, {Ichinohe}  \&
  {Uchiyama}}{{Iwasaki} et~al.}{2019}]{2019MNRAS.488.4106I}
{Iwasaki} H.,  {Ichinohe} Y.,   {Uchiyama} Y.,  2019, \mn@doi [\mnras]
  {10.1093/mnras/stz1990}, \href
  {https://ui.adsabs.harvard.edu/abs/2019MNRAS.488.4106I} {488, 4106}

\bibitem[\protect\citeauthoryear{{Jin}, {Yang}  \& {Chiang}}{{Jin}
  et~al.}{2022}]{Jin_2022}
{Jin} Y.,  {Yang} L.,   {Chiang} C.-E.,  2022, arXiv e-prints, \href
  {https://ui.adsabs.harvard.edu/abs/2022arXiv220400721J} {p. arXiv:2204.00721}

\bibitem[\protect\citeauthoryear{Kaplan \& Winther}{Kaplan \&
  Winther}{2013}]{Kaplan2013-KAPPOA}
Kaplan J.~M.,  Winther R.~G.,  2013, \mn@doi [Biological Theory]
  {10.1007/s13752-012-0048-0}, 7, 401

\bibitem[\protect\citeauthoryear{Karypis, Han  \& Kumar}{Karypis
  et~al.}{1999}]{781637}
Karypis G.,  Han E.-H.,   Kumar V.,  1999, \mn@doi [Computer]
  {10.1109/2.781637}, 32, 68

\bibitem[\protect\citeauthoryear{Kheirdastan \& Bazarghan}{Kheirdastan \&
  Bazarghan}{2016}]{Kheirdastan_2016}
Kheirdastan S.,  Bazarghan M.,  2016, \mn@doi [Astrophysics and Space Science]
  {10.1007/s10509-016-2880-3}, 361, 304

\bibitem[\protect\citeauthoryear{Kiang, Hu, Fisher  \& Chi}{Kiang
  et~al.}{2005}]{1385384}
Kiang M.,  Hu M.,  Fisher D.,   Chi R.,  2005, in Proceedings of the 38th
  Annual Hawaii International Conference on System Sciences. pp 73b--73b,
  \mn@doi{10.1109/HICSS.2005.590}

\bibitem[\protect\citeauthoryear{Kiar, Barmby  \& Hidalgo}{Kiar
  et~al.}{2017}]{M82017}
Kiar A.~K.,  Barmby P.,   Hidalgo A.,  2017, \mn@doi [\mnras]
  {10.1093/mnras/stx2037}, 472, 1074–1087

\bibitem[\protect\citeauthoryear{{Kuhn} \& {Feigelson}}{{Kuhn} \&
  {Feigelson}}{2017}]{10.2017arXiv171111101K}
{Kuhn} M.~A.,  {Feigelson} E.~D.,  2017, arXiv e-prints, \href
  {https://ui.adsabs.harvard.edu/abs/2017arXiv171111101K} {p. arXiv:1711.11101}

\bibitem[\protect\citeauthoryear{Lam \& Wunsch}{Lam \&
  Wunsch}{2014}]{LAM20141115}
Lam D.,  Wunsch D.~C.,  2014, in Diniz P.~S.,  Suykens J.~A.,  Chellappa R.,
  Theodoridis S.,  eds, Academic Press Library in Signal Processing, Vol.~1,
  Academic Press Library in Signal Processing: Volume 1.
Elsevier, pp 1115--1149, \mn@doi{10.1016/B978-0-12-396502-8.00020-6}

\bibitem[\protect\citeauthoryear{{Li}, {Cai}, {Yang}, {Zhang}  \& {Zhao}}{{Li}
  et~al.}{2019}]{8732318}
{Li} Y.,  {Cai} J.,  {Yang} H.,  {Zhang} J.,   {Zhao} X.,  2019, \mn@doi [IEEE
  Access] {10.1109/ACCESS.2019.2921320}, 7, 74683

\bibitem[\protect\citeauthoryear{Liang, Cai  \& Yang}{Liang
  et~al.}{2022}]{LIANG2022116410}
Liang B.,  Cai J.,   Yang H.,  2022, \mn@doi [Expert Systems with Applications]
  {10.1016/j.eswa.2021.116410}, 193, 116410

\bibitem[\protect\citeauthoryear{{Logan} \& {Fotopoulou}}{{Logan} \&
  {Fotopoulou}}{2020}]{2020A&A...633A.154L}
{Logan} C.~H.~A.,  {Fotopoulou} S.,  2020, \mn@doi [\aap]
  {10.1051/0004-6361/201936648}, \href
  {https://ui.adsabs.harvard.edu/abs/2020A&A...633A.154L} {633, A154}

\bibitem[\protect\citeauthoryear{{Luo}, {Zhang}  \& {Zhao}}{{Luo}
  et~al.}{2004}]{2004SPIE.5496..756L}
{Luo} A.~L.,  {Zhang} Y.-X.,   {Zhao} Y.-H.,  2004, in {Lewis} H.,  {Raffi} G.,
   eds,  Society of Photo-Optical Instrumentation Engineers (SPIE) Conference
  Series Vol. 5496, Advanced Software, Control, and Communication Systems for
  Astronomy. pp 756--764, \mn@doi{10.1117/12.548737}

\bibitem[\protect\citeauthoryear{{Luo}, {Wu}, {Zhao}  \& {Zhao}}{{Luo}
  et~al.}{2008}]{2008SPIE.7019E..35L}
{Luo} A.~L.,  {Wu} Y.,  {Zhao} J.,   {Zhao} G.,  2008, in {Bridger} A.,
  {Radziwill} N.~M.,  eds,  Society of Photo-Optical Instrumentation Engineers
  (SPIE) Conference Series Vol. 7019, Advanced Software and Control for
  Astronomy II. p. 701935, \mn@doi{10.1117/12.788251}

\bibitem[\protect\citeauthoryear{Luo et~al.,}{Luo
  et~al.}{2013}]{2014IAUS..298..428L}
Luo A.,  et~al., 2013, \mn@doi [Proceedings of the International Astronomical
  Union] {10.1017/S1743921313006947}, 9, 428–428

\bibitem[\protect\citeauthoryear{{Luo} et~al.,}{{Luo}
  et~al.}{2015}]{2015RAA....15.1095L}
{Luo} A.~L.,  et~al., 2015, \mn@doi [Research in Astronomy and Astrophysics]
  {10.1088/1674-4527/15/8/002}, \href
  {https://ui.adsabs.harvard.edu/abs/2015RAA....15.1095L} {15, 1095}

\bibitem[\protect\citeauthoryear{Madhusudan, Amrutha  \& Venugopal}{Madhusudan
  et~al.}{2017}]{8282521}
Madhusudan P.,  Amrutha A.,   Venugopal K.,  2017, in 2017 International
  Conference on Computing Methodologies and Communication (ICCMC). pp 526--529,
  \mn@doi{10.1109/ICCMC.2017.8282521}

\bibitem[\protect\citeauthoryear{Mahajan, Singh  \& Shobhana}{Mahajan
  et~al.}{2018}]{10.1093/mnras/sty1370}
Mahajan S.,  Singh A.,   Shobhana D.,  2018, \mn@doi [\mnras]
  {10.1093/mnras/sty1370}, 478, 4336

\bibitem[\protect\citeauthoryear{McInnes, Healy  \& Melville}{McInnes
  et~al.}{2018}]{1.umap}
McInnes L.,  Healy J.,   Melville J.,  2018, arXiv preprint arXiv:1802.03426

\bibitem[\protect\citeauthoryear{Merényi, Taylor  \& Isella}{Merényi
  et~al.}{2016}]{7849952}
Merényi E.,  Taylor J.,   Isella A.,  2016, in 2016 IEEE Symposium Series on
  Computational Intelligence (SSCI). pp~1--9,
  \mn@doi{10.1109/SSCI.2016.7849952}

\bibitem[\protect\citeauthoryear{{Meusinger}, {Br{\"u}necke}, {Schalldach}  \&
  {in der Au}}{{Meusinger} et~al.}{2017}]{2017A&A...597A.134M}
{Meusinger} H.,  {Br{\"u}necke} J.,  {Schalldach} P.,   {in der Au} A.,  2017,
  \mn@doi [\aap] {10.1051/0004-6361/201629139}, \href
  {https://ui.adsabs.harvard.edu/abs/2017A&A...597A.134M} {597, A134}

\bibitem[\protect\citeauthoryear{{Morales-Luis}, {S{\'a}nchez Almeida},
  {Aguerri}  \& {Mu{\~n}oz-Tu{\~n}{\'o}n}}{{Morales-Luis}
  et~al.}{2011}]{2011ApJ...743...77M}
{Morales-Luis} A.~B.,  {S{\'a}nchez Almeida} J.,  {Aguerri} J.~A.~L.,
  {Mu{\~n}oz-Tu{\~n}{\'o}n} C.,  2011, \mn@doi [\apj]
  {10.1088/0004-637X/743/1/77}, \href
  {https://ui.adsabs.harvard.edu/abs/2011ApJ...743...77M} {743, 77}

\bibitem[\protect\citeauthoryear{{Mosby}, {Tremonti}, {Hooper}, {Wolf},
  {Sheinis}  \& {Richards}}{{Mosby} et~al.}{2015}]{2015MNRAS.447.1638M}
{Mosby} G.,  {Tremonti} C.~A.,  {Hooper} E.~J.,  {Wolf} M.~J.,  {Sheinis}
  A.~I.,   {Richards} J.~W.,  2015, \mn@doi [\mnras] {10.1093/mnras/stu2531},
  \href {https://ui.adsabs.harvard.edu/abs/2015MNRAS.447.1638M} {447, 1638}

\bibitem[\protect\citeauthoryear{Ng, Krishnan  \& McLachlan}{Ng
  et~al.}{2012}]{ng2012algorithm}
Ng S.~K.,  Krishnan T.,   McLachlan G.~J.,  2012, in , Handbook of
  computational statistics.
Springer, pp 139--172

\bibitem[\protect\citeauthoryear{{Oliver}, {Elahi}, {Lewis}  \&
  {Power}}{{Oliver} et~al.}{2021}]{2021MNRAS.501.4420O}
{Oliver} W.~H.,  {Elahi} P.~J.,  {Lewis} G.~F.,   {Power} C.,  2021, \mn@doi
  [\mnras] {10.1093/mnras/staa3879}, \href
  {https://ui.adsabs.harvard.edu/abs/2021MNRAS.501.4420O} {501, 4420}

\bibitem[\protect\citeauthoryear{Openshaw, Upton  \& Fingleton}{Openshaw
  et~al.}{1985}]{1985sdae.book.....U}
Openshaw S.,  Upton G. J.~G.,   Fingleton B.,  1985, Journal of Ecology, 74,
  313

\bibitem[\protect\citeauthoryear{Ordonez \& Stripling}{Ordonez \&
  Stripling}{2022}]{Ordonez_2022}
Ordonez J.,  Stripling C.,  2022, \mn@doi [Research Notes of the {AAS}]
  {10.3847/2515-5172/ac6b41}, 6, 90

\bibitem[\protect\citeauthoryear{{Ordov{\'a}s-Pascual} \& {S{\'a}nchez
  Almeida}}{{Ordov{\'a}s-Pascual} \& {S{\'a}nchez
  Almeida}}{2014}]{2014A&A...565A..53O}
{Ordov{\'a}s-Pascual} I.,  {S{\'a}nchez Almeida} J.,  2014, \mn@doi [\aap]
  {10.1051/0004-6361/201423806}, \href
  {https://ui.adsabs.harvard.edu/abs/2014A&A...565A..53O} {565, A53}

\bibitem[\protect\citeauthoryear{Ordóñez, Dafonte, Arcay  \&
  Manteiga}{Ordóñez et~al.}{2012}]{ORDONEZ2012204}
Ordóñez D.,  Dafonte C.,  Arcay B.,   Manteiga M.,  2012, \mn@doi [Applied
  Soft Computing] {https://doi.org/10.1016/j.asoc.2011.08.052}, 12, 204

\bibitem[\protect\citeauthoryear{{Panos}, {Kleint}, {Huwyler}, {Krucker},
  {Melchior}, {Ullmann}  \& {Voloshynovskiy}}{{Panos}
  et~al.}{2018}]{2018ApJ...861...62P}
{Panos} B.,  {Kleint} L.,  {Huwyler} C.,  {Krucker} S.,  {Melchior} M.,
  {Ullmann} D.,   {Voloshynovskiy} S.,  2018, \mn@doi [\apj]
  {10.3847/1538-4357/aac779}, \href
  {https://ui.adsabs.harvard.edu/abs/2018ApJ...861...62P} {861, 62}

\bibitem[\protect\citeauthoryear{Price-Jones \& Bovy}{Price-Jones \&
  Bovy}{2019}]{10.1093/mnras/stz1260}
Price-Jones N.,  Bovy J.,  2019, \mn@doi [\mnras] {10.1093/mnras/stz1260}, 487,
  871

\bibitem[\protect\citeauthoryear{{Price-Jones} et~al.,}{{Price-Jones}
  et~al.}{2020}]{2020MNRAS.496.5101P}
{Price-Jones} N.,  et~al., 2020, \mn@doi [\mnras] {10.1093/mnras/staa1905},
  \href {https://ui.adsabs.harvard.edu/abs/2020MNRAS.496.5101P} {496, 5101}

\bibitem[\protect\citeauthoryear{{Rahmani}, {Teimoorinia}  \&
  {Barmby}}{{Rahmani} et~al.}{2018}]{Rahmani_2018}
{Rahmani} S.,  {Teimoorinia} H.,   {Barmby} P.,  2018, \mn@doi [\mnras]
  {10.1093/mnras/sty1291}, \href
  {https://ui.adsabs.harvard.edu/abs/2018MNRAS.478.4416R} {478, 4416}

\bibitem[\protect\citeauthoryear{Rebbapragada, Protopapas, Brodley  \&
  Alcock}{Rebbapragada et~al.}{2008}]{Rebbapragada_2008}
Rebbapragada U.,  Protopapas P.,  Brodley C.~E.,   Alcock C.,  2008, \mn@doi
  [Machine Learning] {10.1007/s10994-008-5093-3}, 74, 281

\bibitem[\protect\citeauthoryear{Reynolds}{Reynolds}{2009}]{reynolds2009gaussian}
Reynolds D.~A.,  2009, Encyclopedia of biometrics, 741

\bibitem[\protect\citeauthoryear{Ricci, Rokach  \& Shapira}{Ricci
  et~al.}{2011}]{2011Introduction}
Ricci F.,  Rokach L.,   Shapira B.,  2011, Introduction to Recommender Systems
  Handbook.
Springer US, Boston, MA, pp 1--35

\bibitem[\protect\citeauthoryear{Rubin \& Gal-Yam}{Rubin \&
  Gal-Yam}{2016}]{Rubin_2016}
Rubin A.,  Gal-Yam A.,  2016, \mn@doi [\apj] {10.3847/0004-637x/828/2/111},
  828, 111

\bibitem[\protect\citeauthoryear{{S{\'a}nchez Almeida} \& {Allende
  Prieto}}{{S{\'a}nchez Almeida} \& {Allende
  Prieto}}{2013}]{2013ApJ...763...50S}
{S{\'a}nchez Almeida} J.,  {Allende Prieto} C.,  2013, \mn@doi [\apj]
  {10.1088/0004-637X/763/1/50}, \href
  {https://ui.adsabs.harvard.edu/abs/2013ApJ...763...50S} {763, 50}

\bibitem[\protect\citeauthoryear{{S{\'a}nchez Almeida} \& {Lites}}{{S{\'a}nchez
  Almeida} \& {Lites}}{2000}]{2000ApJ...532.1215S}
{S{\'a}nchez Almeida} J.,  {Lites} B.~W.,  2000, \mn@doi [\apj]
  {10.1086/308603}, \href
  {https://ui.adsabs.harvard.edu/abs/2000ApJ...532.1215S} {532, 1215}

\bibitem[\protect\citeauthoryear{{S{\'a}nchez Almeida}, {Aguerri},
  {Mu{\~n}oz-Tu{\~n}{\'o}n}  \& {de Vicente}}{{S{\'a}nchez Almeida}
  et~al.}{2010}]{2010ApJ...714..487S}
{S{\'a}nchez Almeida} J.,  {Aguerri} J.~A.~L.,  {Mu{\~n}oz-Tu{\~n}{\'o}n} C.,
  {de Vicente} A.,  2010, \mn@doi [\apj] {10.1088/0004-637X/714/1/487}, \href
  {https://ui.adsabs.harvard.edu/abs/2010ApJ...714..487S} {714, 487}

\bibitem[\protect\citeauthoryear{{S{\'a}nchez Almeida}, {Terlevich},
  {Terlevich}, {Cid Fernandes}  \& {Morales-Luis}}{{S{\'a}nchez Almeida}
  et~al.}{2012}]{2012ApJ...756..163S}
{S{\'a}nchez Almeida} J.,  {Terlevich} R.,  {Terlevich} E.,  {Cid Fernandes}
  R.,   {Morales-Luis} A.~B.,  2012, \mn@doi [\apj]
  {10.1088/0004-637X/756/2/163}, \href
  {https://ui.adsabs.harvard.edu/abs/2012ApJ...756..163S} {756, 163}

\bibitem[\protect\citeauthoryear{{Sans Fuentes}, {De Ridder}  \&
  {Debosscher}}{{Sans Fuentes} et~al.}{2017}]{2017A&A...599A.143S}
{Sans Fuentes} S.~A.,  {De Ridder} J.,   {Debosscher} J.,  2017, \mn@doi [\aap]
  {10.1051/0004-6361/201629719}, \href
  {https://ui.adsabs.harvard.edu/abs/2017A&A...599A.143S} {599, A143}

\bibitem[\protect\citeauthoryear{Sasdelli et~al.,}{Sasdelli
  et~al.}{2016}]{10.1093/mnras/stw1228}
Sasdelli M.,  et~al., 2016, \mn@doi [\mnras] {10.1093/mnras/stw1228}, 461, 2044

\bibitem[\protect\citeauthoryear{Saxena et~al.,}{Saxena
  et~al.}{2017}]{SAXENA2017664}
Saxena A.,  et~al., 2017, \mn@doi [Neurocomputing]
  {https://doi.org/10.1016/j.neucom.2017.06.053}, 267, 664

\bibitem[\protect\citeauthoryear{Seo, Goldsmith, Tolls, Shipman, Kulesa,
  Peters, Walker  \& Melnick}{Seo et~al.}{2020}]{seo2020applications}
Seo Y.~M.,  Goldsmith P.~F.,  Tolls V.,  Shipman R.,  Kulesa C.,  Peters W.,
  Walker C.,   Melnick G.,  2020, Applications of Machine Learning Algorithms
  In Processing Terahertz Spectroscopic Data (\mn@eprint {arXiv} {2009.01203})

\bibitem[\protect\citeauthoryear{{Shang} \& {Oh}}{{Shang} \&
  {Oh}}{2012}]{4.2012MNRAS.426.3435S}
{Shang} C.,  {Oh} S.~P.,  2012, \mn@doi [\mnras]
  {10.1111/j.1365-2966.2012.21897.x}, \href
  {https://ui.adsabs.harvard.edu/abs/2012MNRAS.426.3435S} {426, 3435}

\bibitem[\protect\citeauthoryear{Shin, Chang, Yi, Kim, Kim  \& Byun}{Shin
  et~al.}{2018}]{shin2018detecting}
Shin M.-S.,  Chang S.-W.,  Yi H.,  Kim D.-W.,  Kim M.-J.,   Byun Y.-I.,  2018,
  The Astronomical Journal, 156, 201

\bibitem[\protect\citeauthoryear{Shuxin \& Weimin}{Shuxin \&
  Weimin}{2017}]{7984705}
Shuxin C.,  Weimin S.,  2017, in 2017 2nd International Conference on Image,
  Vision and Computing (ICIVC). pp 1002--1005,
  \mn@doi{10.1109/ICIVC.2017.7984705}

\bibitem[\protect\citeauthoryear{Siemens \& Baker}{Siemens \&
  Baker}{2012}]{2012Learning}
Siemens G.,  Baker R. S. J.~d.,  2012, in Proceedings of the 2nd International
  Conference on Learning Analytics and Knowledge. LAK '12.
Association for Computing Machinery, New York, NY, USA, p. 252–254,
  \mn@doi{10.1145/2330601.2330661}

\bibitem[\protect\citeauthoryear{{Simpson}, {Cottrell}  \& {Worley}}{{Simpson}
  et~al.}{2012}]{2012MNRAS.427.1153S}
{Simpson} J.~D.,  {Cottrell} P.~L.,   {Worley} C.~C.,  2012, \mn@doi [\mnras]
  {10.1111/j.1365-2966.2012.22012.x}, \href
  {https://ui.adsabs.harvard.edu/abs/2012MNRAS.427.1153S} {427, 1153}

\bibitem[\protect\citeauthoryear{Tahmasebi, Hezarkhani  \& Sahimi}{Tahmasebi
  et~al.}{2012}]{2012Multiple}
Tahmasebi P.,  Hezarkhani A.,   Sahimi M.,  2012, Computational Geosciences,
  16, 779

\bibitem[\protect\citeauthoryear{{Tammour}, {Gallagher}, {Daley}  \&
  {Richards}}{{Tammour} et~al.}{2016}]{2016MNRAS.459.1659T}
{Tammour} A.,  {Gallagher} S.~C.,  {Daley} M.,   {Richards} G.~T.,  2016,
  \mn@doi [\mnras] {10.1093/mnras/stw586}, \href
  {https://ui.adsabs.harvard.edu/abs/2016MNRAS.459.1659T} {459, 1659}

\bibitem[\protect\citeauthoryear{{Tarricq}, {Soubiran}, {Casamiquela},
  {Castro-Ginard}, {Olivares}, {Miret-Roig}  \& {Galli}}{{Tarricq}
  et~al.}{2022}]{Tarricq_2022}
{Tarricq} Y.,  {Soubiran} C.,  {Casamiquela} L.,  {Castro-Ginard} A.,
  {Olivares} J.,  {Miret-Roig} N.,   {Galli} P.~A.~B.,  2022, \mn@doi [\aap]
  {10.1051/0004-6361/202142186}, \href
  {https://ui.adsabs.harvard.edu/abs/2022A&A...659A..59T} {659, A59}

\bibitem[\protect\citeauthoryear{{T{\'o}th}, {R{\'a}cz}  \&
  {Horv{\'a}th}}{{T{\'o}th} et~al.}{2019}]{6.2019MNRAS.486.4823T}
{T{\'o}th} B.~G.,  {R{\'a}cz} I.~I.,   {Horv{\'a}th} I.,  2019, \mn@doi
  [\mnras] {10.1093/mnras/stz1188}, \href
  {https://ui.adsabs.harvard.edu/abs/2019MNRAS.486.4823T} {486, 4823}

\bibitem[\protect\citeauthoryear{{Tramacere}, {Paraficz}, {Dubath}, {Kneib}  \&
  {Courbin}}{{Tramacere} et~al.}{2016}]{2016MNRAS.463.2939T}
{Tramacere} A.,  {Paraficz} D.,  {Dubath} P.,  {Kneib} J.~P.,   {Courbin} F.,
  2016, \mn@doi [\mnras] {10.1093/mnras/stw2103}, \href
  {https://ui.adsabs.harvard.edu/abs/2016MNRAS.463.2939T} {463, 2939}

\bibitem[\protect\citeauthoryear{Traven et~al.,}{Traven
  et~al.}{2017}]{Traven_2017}
Traven G.,  et~al., 2017, \mn@doi [The Astrophysical Journal Supplement Series]
  {10.3847/1538-4365/228/2/24}, 228, 24

\bibitem[\protect\citeauthoryear{Van~der Maaten \& Hinton}{Van~der Maaten \&
  Hinton}{2008}]{van2008visualizing}
Van~der Maaten L.,  Hinton G.,  2008, Journal of machine learning research, 9

\bibitem[\protect\citeauthoryear{{Wagenveld}, {Saxena}, {Duncan},
  {R{\"o}ttgering}  \& {Zhang}}{{Wagenveld} et~al.}{2022}]{Wagenveld_2022}
{Wagenveld} J.~D.,  {Saxena} A.,  {Duncan} K.~J.,  {R{\"o}ttgering} H.~J.~A.,
  {Zhang} M.,  2022, \mn@doi [\aap] {10.1051/0004-6361/202142445}, \href
  {https://ui.adsabs.harvard.edu/abs/2022A&A...660A..22W} {660, A22}

\bibitem[\protect\citeauthoryear{Wang, Guo  \& Luo}{Wang
  et~al.}{2015}]{10.1109/BigData.2015.7363804}
Wang K.,  Guo P.,   Luo A.-L.,  2015, in Proceedings of the 2015 IEEE
  International Conference on Big Data (Big Data). BIG DATA '15.
IEEE Computer Society, USA, p. 601–608

\bibitem[\protect\citeauthoryear{Wattenberg, Vi{\'e}gas  \& Johnson}{Wattenberg
  et~al.}{2016}]{wattenberg2016use}
Wattenberg M.,  Vi{\'e}gas F.,   Johnson I.,  2016, Distill, 1, e2

\bibitem[\protect\citeauthoryear{Wu, Pan, Yi  \& Wei}{Wu
  et~al.}{2020}]{9049419}
Wu M.,  Pan J.,  Yi Z.,   Wei P.,  2020, \mn@doi [IEEE Access]
  {10.1109/ACCESS.2020.2983745}, 8, 66475

\bibitem[\protect\citeauthoryear{Xu \& Tian}{Xu \& Tian}{2015}]{2015A}
Xu D.,  Tian Y.,  2015, Annals of Data Science, 2, 165

\bibitem[\protect\citeauthoryear{{Yan}, {Yang}, {Su}, {Sun}  \& {Wang}}{{Yan}
  et~al.}{2020}]{2020ApJ...898...80Y}
{Yan} Q.-Z.,  {Yang} J.,  {Su} Y.,  {Sun} Y.,   {Wang} C.,  2020, \mn@doi
  [\apj] {10.3847/1538-4357/ab9f9c}, \href
  {https://ui.adsabs.harvard.edu/abs/2020ApJ...898...80Y} {898, 80}

\bibitem[\protect\citeauthoryear{Yang, Cai, Yang, Zhang  \& Zhao}{Yang
  et~al.}{2020}]{YANG2020112846}
Yang Y.,  Cai J.,  Yang H.,  Zhang J.,   Zhao X.,  2020, \mn@doi [Expert
  Systems with Applications] {https://doi.org/10.1016/j.eswa.2019.112846}, 139,
  112846

\bibitem[\protect\citeauthoryear{Yang, Cai, Yang, Li  \& Zhao}{Yang
  et~al.}{2022a}]{YANG2022117018}
Yang Y.,  Cai J.,  Yang H.,  Li Y.,   Zhao X.,  2022a, \mn@doi [Expert Systems
  with Applications] {10.1016/j.eswa.2022.117018}, 201, 117018

\bibitem[\protect\citeauthoryear{Yang, Cai, Yang  \& Zhao}{Yang
  et~al.}{2022b}]{YANG2022414}
Yang Y.,  Cai J.,  Yang H.,   Zhao X.,  2022b, \mn@doi [Information Sciences]
  {https://doi.org/10.1016/j.ins.2022.03.027}, 596, 414

\bibitem[\protect\citeauthoryear{{Zari}, {Brown}  \& {de Zeeuw}}{{Zari}
  et~al.}{2019}]{2019A&A...628A.123Z}
{Zari} E.,  {Brown} A.~G.~A.,   {de Zeeuw} P.~T.,  2019, \mn@doi [\aap]
  {10.1051/0004-6361/201935781}, \href
  {https://ui.adsabs.harvard.edu/abs/2019A&A...628A.123Z} {628, A123}

\bibitem[\protect\citeauthoryear{Zhang, Ramakrishnan  \& Livny}{Zhang
  et~al.}{1996}]{10.1145/233269.233324}
Zhang T.,  Ramakrishnan R.,   Livny M.,  1996, in Proceedings of the 1996 ACM
  SIGMOD International Conference on Management of Data. SIGMOD '96.
Association for Computing Machinery, New York, NY, USA, p. 103–114

\makeatother
\end{thebibliography}

\bsp	
\label{lastpage}
\end{document}